\documentclass{aa}  

\usepackage{graphicx}
\usepackage{txfonts}
\usepackage{lipsum}
\usepackage{subcaption} 
\usepackage{tablefootnote}
\usepackage{lscape}             
\usepackage{placeins}           

\usepackage{multirow}
\usepackage{makecell}
\usepackage{amsmath}
\usepackage{stfloats}

\usepackage[pdftex, colorlinks=true, linkcolor=blue, citecolor=blue, urlcolor=blue]{hyperref}

\begin{document}



    \title{
A half-ring of ionized circumstellar material trapped in the magnetosphere of a white dwarf merger remnant}

\titlerunning{ZTF\,J2008+4449 - A merger remnant with a half-ring of circumstellar material}

\authorrunning{Cristea A. A. et al.}


    \subtitle{A new class of white dwarf merger remnants with X-ray emission}


   \author{
        Andrei A. Cristea\inst{1}
        \and Ilaria Caiazzo\inst{1,2}
        \and Tim Cunningham\inst{3}
        \and John C. Raymond\inst{3}
        \and Stephane Vennes\inst{4}
        \and Adela Kawka\inst{5}
        \and Aayush Desai\inst{1}
        \and David R. Miller\inst{6}
        \and J. J. Hermes\inst{7}
        \and Jim Fuller\inst{8}
        \and Jeremy Heyl\inst{6}
        \and Jan van Roestel\inst{9}
        \and Kevin B. Burdge\inst{10}
        \and Antonio C. Rodriguez\inst{2}
        \and Ingrid Pelisoli\inst{11}
        \and Boris T. G\"ansicke\inst{11,12} 
        \and Paula Szkody\inst{13}
        \and Scott J. Kenyon\inst{3}
        \and Zach Vanderbosch\inst{2}
        \and Andrew Drake\inst{2}
        \and Lilia Ferrario\inst{4}
        \and Dayal Wickramasinghe\inst{4}
        \and Viraj R. Karambelkar\inst{2}
        \and Stephen Justham\inst{14}
        \and Ruediger Pakmor\inst{14}
        \and Kareem El-Badry\inst{2}
        \and Thomas Prince\inst{2}
        \and S. R. Kulkarni\inst{2}
        \and Matthew J. Graham\inst{2}
        \and Frank J. Masci\inst{15}
        \and Steven L. Groom\inst{15}
        \and Josiah Purdum\inst{16}
        \and Richard Dekany\inst{16}
        \and Eric C. Bellm\inst{17}
        }

   \institute{
        Institute of Science and Technology Austria, 
        Am Campus 1, 3400, Klosterneuburg, Austria
        \and Division of Physics, Mathematics and Astronomy, California Institute of Technology, Pasadena, CA91125, USA
        \and Center for Astrophysics — Harvard \& Smithsonian, 60 Garden St., Cambridge, MA 02138, USA
        \and Mathematical Sciences Institute, The Australian National University, Hanna Neumann Building 145, ACT2601, Canberra, Australia
        \and International Center for Radio Astronomy Research, Curtin University, GPO Box U1987, Perth, WA 6845, Australia
        \and Department of Physics and Astronomy, University of British Columbia, Vancouver, BC V6T 1Z1, Canada
        \and Department of Astronomy \& Institute for Astrophysical Research, Boston University, 725 Commonwealth Ave., Boston, MA 02215, USA
        \and TAPIR, Mailcode 350-17, California Institute of Technology, Pasadena, CA 91125, USA
        \and Anton Pannekoek Institute for Astronomy, University of Amsterdam, NL-1090 GE Amsterdam, the Netherlands
        \and Department of Physics, Massachusetts Institute of Technology, Cambridge, MA 02139, USA
        \and Department of Physics, University of Warwick, Gibbet Hill Road, Coventry CV4 7AL, UK
        \and Centre for Exoplanets and Habitability, University of Warwick, Gibbet Hill Road, Coventry CV4 7AL, UK
        \and University of Washington, Department of Astronomy, Box 351580, Seattle, WA 98195, USA
        \and Max-Planck-Institut für Astrophysik, Karl-Schwarzschild-Str 1, D-85748 Garching, Germany
        \and IPAC, California Institute of Technology, 1200 E. California Blvd, Pasadena, CA 91125, USA
        \and Caltech Optical Observatories, California Institute of Technology, Pasadena, CA  91125
        \and DIRAC Institute, Department of Astronomy, University of Washington, 3910 15th Avenue NE, Seattle, WA 98195, USA
            }

   \date{Received July 15, 2025}

 
  \abstract
   {Many white dwarfs are observed in compact double white dwarf binaries and, through the emission of gravitational waves, a large fraction are destined to merge. The merger remnants that do not explode in a Type Ia supernova are expected to initially be rapidly rotating and highly magnetized. We here present our discovery of the variable white dwarf ZTF~J200832.79+444939.67, hereafter ZTF\,J2008+4449, as a likely merger remnant showing signs of circumstellar material without a stellar or substellar companion. The nature of ZTF\,J2008+4449 as a merger remnant is supported by its physical properties: hot ($35,500\pm300$~K) and massive ($1.12\pm0.03$~M$_\odot$), the white dwarf is rapidly rotating with a period of $\approx$\,6.6 minutes and likely possesses exceptionally strong magnetic fields ($\sim$\,400--600\,MG) at its surface. Remarkably, we detect a significant period derivative of $(1.80\pm0.09)\times10^{-12}$\,s/s, indicating that the white dwarf is spinning down, and a soft X-ray emission that is inconsistent with photospheric emission. As the presence of a mass-transferring stellar or brown dwarf companion is excluded by infrared photometry, the detected spin down and X-ray emission could be tell-tale signs of a magnetically driven wind or of 
   interaction with circumstellar material, possibly originating from the fallback of gravitationally bound merger ejecta or from the tidal disruption of a planetary object. We also detect Balmer emission, which requires the presence of ionized hydrogen in the vicinity of the white dwarf, showing Doppler shifts as high as $\approx$\,2000\,km\,s$^{-1}$. The unusual variability of the Balmer emission on the spin period of the white dwarf is consistent with the trapping of a half ring of ionised gas in the magnetosphere of the white dwarf.
   }


   \maketitle

\section{Introduction}

The loss of orbital energy through gravitational wave emission by low-mass compact binaries, such as e.g., binary neutron stars, neutron star-black hole systems or double white dwarf binaries, causes the orbits of these systems to gradually tighten, leading many of them to conclude their evolution in a merger event \citep{nelemans2001,2002LRR.....5....3B,2015ApJ...805L...6S}.
Depending on the masses of the two inspiralling objects, the merger of two white dwarfs can result in different outcomes, including a type Ia supernova \citep{1984ApJ...277..355W}, a collapse into a neutron star \citep{nomoto1985}, the formation of a hot subdwarf \citep{heber2009} or of a R Coronae Borealis star \citep{clayton2007}, with the most common outcome being the formation of another, more massive, white dwarf \citep{toonen2012,kilic2024}.

For mergers involving a black hole or neutron star, several studies \citep[see e.g.,][]{rosswog2007} have shown that the late-time transient behaviour of the merger products, namely the long-lasting high-energy (X-ray) emission which follows the short gamma-ray bursts produced in the process of merging, is likely due to the fallback accretion of gravitationally bound post-merger ejecta. In the case of GW170817, the first detected gravitational counterpart of a binary neutron star merger event \citep{GW1708172017}, several studies \citep[see e.g.][]{metzger2021,ishizaki2021} have shown that this process neatly explains the year-scale X-ray excess. Double white dwarf mergers have also been theorised to be capable of hosting fallback accretion, with ejecta masses of the order of $\sim$\,$10^{-3}\,$M$_{\odot}$ \citep[see e.g.,][]{guerrero2004,loren2009,dan2014} being gravitationally bound and eventually re-accreted onto the newly formed remnant. Similarly to the case of binary neutron star or neutron star-black hole mergers \citep{musolino2024}, the fallback of the ejecta on a double white dwarf merger remnant is expected to drive the emission of soft X-rays \citep{rueda2019}, but also optical and infrared emission \citep{rueda2019,sousa2023}. However, these predictions remain largely unconstrained, as no non-explosive double white dwarf merger events have yet been detected in transient surveys. 

Potential white dwarf merger remnants may still act as a valuable source of information regarding the merging process. Models of white dwarf merger products tend to be massive and rapidly rotating, as a consequence of mass and angular momentum conservation during the merging process \citep{2021ApJ...906...53S}. Additionally, due to the strong dynamos that arise during the merger, the remnants are predicted to be highly magnetized \citep{tout2008,2012ApJ...749...25G,2015MNRAS.447.1713B}. 
Thus, the combination of high mass, short spin period and high magnetic field in a white dwarf are plausible fingerprints of a double white dwarf merger product. Examples of this are systems such as EUVE\,J0317-855
\citep{ferrario1997,burleigh1999,vennes2003}, ZTF\,J1901+1458 \citep{caiazzo2021,desai2025} and SDSS\,J2211+1136 \citep{kilic2021}. Variations in the magnetic field strength and structure over the surface of these white dwarfs can be revealed by their spectroscopic and photometric variability \citep[see e.g.][]{vennes2003}; therefore, time-resolved observations on the spin period are key in the study of these systems. 

In this paper, we present the discovery and characterisation of the white dwarf ZTF\,J200832.79+444939.67, hereafter ZTF\,J2008+4449, as a double white dwarf merger remnant showing the presence of circumstellar material. The existence of such material is supported by a host of evidence: the emission of non-photospheric X-rays, a significant increase in the  short rotation period of the white dwarf (indicating that material is being ejected from the system, possibly in a propeller mechanism, carrying away angular momentum), and the intriguing presence of a half-ring of ionised gas trapped in the magnetosphere of the white dwarf. We offer particular attention to the analysis and modelling of the latter, due to the unique nature of such a structure of circumstellar gas. Moreover, the absence of a stellar or brown dwarf companion (constrained by infrared observations) raises the question of the origin of this circumstellar material, to which we propose a handful of possible answers: the fallback of post-merger ejecta, the disruption of a planetary body, or a magnetically driven wind from the white dwarf itself. Given the unusual characteristics of this system, we believe it may be an invaluable probe into both the evolution and phenomenology of double white dwarf merger remnants. Additionally, the recent analysis of the merger remnant ZTF\,J1901+1458 by \citet{desai2025}, shows striking similarities to the white dwarf analysed in this paper, as the object is a fast spinning, highly magnetized with dwarf with similar X-ray emission properties to ZTF\,J2008+4449. This suggests their likely membership to the same new class of merger remnants.

We structure the paper as follows.
In Section~\ref{sec:discovery}, we catalogue and present all of the relevant available data on ZTF\,J2008+4449 and preliminary describe the distinguishing features in this data. Section~\ref{sec:analysis} is dedicated to a detailed account of our scientific methods and analysis of the data, characterising the WD. In section~\ref{sec:discussion}, we present the significance of our results and provide an extensive discussion of their physical implications for this system.  Finally, in section~\ref{sec:summary}, we offer a summary of our findings and an outlook on future studies needed to fully understand this fascinating object.


\section{Discovery and observations}
\label{sec:discovery}
ZTF\,J2008+4449 (Gaia\,DR3\,2082008971824158720) 
was discovered as part of a search for periodic variability among Gaia white dwarfs \citep{gentile-fusillo2019,gentile-fusillo2021} within the Zwicky Transient Facility archive \citep[ZTF,][]{bellm2019,graham2019,masci2019,dekany2020}, which has already yielded several results, such as finding a large number of variable, magnetic white dwarfs and compact binaries \citep{burdge2019a,burdge2019b,burdge2020a,burdge2020b,caiazzo2021,burdge2022b,burdge2022a,caiazzo2023}.

\begin{figure}[tb!]
\centering
\includegraphics[width =0.9\columnwidth]{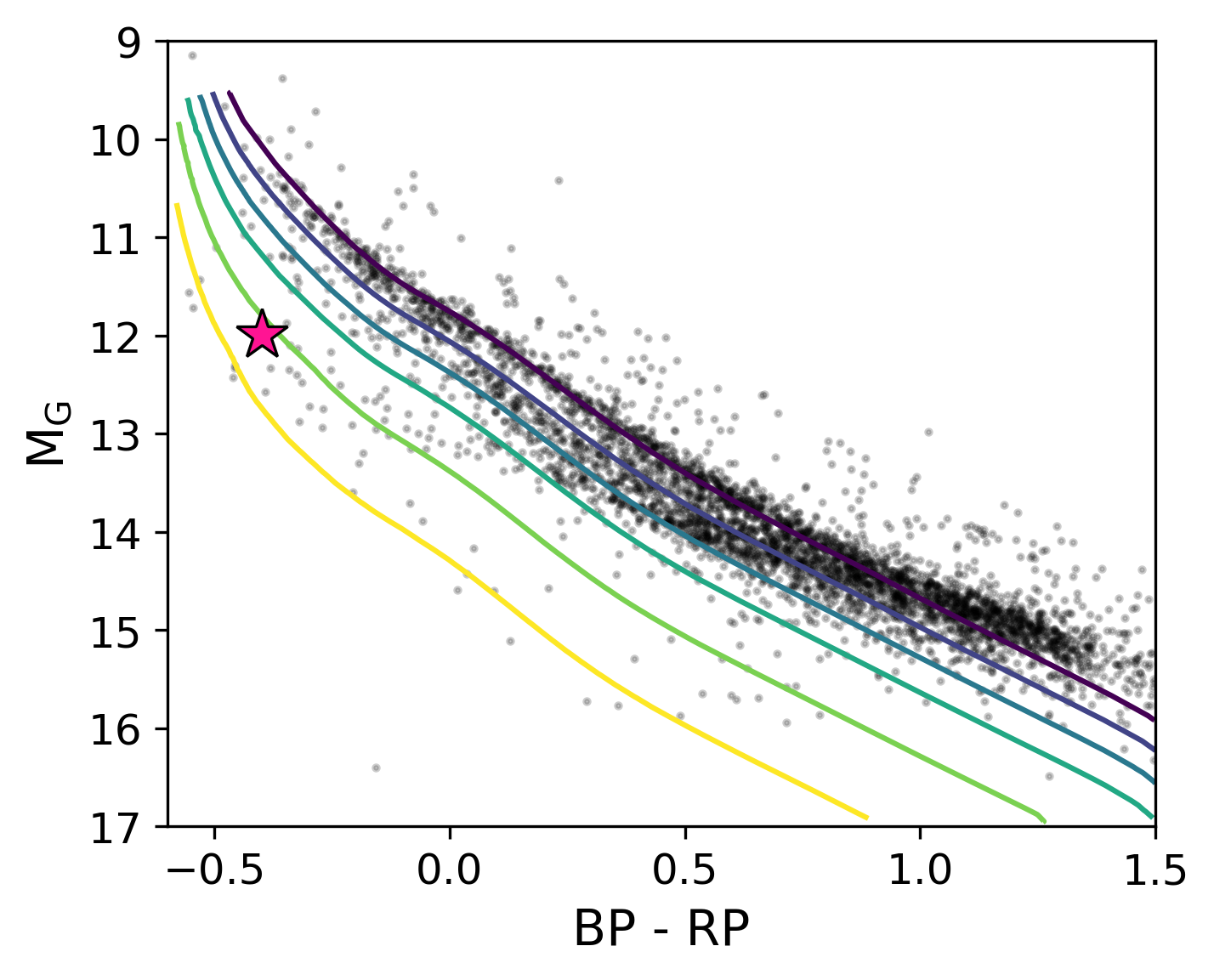}

\caption{Gaia color-magnitude diagram for white dwarfs within 100~pc from Earth and within the SDSS footprint \citep{kilic2020}, where the x-axis depicts the difference between the Gaia BP and RP bands, and the y-axis the absolute magnitude in the Gaia G filter. Solid lines show theoretical cooling tracks for white dwarfs with masses between 0.6~M$_\odot$ (top) and 1.28~M$_\odot$ (bottom), equally spaced in mass; the atmosphere is assumed to be hydrogen-dominated \citep{holberg2006} and the interior composition to be carbon-oxygen \citep{fontaine2001} for $M<1.1$~M$_\odot$ and oxygen-neon \citep{camisassa2019} for $M>1.1$~M$_\odot$. ZTF~J2008+4449 is shown as a red star, and its location in the colour-magnitude diagram reveals its high mass and high temperature. Reddening corrections were applied only to ZTF~J2008+4449; as the objects in the sample are close to the Solar System, reddening is expected to be small. 1$\sigma$ error bars are smaller than the size of the coloured marker, and are omitted for the black background dots for clarity.
}
\label{fig:CMD}
\end{figure}

In the Gaia colour-magnitude diagram (CMD), ZTF\,J2008+4449 appears as a hot ($\approx$\,30,000\,K) and massive ($\approx$\,1\,M$_\odot$) isolated white dwarf (see Fig.\,\ref{fig:CMD}). In our search, the object stood out because of its strong photometric variability in the optical with a very short period of $\approx$\,6.6\,minutes. To understand the origin of the variability, we acquired time-resolved spectroscopic and photometric data necessary for the accurate characterisation of this system. The array of available data is described in the following Sections.

\subsection{Optical light curve}

The ZTF archive currently holds optical photometric data on ZTF\,J2008+4449 in the \textit{g} and \textit{r} bands throughout a roughly 6-year timespan. Additionally, we obtained high-speed photometric data using the Caltech HIgh-speed Multi-color camERA \citep[CHIMERA;][]{harding2016} on the 200 inch Hale telescope at Palomar Observatory. We obtained these observations on eight separate nights in the $g$ and $r$ filters. The full list of observations and exposures is shown in Table\,\ref{tab:data}. The absolute local time of each exposure was measured to better than millisecond accuracy using a GPS receiver. 
We applied bias and flatfield corrections to the images and to perform aperture photometry we used the ULTRACAM pipeline \citep{dhillon2007}. We used one reference star to construct differential light curves. In all observations, including spectroscopic and photometric data, we converted times into modified Barycentric Julian Date (MBJD) in the Barycentric Dynamical Time (TDB) scale using the python library \texttt{astropy.time} \citep{astropy:2022}.

\begin{figure}[tb!]
\centering
\includegraphics[width = \columnwidth]{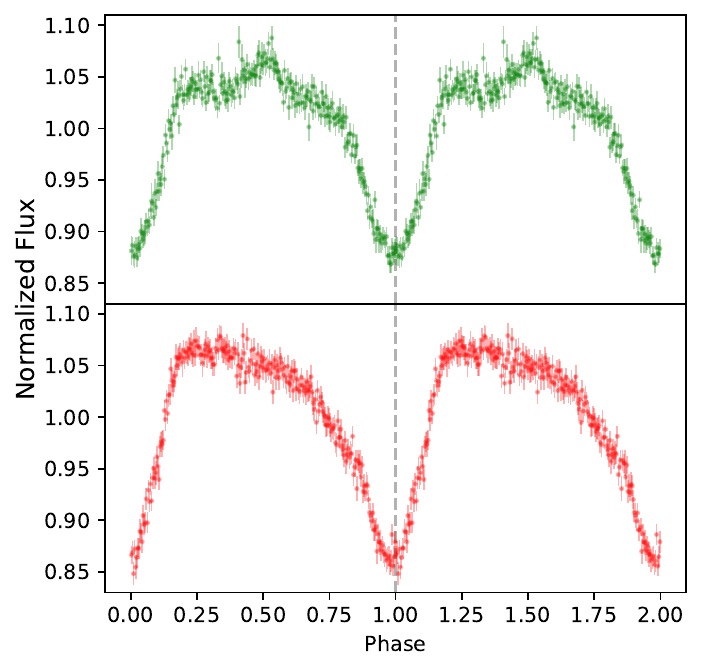}

\caption{Binned CHIMERA light curve with 3-second exposures, phase-folded at the correct period for each epoch as derived in Section\,\ref{sec:period} and normalized to the mean of the light curve in the $g^\prime$-band (upper) and in the $r^\prime$-band (lower).
}
\label{fig:lc}
\end{figure}

The phase-binned CHIMERA light curve, phase-folded using the 
period and period derivative for each epoch that we derive in 
Section\,\ref{sec:period}, is shown in Fig.\,\ref{fig:lc}. The shape of the light curve is quite unusual compared to typical magnetic rotating white dwarfs, which usually show roughly sinusoidal light curves. Furthermore, the data shows some aperiodic variations in the bright flat portion of the light curve (between phases 0.1 and 0.7), which results in 
increased scatter in the phase-folded light curve (we can see that, although it is brighter, the flat portion of the light curve is noisier than in the rest). The un-binned light curves are shown in the Appendix in Fig.\,~\ref{fig:lc-unbinned}. 

\begin{table}
	\centering
	\caption{Phase-resolved Data.}
	\label{tab:data}
	\begin{tabular}{l|ccc} 
		\hline
            \hline
		Instrument & Date & Single Exp. & Total Exp. \\
		\hline
            \hline
		\multirow{7}{*}{CHIMERA} 
        & 2021-05-08 & 3\,s & 2131 s \\ 
        & 2021-07-06 & 3\,s & 3035 s \\ 
        & 2023-05-24 & 3\,s & 3542 s \\ 
        & 2023-06-18 & 3\,s & 1818 s \\ 
        & 2023-09-08 & 3\,s & 1815 s \\ 
        & 2024-07-13 & 3\,s & 2604 s \\ 
        & 2024-08-29 & 3\,s & 5763 s \\
        & 2025-05-30 & 3\,s & 1815 s \\
        \hline
        \multirow{3}{*}{LRIS} 
        & 2019-05-07 & 52\,s & 2998 s \\ 
        & 2021-07-15 & 50\,s & 7384 s \\ 
        & 2024-07-06 & 40\,s & 4767 s \\ 
        \hline
        \multirow{5}{*}{COS} 
        & 2025-02-08 & TIME-TAG & 1478.8 s \\ 
        & 2025-02-08 & TIME-TAG & 2235.4 s \\ 
        & 2025-02-09 & TIME-TAG & 2296.6 s \\  
        & 2025-02-09 & TIME-TAG & 2225.6 s \\ 
        & 2025-02-09 & TIME-TAG & 2244.1 s \\ 
        \hline
        \multirow{4}{*}{STIS} 
        & 2025-05-20 & TIME-TAG & 1998.2 s \\ 
        & 2025-05-20 & TIME-TAG & 2641.2 s \\ 
        & 2025-05-20 & TIME-TAG & 2616.2 s \\  
        & 2025-05-20 & TIME-TAG & 2616.2 s \\
	\hline
        \hline
	\end{tabular}
\end{table}

\begin{figure*}[tb!]
\centering

\includegraphics[width = \textwidth]{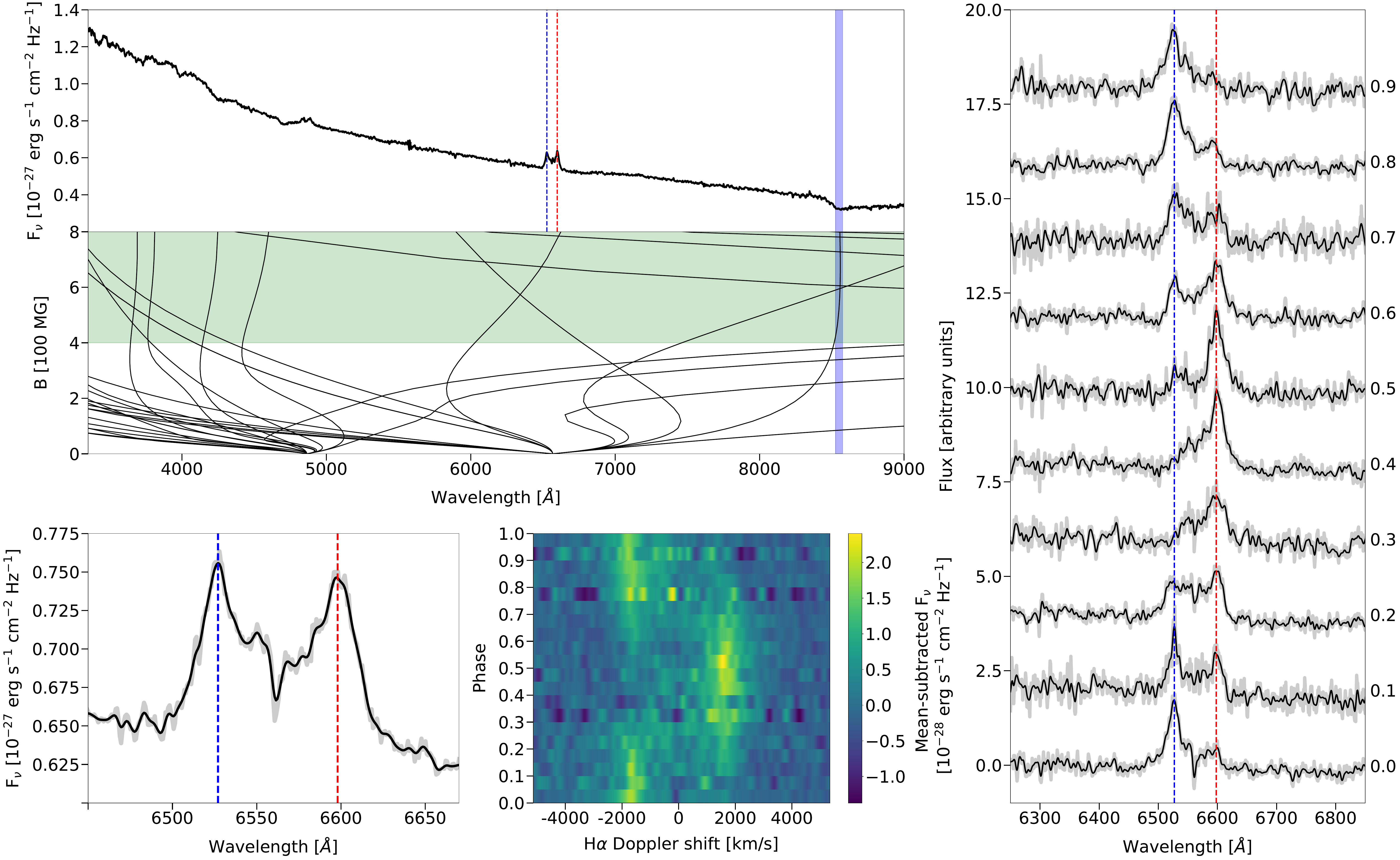}
\caption{{\bf Top left}: Phase-averaged Keck/LRIS spectrum of ZTF\,J2008+4449 and, below, the predicted positions of H$\alpha$ and H$\beta$ absorption lines as a function of magnetic field. The broad ans shallow absorption lines, in particular at $\approx$\,$8,500$\,\AA\, marked in blue, 
indicate a surface magnetic field $B$\,$>$\,400\,MG, while the Balmer emission lines suggest ongoing accretion. The red and blue dashed lines mark the maximal Doppler shifted wavelengths of H$\alpha$ and have the same significance in all panels in this plot. {\bf Bottom left}: Enlarged view of the phase-averaged H$\alpha$ emission line, binned (unbinned) according to the instrument resolution shown as the solid black (faded gray) line. {\bf Bottom middle}: Phase evolution of the trailed spectrum around H$\alpha$ over the 6.6-minute period. The heat map reflects the mean-subtracted flux. {\bf Right}: Trailed spectra of the H$\alpha$ line, with spin phase progressing from bottom to top, as denoted by the numbers on the right hand side of the plot.
}
\label{fig:optspectrum}
\end{figure*}

\subsection{Optical spectra}

We acquired time-resolved optical spectra with the Low Resolution Imaging Spectrometer \citep[LRIS;][]{oke1995} on the Keck I telescope on Maunakea, on six separate nights over a period of $\sim$\,5 years (see Table\,\ref{tab:data}). We reduced all LRIS observations using the publicly available LPIPE automated data reduction pipeline \citep{perley2019}. 
The full optical spectrum averaged over all exposures is shown in the top-left panel of Fig.\,\ref{fig:optspectrum} and shows broad and shallow absorption features which are typical of highly magnetised white dwarfs. In particular, the absorption feature at $\approx$\,8500\,\AA, with its peculiar step-like shape, 
is often observed in hydrogen-dominated white dwarfs with magnetic fields in excess of a few hundred MG \citep[see e.g.,][]{caiazzo2021}, as it corresponds to a blend of two strong Zeeman components of H$\alpha$,  as illustrated in the lower section of the same plot in Fig\,\ref{fig:optspectrum}. 
Lastly, the spectrum shows weak emission in the Balmer lines H$\alpha$ and H$\beta$. The emission lines are double-peaked, indicating the presence of Doppler shifts as high as $\approx$\,2000\,km\,s$^{-1}$, but no Zeeman splitting due to a strong magnetic field. It is therefore likely that the narrow emission comes from a region far from the surface of the white dwarf, where the local magnetic field is weak.

Trailed optical spectra were obtained by binning the data in ten phase bins over the optical modulation period, corrected for the epoch considered using the period and period derivative that we obtain in Section~\ref{sec:period} (Fig.\,\ref{fig:optspectrum}). The final spectrum of each phase bin was obtained by averaging all spectra in that bin and propagating their errors in quadrature. Upon inspecting the behaviour of the Balmer emission lines in these phase-resolved spectra, it is clear that their shape in the averaged spectrum is caused by the periodic transition of one emission peak between maximally blue-shifted and maximally red-shifted wavelengths, as shown in the right and bottom-middle panels of Fig.\,\ref{fig:optspectrum} for the case of $H\alpha$. Interestingly, the variation in wavelength does not look like an S-curve, as typically seen for Doppler shifts caused by rotational or orbital velocities, but instead the transition between these states looks almost abrupt, each one lasting roughly half of the period cycle (see the middle low panel of Fig.\,\ref{fig:optspectrum}). The maximally blue-shifted emission corresponds to the minimum brightness in the ZTF and CHIMERA light curves (phase 0 in the right panel of Fig.\,\ref{fig:lc}). As a note to the reader, throughout this paper, all figures use a common zero phase. 

\begin{figure*}[tb!]
\centering
\includegraphics[width = 0.84\textwidth]{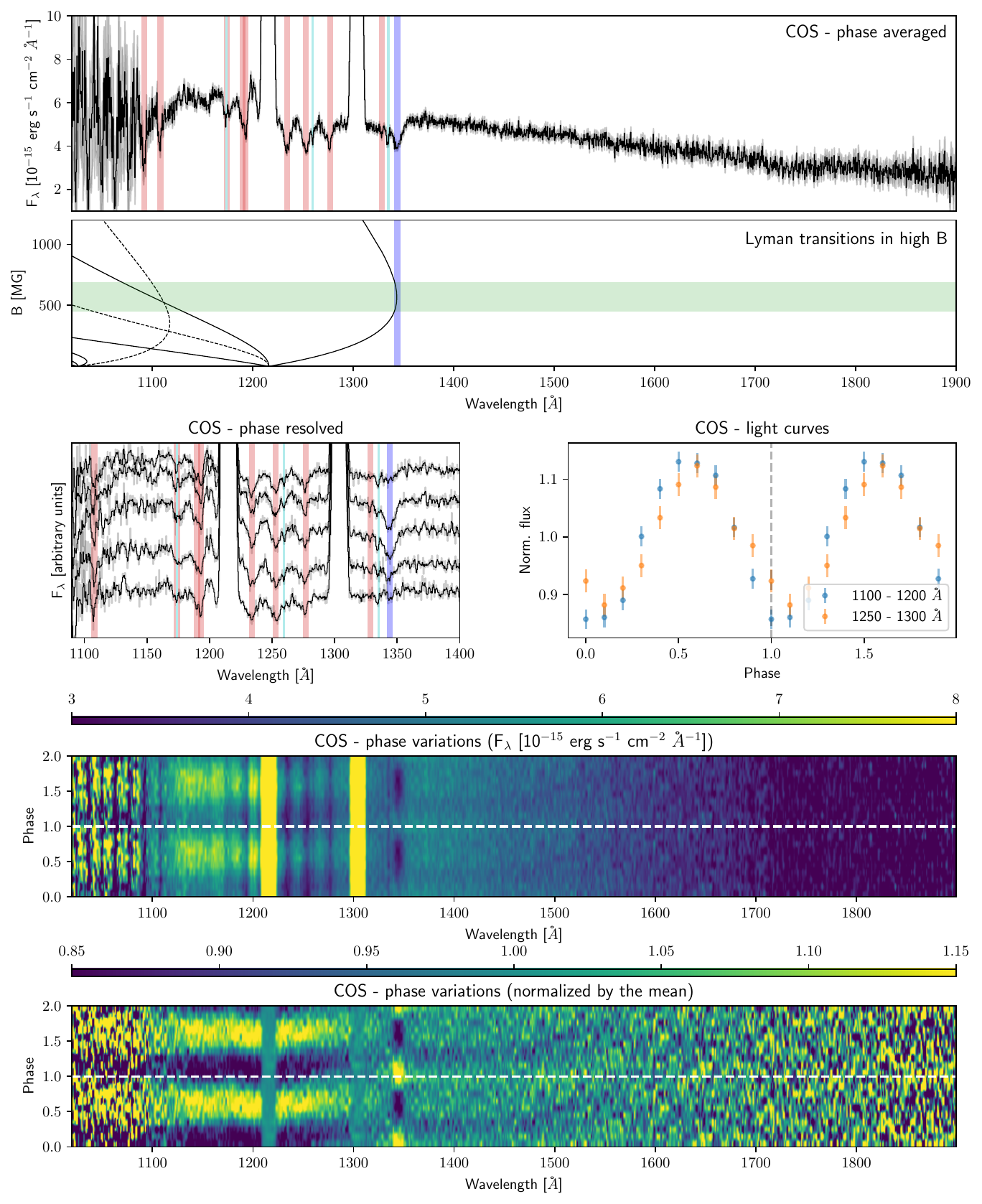}\vspace{-10pt}
\caption{{\bf Upper Panel}: Phase-averaged COS spectrum in F$_\lambda$ (black solid line, errors in grey). Strong absorption lines in the spectrum are highlighted: a possible Ly$\alpha$ component in blue and non identified metal lines in red (see Fig.\ref{fig:metal_lines}). Additionally, narrow carbon and silicon absorption lines (possibly from the ISM or circumstellar material) are highlighted in cyan. In the lower part, the Ly$\alpha$ component is compared to predicted line positions of Ly$\alpha$ and Ly$\beta$ as a function of magnetic field, where dashed lines indicate forbidden components \citep{1994asmf.book.....R}. {\bf Middle Left}: COS phase-resolved spectrum (5 phase bins) over one period, zoomed in on the region with absorption lines. Each phase is shifted vertically by the same amount. Most of the absorption lines do not change significantly with phase, with the exception of the Ly$\alpha$ line, which disappears for half the period. {\bf Middle Right}: COS light curves in two wavelength ranges: 1100--1200\,\AA\ (blue) and 1250--1300\,\AA\ (orange). {\bf Bottom Two Panels}: Phase resolved COS spectra plotted over two periods. In the upper panel, the colorbar indicates the flux in units of $10^{-15}$\,erg\,s$^{-1}$\,cm$^{-2}$\,\AA$^{-1}$. In the lower panel, the phase-resolved spectra have been divided by the phase-averaged spectrum, so the colorbar shows the variation with respect to the mean. We can clearly see two sources of variability: a strong variation of the continuum between 1100 and 1300 \AA, and a variation in the strength of the line at 1344 \AA. The absorption is most pronounced when the continuum is at maximum brightness.}\vspace{-10pt}
\label{fig:uv-mosaic}
\end{figure*}

\begin{figure*}[tb!]
\centering
\includegraphics[width = 0.88\textwidth]{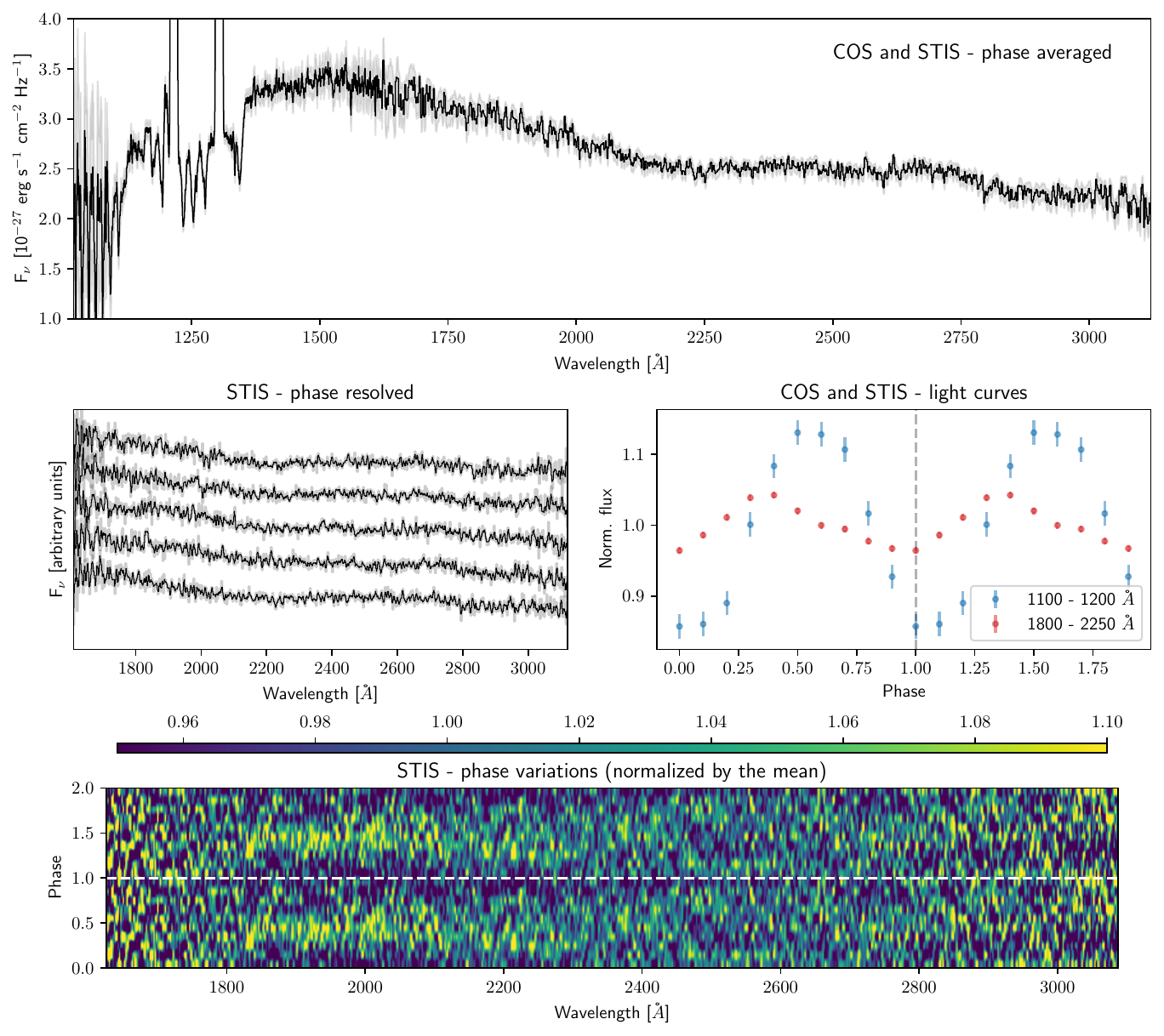}\vspace{-5pt}
\caption{STIS and COS UV data for ZTF\,J2008+4449. {\bf Upper Panel}: The full phase-averaged UV spectrum in $F_\nu$ (black solid line, errors in grey). {\bf Middle Left}: Phase-resolved spectrum (5 phase bins) in the STIS band over one period. Each phase is shifted vertically by the same amount. There is very little variability with phase in the STIS band. {\bf Middle Right}: Comparison between the COS light curve in the 1100--1200\,\AA\ range (blue, same as Fig.\,\ref{fig:uv-mosaic}) and the STIS light curve in the  1800--2250\,\AA\ range (red), which is where the STIS spectrum is most variable. The STIS light curve is also roughly sinusoidal, with a lower amplitude than the COS one, and slightly out of phase. {\bf Bottom Panel}: STIS phase resolved spectra plotted over two periods and divided by the phase-averaged spectrum (the colorbar shows the variation with respect to the mean). We can see a weak variability, mostly in the 1800--2250 \AA\ range.}\vspace{-5pt}
\label{fig:STIS-mosaic}
\end{figure*}

\subsection{UV spectra}
\label{sec:UV}

We obtained UV spectra of ZTF\,J2008+4449 with the Cosmic Origins Spectrograph \citep[COS;][]{green2011} and the Space Telescope Imaging Spectrograph \citep[STIS;][]{woodgate1998} on the Hubble Space Telescope \citep[HST, GO program 17720;][]{2024hst..prop17720C}. We asked for the low-resolution grating G140L on COS, with central wavelength 800\,\AA, which provides a spectral coverage over the range $\simeq$\,900--1900$\,$\AA\ at a resolving power of $\approx$\,2000. For STIS, we requested the G230L NUV MAMA grating, covering the range $\simeq$\,1600--3000\,\AA. The COS observations spanned 5 orbits, for a total of $\sim$\,10,000 seconds, and were acquired in TIME-TAG mode, thus recording the arrival time of each individual photon. This event stream allowed us to construct the phase-binned spectra for 10 phase bins over the 6.6-minute optical modulation cycle. 

In Fig.\,\ref{fig:uv-mosaic} we show the COS data, while in Fig.\,\ref{fig:STIS-mosaic} we show the full UV spectrum and the STIS data. The phase-averaged COS spectrum is depicted in the upper panel of Fig.\,\ref{fig:uv-mosaic}. The two strong emission features at $\approx$\,1210 and $\approx$\,1300\,\AA\ that exceed the upper limit of the plot are due to geocoronal airglow. The narrow absorption lines indicated in cyan are likely due to either interstellar or circumstellar carbon and silicon absorption. The phase-averaged UV spectrum of ZTF\,J2008+4449 shows an array of deep and broad absorption features clustered in the wavelength range between 1000--1350 \,\AA. In blue in the figure, we indicate a feature at 1344\,\AA\ that could correspond to a hydrogen absorption line (a Zeeman Lyman-$\alpha$ component) in a strong magnetic field (in agreement with the Zeeman H$\alpha$ component detected in the optical at 8500\,\AA). In the lower part of the same panel, we show the Lyman transitions as a function of magnetic field: the stationary Zeeman component at $\approx1344$\AA\ coincides with the observed absorption line. We do not have a clear identification for the other absorption lines, that we highlight in red. As we discuss in Section~\ref{sec:SED}, these lines are likely absorption lines of metals. In the top panel of Fig.\,\ref{fig:STIS-mosaic} we show the full COS+STIS phase-averaged spectrum. In the STIS range, the spectrum does not show any additional strong absorption features.

Time-binning the UV spectra once again reveals variability over the 6.6\,min period. As illustrated in the other panels of Fig.\,\ref{fig:uv-mosaic} for COS, two types of variability stand out in the phase-resolved COS spectra. Firstly, the absorption line at 1344\,\AA\ (the possible Lyman-$\alpha$ component, highlighted in blue) varies in strength with phase, completely disappearing for half of the period. Secondly, we can see a strong sinusoidal modulation (with a peak-to-peak amplitude of 30\%) in the continuum blueward of 1300 \AA, with the brightest continuum phase corresponding to the strongest absorption in the 1344\,\AA\ line. We can also see a slight shift in phase at different wavelengths; the minimum in the 1100--1200\,\AA\ range coincides with the minimum in the optical light curve (that we define as phase 0, see Fig.\,\ref{fig:lc}), while the minimum in the 1250--1300\,\AA\ is shifted later in phase by approximately 0.1 (we compare all light curves in all wavelengths in Fig.\,\ref{fig:all_lc}). The variability in STIS is more subtle. Sinusoidal variability can be seen in the 1800--2250\,\AA\ range, with an amplitude peak-to-peak of about 10\% (see the middle right panel of Fig.\,\ref{fig:STIS-mosaic}).

\subsection{Photometry}
\label{sec:phot}
We acquired new photometric data for ZTF\,J2008+4449 and compiled it together with pre-existing photometry. The photometric data is presented in the top panel of Fig.\,\ref{fig:SED_BD} and catalogued in Table\,\ref{tab:phot_data}. In the optical wavelength range, we used the photometry available in the PanSTARRS catalogue \citep{panstarrs1}. We also employed the Wide-field InfraRed Camera (WIRC) \citep{WIRC} on the Hale Telescope at Palomar Observatory to obtain infrared photometry on two separate nights, in WIRC filters; J, H and K (shown in light blue). The NUV photometry (dark blue) was acquired with the Neil Gehrels Swift Observatory (Swift) \citep{Swift} through the Target of Opportunity proposals with numbers 14960 and 17885, target ID 13925. Finally, we converted the previously described COS UV spectrum to AB magnitudes to fill in the FUV end of the photometric data (this is shown in cyan in Fig.\,\ref{fig:SED_BD}).

\begin{table}[tb]
	\centering
	\caption{Photometric Data.}
	\label{tab:phot_data}
	\begin{tabular}{lc|cc} 
		\hline
            \hline
		Instrument & Filter & Exp. time & AB magnitude \\
		\hline
            \hline 
        & UVW2 & 443 s & $17.80\pm0.07$ \\ 
        Swift/& UVM2 & 340 s & $17.83\pm0.09$ \\
        UVOT& UVW1 & 221 s & $17.99\pm0.09$\\
        & U & 111 s & $18.18\pm0.11$ \\
        \hline
        \hline
		\multirow{5}{*}{PanSTARRS} 
        & PS-g & -- & $18.73\pm0.02$ \\ 
        & PS-r & -- & $19.04\pm0.01$ \\
        & PS-i & -- & $19.32\pm0.01$\\
        & PS-y & -- & $19.73\pm0.04$ \\
        & PS-z & -- & $19.80\pm0.02$ \\
	\hline
        \hline 
        & \multirow{2}{*}{J} 
            & 1 hour & $20.46\pm0.06$ \\
        Palomar/&    & 1 hour & $20.48\pm0.05$ \\
        WIRC& H & 1 hour & $21.06\pm0.09$ \\
        & Ks & 1 hour & $21.59\pm0.12$\\
	\hline
        \hline
	\end{tabular}
\end{table}

\begin{figure}[tb!]
\centering
\includegraphics[width = 0.9\columnwidth]{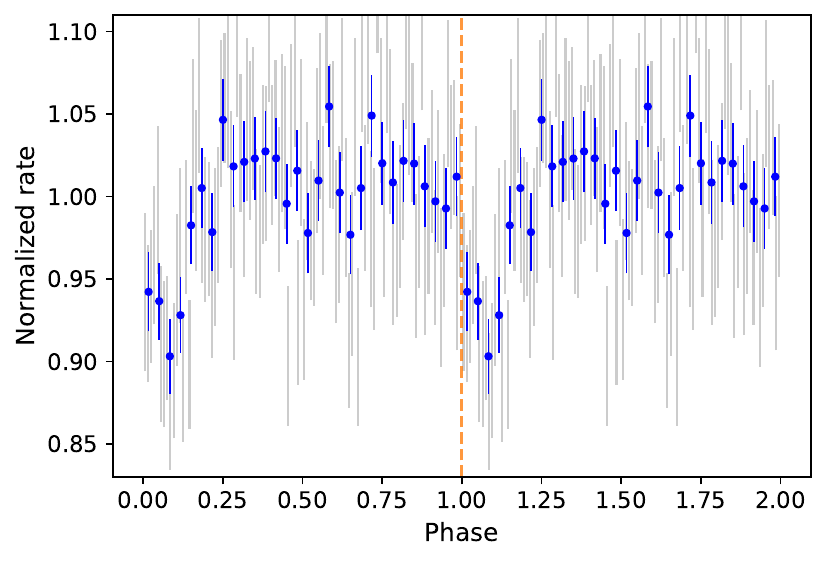}\vspace{-5pt}
\caption{Binned XMM/OM light curve, shown with less (more) bins in blue (gray), phase-folded at the correct period for each epoch as derived in Section\,\ref{sec:period} and normalized to the mean of the light curve in the $U$-band.
}\vspace{-10pt}
\label{fig:lc-U}
\end{figure}

\subsection{XMM-Newton}
The program that provided the HST observations \citep{2024hst..prop17720C} was a joint program with XMM-Newton \citep{jansen2001}. We requested 50\,ks of non-continuous observation, with simultaneous Fast-Mode photometry with the Optical Monitor \citep[OM;][]{mason2001}. The observation was carried out in two separate visits (Obs IDs 0952990101 and 0952990201), each with a duration of approximately 32\,ks. The source was clearly detected in all three cameras of the European Photon Imaging Camera (EPIC) in both visits: PN \citep{struder2001}, MOS 1 and MOS 2 \citep[M1 and M2,][]{turner2001}. The EPIC data were reduced using v21.0.0 of the Science Analysis Software (SAS) designed for the reduction of XMM-Newton data \citep{gabriel2004-sas}. We filtered the event files using the routine \texttt{evselect} and Good Time Intervals (GTI) that were defined with standard defaults and thresholds on the count rate chosen by visual inspection of the total light curve for each observation; specifically, for Obs\,01 we used 0.3, 0.1, and 0.15 cts/s for PN, M1, and M2, repectively, whilst for Obs\,02 we used 0.26, 0.15, and 0.15 cts/s . This procedure resulted in total effective exposure times of 17.9, 21.4, and 21.9\,ks for PN, M1, and M2, respectively, in Obs\,01, and 19.3, 28.6, 28.9\,ks for Obs\,02. We extracted spectra for the sources and adjacent background regions using \texttt{evselect}, with background regions placed on the same chip and pixel columns as the source. The source and background circular apertures had radii of 20\,arcsec and 80\,arcsec, respectively. A source aperture of 20\,arcsec should enclose 75--80\,\% of source photons for an on-axis pointing in all three EPIC cameras\footnote{See e.g., Section 3.2.1.1 of the XMM-Newton Users Handbook: \url{https://xmm-tools.cosmos.esa.int/external/xmm_user_support/documentation/uhb/onaxisxraypsf.html}}. We list the total source and background counts in Table\,\ref{tab:allinst_counts_both}, as well as the significance of detection in each band. In the PN camera, the source was detected with high significance ($>3\sigma$) in all standard science bands below 4.5 keV (0.2--0.5, 0.5--2.0, and 2.0--4.5\,\,keV) and in the broad band (0.2--10.0\,keV), while M1 and M2 only detected the source below 2.0 keV and in the broad band. The X-ray spectral and timing analysis is presented in Section\,\ref{sec:xray_analysis}.

Light curves were obtained simultaneously to the X-ray observations with the Optical/UV Monitor (XMM-OM) in Fast Mode in the U and the UVM2 bands, in Obs\,01 and 02, respectively. The signal-to-noise in the UVM2 band is very low and the light curve does not show any significant variability. On the other hand, the phase-folded U-band light curve, shown in Fig.\ref{fig:lc-U}, presents a similar shape as the optical light curves, albeit with a smaller amplitude and a small shift in phase (the orange dashed line indicates the minimum in the optical light curve, phase 0 in all plots).

\section{Analysis}
\label{sec:analysis}

\begin{figure*}[tb!]
\centering
\vspace{10pt}
\includegraphics[width = 0.46\textwidth]{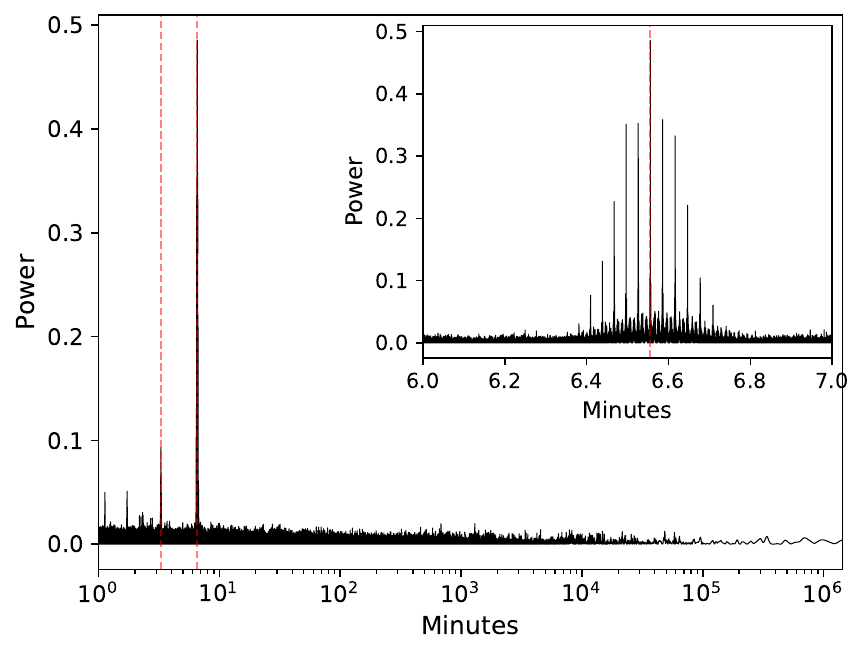}
\includegraphics[width = 0.53\textwidth]{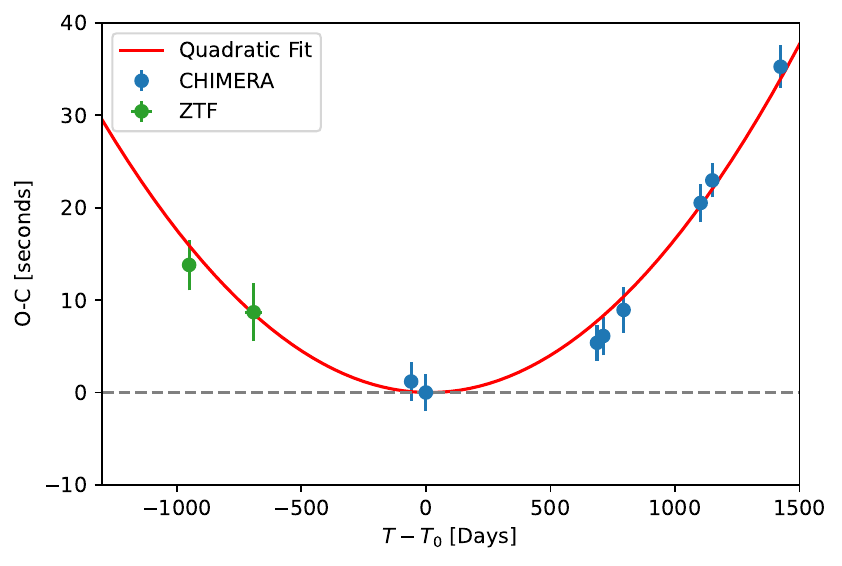}
\caption{\label{fig:LS_o-c} {\bf Left}: ZTF periodogram for ZTF\,J2008+4449 in the $r$-band. The main period and first harmonic (half the period) are highlighted by vertical red dashed lines. The inset shows the complex structure of the main peak, due to a combination of the light curve windowing function and the fact that the period is changing significantly over the ZTF baseline. {\bf Right}: O-C diagram for ZTF\,J2008+4449 for two high cadence ZTF observations (in green) and the 8 CHIMERA observations (in blue), all in the $r$ band. On the y-axis, we plot the difference in seconds between the observed time of the minimum in the light curve and the expected time calculated assuming a constant period, while on the x-axis, we plot the elapsed time since our reference time $T_0=59401.76$. For the second ZTF observation, the error in the x direction indicates the total duration of the ZTF data used in the fit (70 nights); for the other ones the duration is too small to show (3 nights for the first ZTF observation, <1 night for the CHIMERA ones). The quadratic behaviour indicates the presence of a significant period derivative.}
\end{figure*}

\subsection{Period and period derivative}
\label{sec:period}
The Lomb-Scargle periodogram of the ZTF light curves reveals a strong optical modulation period at $6.55577$\,$\pm$\,$0.00002$\,minutes and no additional significant periodicities (see Fig.\,\ref{fig:LS_o-c}).
When phase-folded at the ZTF period, the CHIMERA light curve shows a significant shift in phase between each epoch, indicating a possible period variation. By measuring the shift in phase of the minimum in the light curve, we can measure the period derivative. 
For a small and constant period derivative ($\dot{P}\ll P$), the shift in the time of arrival of the minimum in the light curve after a number of cycles $n$ can be expressed as:
\begin{equation}
    T_n - T_0 \approx nP_0 + \dot{P}P_0\frac{n^2}{2}
\end{equation}
where $T_n$ is the time of the minimum at the cycle $n$, $T_0$ is the time of the minimum at cycle $n$=0, $P_0$ is the period at the time $T_0$ and $\dot{P}$ is the period derivative, assumed constant. 

We took the eight CHIMERA observations in the $r$ band and measured the time of the first minimum in each observation. This time was determined by period-folding the entirety of the observation and fitting a skewed inverted Gaussian profile to the dip-like feature in the light curve. As the change in period over the course of one individual observation is negligible, this method does not bias our measurement while it increases its precision. Fortunately, ZTF observed the white dwarf in high-cadence mode in two instances in 2018 and in 2019, allowing us to determine the phase shift in those times and therefore extend our baseline by three years. In particular, in November 2018, ZTF observed the target in the $r$ band in ``deep-drilling mode'' \citep[248 epochs in 3 nights, compared to the normal cadence of about 1 epoch every 2 nights,][]{kupfer2021}, while between August and November 2019, ZTF observed the target in a high-cadence mode (310 epochs in 70 nights, an average of 4.4 epochs per night).

In the right panel of Fig.\,\ref{fig:LS_o-c} we show the O$-$C (observed minus calculated) diagram for different observations in the $r$ band of ZTF\,J2008+4449. On the y-axis, we plot the phase shift $T_n-T_0$ in seconds, while on the x-axis, we plot the elapsed time since our reference time $T_0=59401.76$ in days (in Barycentric Dynamical Time). If there was no period derivative, the O-C should follow a constant or linear trend, whilst for a (constant) period derivate, the O-C should follow a quadratic relation. The quadratic fit to the data shows a clear quadratic relation in the O-C, yielding a 
period derivative of $\dot{P}=(1.80\pm0.09)\times10^{-12}$\,s/s, and a refined period estimate of $P_0=6.55576688\pm0.00000008$ minutes at the epoch $T_0$. The positive period derivative provides clear evidence that the period is increasing. The CHIMERA optical light curves, phase-folded with the correct period at each epoch, are shown in Fig.\,\ref{fig:lc}.

\subsection{Physical properties of the white dwarf}
\label{sec:SED}

\begin{figure*}[tb!]
\centering
\includegraphics[width = \textwidth]{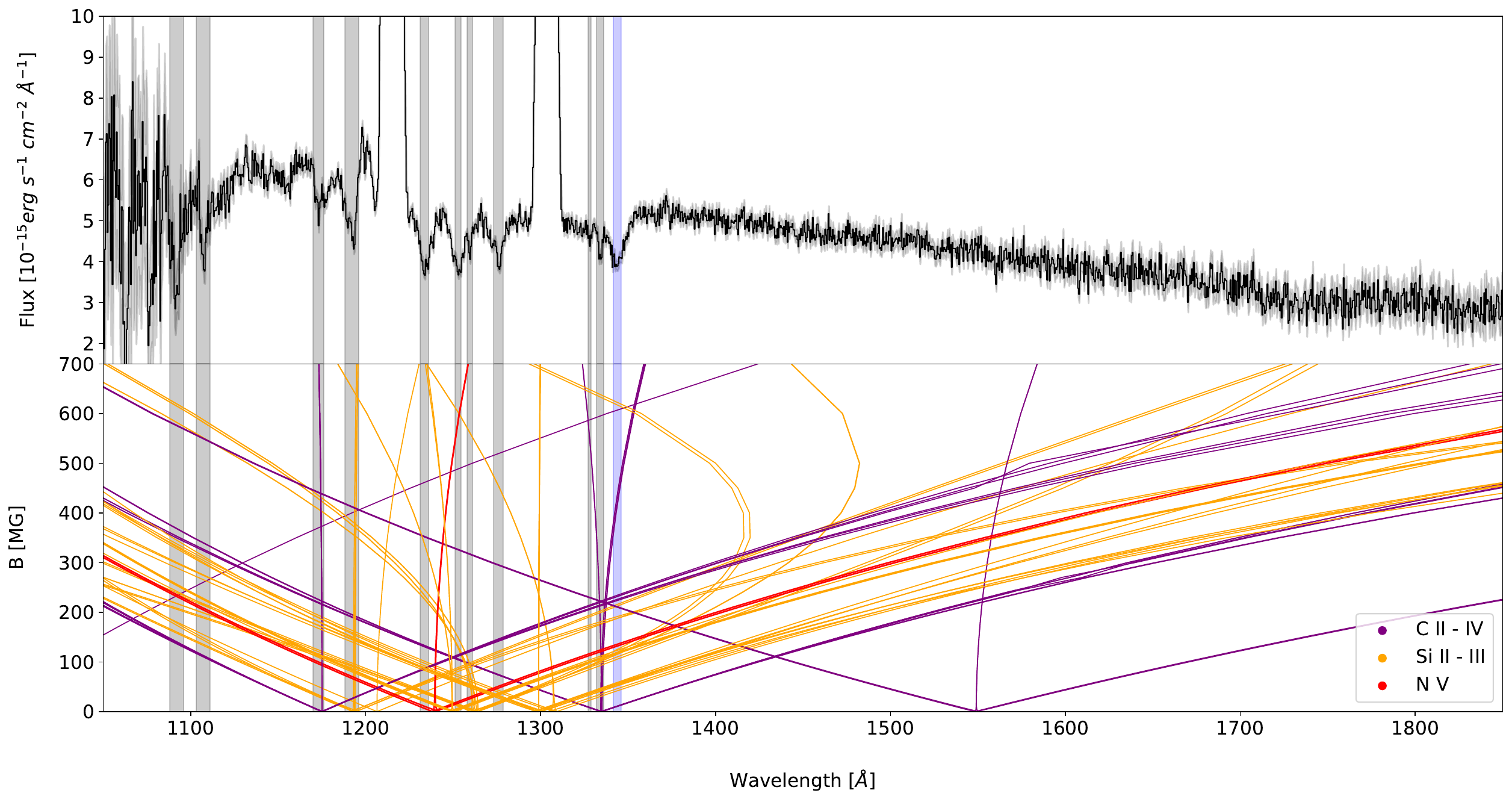}
\caption{\textbf{Top}:
The phase-averaged COS spectrum in F$_\lambda$ is shown in black, with errors in light gray. The vertical bands in both panels mark the position of the absorption lines, with the blue one corresponding to the absorption line at $1344\,$\AA, as in the top panel of Fig.\,\ref{fig:uv-mosaic}. \textbf{Bottom}: Positions of a suite of C II-IV (black), Si II-III (purple) and N V (red) ionic transition components as a function of magnetic field strengths. Only transitions with a relatively large oscillator strength at 500\,MG have been plotted. It can be seen that the \textit{$\pi$} components of the Zeeman-split lines do not deviate far from the wavelength range where most of the absorption lines in the UV spectrum are located.
}
\label{fig:metal_lines}
\end{figure*}

The broad and shallow absorption features in the optical spectrum of ZTF\,J2008+4449 (see left panel of Fig.\,\ref{fig:optspectrum}) are characteristic of highly-magnetised white dwarfs ($>$\,100\,MG). Similarly, the broad and shallow absorption lines in the ultraviolet spectrum are likely high-field Zeeman components of hydrogen and other metals. As we show in Figs.\,\ref{fig:optspectrum} and\,\ref{fig:uv-mosaic},  the absorption lines at 8500\,\AA\, and at 1344\,\AA\, are likely Zeeman components of H$\alpha$ and Ly$\alpha$, respectively. If this is the case, they indicate an average magnetic field on the surface of the white dwarf in the range 500--600\,MG (see Fig.\,\ref{fig:uv-mosaic}). To explore the origin of the other lines, we calculated the transition energies in the Paschen-Back regime for a few metals (carbon, silicon and nitrogen) for which absorption lines have been observed in the spectra of dwarf novae in quiescence \citep[see e.g.,][]{godon2022} and for which lines can be expected at the effective temperature of the white dwarf. We explain our calculations in more detail in Appendix~\ref{sec:metalsUV} and we show the different transitions in Fig.\,\ref{fig:metal_lines}. Although our calculations are approximate (the perturbation theory that we employ might break down at such high field) and a more careful approach is needed to obtain confident identification in the lines, we can see that the $\pi$ Zeeman components (the line transitions with no change in the $m$ quantum number, $\Delta m=0$) at high magnetic field cluster in the same wavelength range where we detect the absorption lines. The other Zeeman components, on the other hand, vary rapidly with field and can be strongly affected by magnetic broadening, producing extremely shallow and broad features that would be hard to detect. 

\begin{figure*}[tb!]
\centering
\includegraphics[width = 0.8\textwidth]{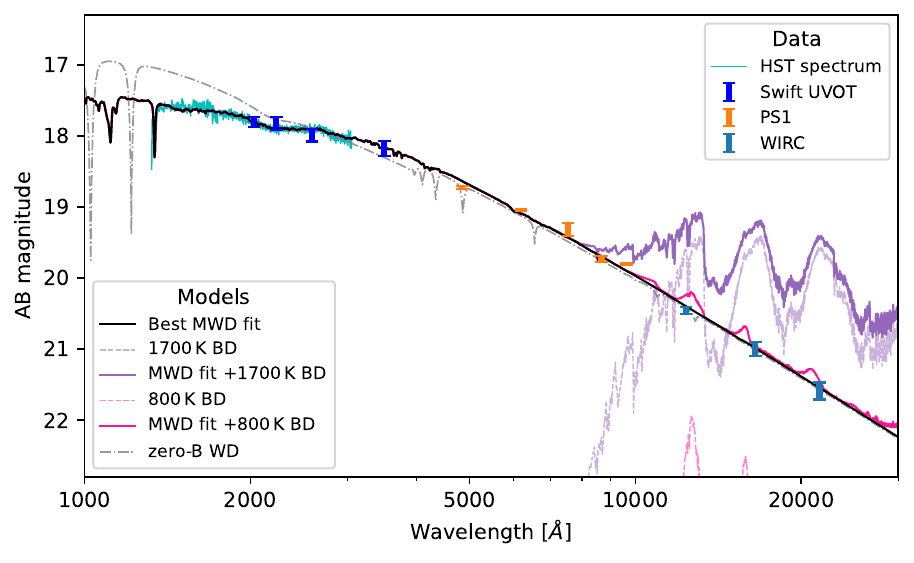}
\includegraphics[width = 0.8\textwidth]{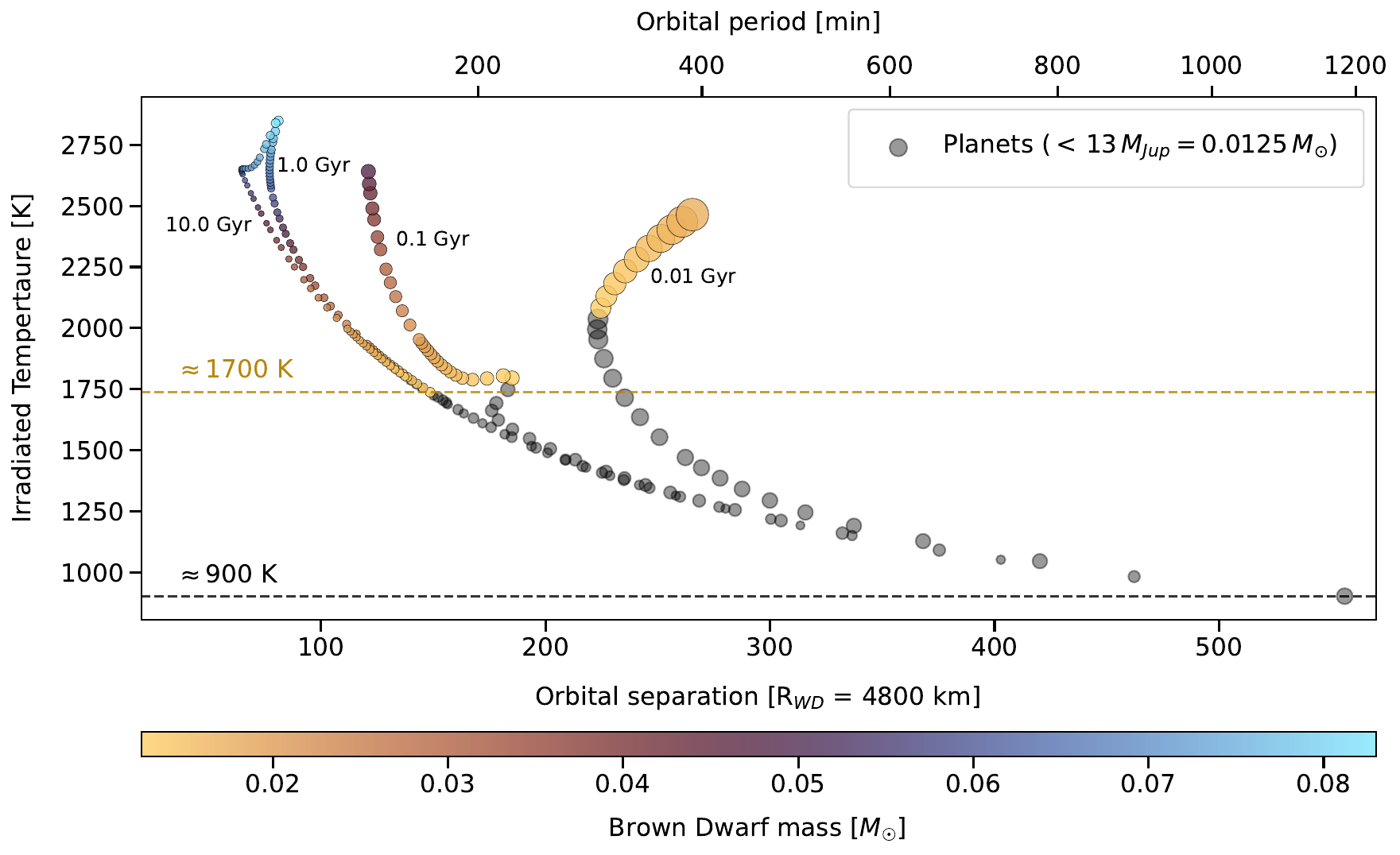}
\caption{\textbf{Top:} Spectral energy distribution (SED) of ZTF\,J2008+4449, plotting AB magnitude against wavelength. The light blue, orange data points respectively show the IR (WIRC), optical (PanSTARRS) and NUV (Swift) photometry. The cyan line shows HST STIS and COS concatenated spectra spanning the NUV and FUV, with the COS spectrum truncated at the lowest wavelengths. The solid black and dashed gray lines show the best fitting pure Hydrogen magnetic and non-magnetic atmosphere models respectively. Lastly, a 1700\,K (800\,K) brown dwarf model is shown in purple (pink), by itself (faded line) and added to the best fitting magnetic white dwarf model (solid line). \textbf{Bottom:} Irradiated face temperatures for a series of Roche lobe-filling brown dwarf and planetary companions of varying ages, against their orbital separation from the white dwarf primary. The upper \textit{y}-axis shows the orbital period given by each respective orbital separation. The four different tracks correspond to four different brown dwarf/planet ages (0.01, 0.1, 1 and 10\, Gyr), marked next to each track. Brown dwarf models ($\geq13\,M_{\rm Jup}$) are plotted in colours corresponding to their respective masses; all planetary models ($\leq13\,M_{\rm Jup}$) are plotted in black. The size of each data point is representative of (but not linearly proportional to) the model radius. The golden (black) horizontal dashed line, marks the minimal irradiation temperature for a brown dwarf (planet) among the evolutionary models utilised, i.e., $\approx1700\,$K ($\approx900\,$K). The irradiated temperatures and corresponding orbital separations were calculated using the Sonora Bobcat brown dwarf evolutionary models grid \citep{sonorabobcat}.
}
\label{fig:SED_BD}
\end{figure*}

Another indication that the star is highly magnetized is in the shape of its spectral energy distribution (SED). As can be seen in Fig.\,\ref{fig:SED_BD}, the SED shows a break between the optical and the ultraviolet which is often seen in highly magnetized white dwarfs \citep{green1981,schmidt1986,gaensicke2001,desai2025}, but which is hard to explain for non-magnetic stars. As we explain in more detail in our companion paper \citep{desai2025}, this flattening of the UV spectrum in high-field white dwarfs is due to the effect of the strong field on the continuum opacities. The 550\,MG magnetic hydrogen-dominated spectral model shown in black in Fig.\,\ref{fig:SED_BD}, described below in this section, captures nicely the SED of the white dwarf across all wavelengths (with the exception of the unidentified absorption in the FUV). For comparison, we also show the non-magnetic, pure-hydrogen model \citep[dot-dashed line][]{tremblay2011} at the same physical parameters, demonstrating that a model which does not account for the strong magnetic field will either be too shallow in the optical, or too steep in the UV. 

\begin{figure}[tb!]
\centering
\includegraphics[width=1.0\columnwidth]{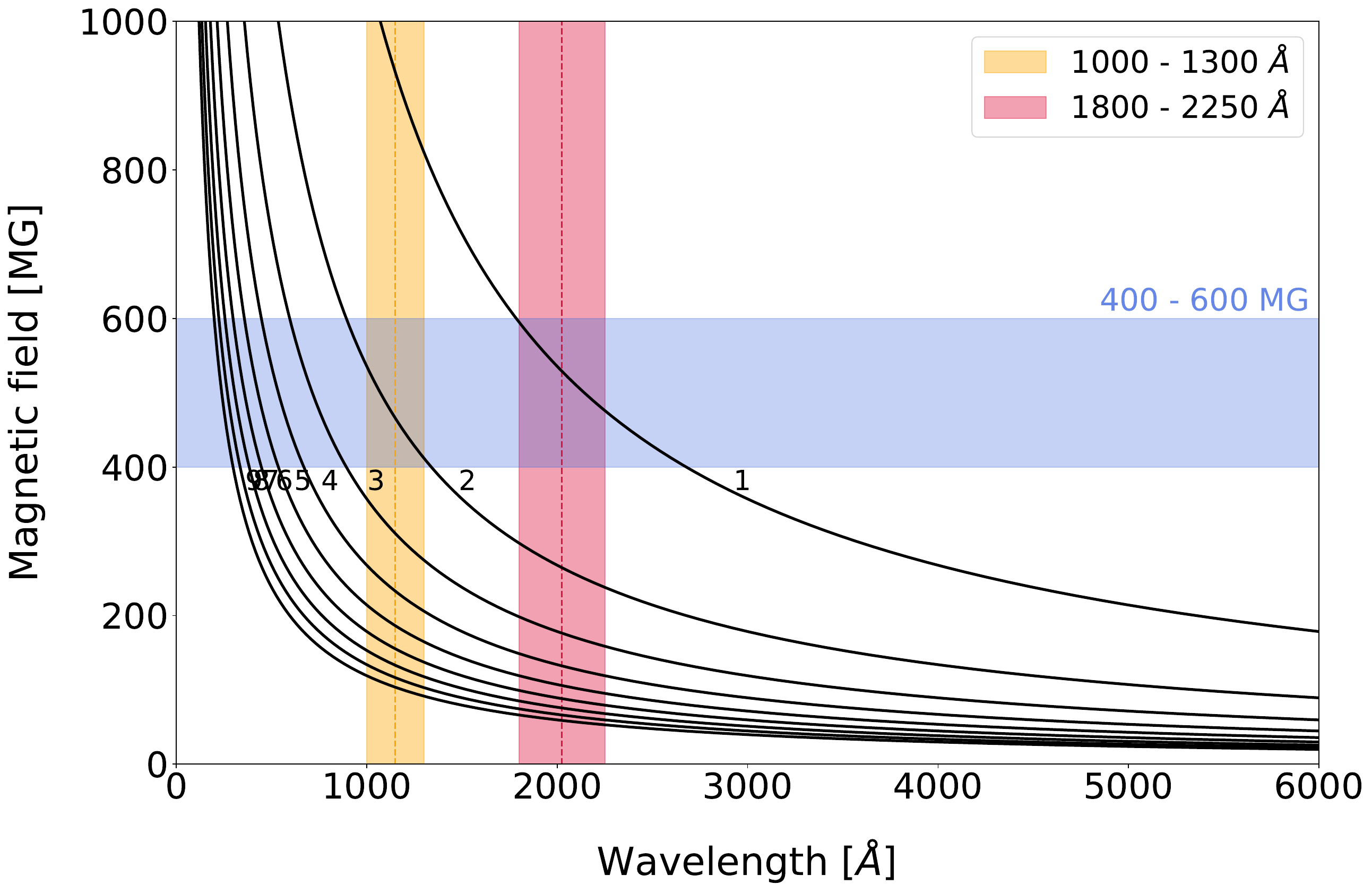}
\caption{Central wavelengths of the first nine cyclotron harmonics (labelled 1 to 9) as a function of magnetic field are depicted by the black lines. The faded red and orange regions mark the ranges covered by the variable humps seen in STIS and COS respectively. The dashed lines mark the centroid of each hump. For magnetic fields ranging from $\approx400-600\,$MG (faded blue region), the positions of the two UV humps are consistent with the first and second cyclotron harmonics.}
\label{fig:cyclotron_res}
\end{figure}

Lastly, the continuum variability in the UV spectrum of ZTF\,J2008+4449 manifests as two broad humps: a strong hump in the FUV, blueward of 1300\,\AA\ (see bottom panel of Fig.\,\ref{fig:uv-mosaic}) and a weaker one between $\approx$\,1800--2250\,\AA\ (see bottom panel Fig.\,\ref{fig:STIS-mosaic}). Their appearance as broad humps and the stark contrast between their large-amplitude modulation on the rotation period of the white dwarf and the rest of the weakly-variable UV continuum (see Fig.\,\ref{fig:all_lc}), suggest that the emission in these ranges may be caused by cyclotron emission close to the surface of the white dwarf. As we do not have a good explanation for the rest of the photometric and spectroscopic variability of this white dwarf (see Section~\ref{sec:var}), we remain agnostic; however, we explore the possibility that indeed these humps are caused by cyclotron emission. For a magnetic field of $\approx$\,400--600\,MG, the location of the observed humps would correspond to the energies of the second and first cyclotron harmonics, as shown in Fig.\,\ref{fig:cyclotron_res}. Alternatively, the two ranges in wavelength could also correspond to the second and third harmonic for a lower field of $\approx$\,300\,MG (in this second case, the brightening in the optical $g$-band at a phase of $\sim0.5$ could be due to emission in the first harmonic). Assuming that there is no contribution from cyclotron emission to the minimum of the light curve, we can extract a lower bound on the cyclotron luminosity from the difference between the minimum and the maximum fluxes of the strongest of the two humps; we find $L_{\rm cyc}$\,$ \gtrsim$\,$3\times10^{30}$\,erg\,s$^{-1}$\,=\,0.001\,L$_{\odot}$. If the cyclotron emission is due to inflowing material in free-fall close to the surface of the white dwarf, this luminosity would correspond to an accretion rate of $\dot{M}_{\rm cyc} \approx 2 \times10^{-13}$\,M$_{\odot}$/yr.

As we just showed, non-magnetic atmosphere models are insufficient in constraining the physical properties of ZTF\,J2008+4449 from its optical and UV SED. We therefore developed state-of-the-art atmosphere models that include continuum and line opacity sets \citep[e.g.,][]{1974IAUS...53..265L,1992A&A...265..570J,1995A&A...298..193M,2014ApJS..212...26S} under
a variable magnetic field strength. For simplicity, we employ a pure hydrogen composition, as we do not have a clear identification of the other metal lines. We describe the models in detail in our companion paper \citep{desai2025}; the models employed here assume a simple dipolar structure with an average magnetic field across the surface of 550\,MG and a field at the magnetic pole of 884\,MG (this is quite arbitrary as we do not have constraints on the structure of the field on the surface of the star). We employ these models to fit the photometric and HST data and constrain the radius and temperature of the white dwarf, as well as the amount of interstellar reddening towards the star. Since we are using hydrogen-dominated models, we choose the spectral phase in which the likely hydrogen absorption is stronger in the UV spectrum (around phase 0.6) and we exclude from the fit wavelengths bluer than 1300\,\AA, where absorption by other elements dominates the spectrum and where the continuum variability is stronger. We find the effective temperature, stellar radius and interstellar reddening to be  $T_{\rm{eff}}$\,=\,35,500\,$\pm$\,300\,K, $R_{\rm WD}$\,=\,4,800\,$\pm$\,300\,km and E(B$-$V)\,=\,0.042\,$\pm$\,0.001, respectively, employing the photogeometric distance estimate from \citet[$350\pm20$\,pc]{bailer-jones2021}. The best-fit model is shown as a black solid line in the upper plot of Fig.\,\ref{fig:SED_BD}, while we show the corner plot for the fitting in Fig.\,\ref{fig:corner} in the Appendix. The derived reddening agrees perfectly with the value reported by the Bayestar19 dust map \citep{green2019}.  From these estimates, we employ the carbon-oxygen core evolutionary models of \citet{bedard2020} to derive a mass of $M_{\rm WD}$\,=\,1.12\,$\pm$\,0.03\,M$_\odot$ and a cooling age of 60\,$\pm$\,10\,Myr (the cooling age is defined as the time since the birth of the object as a white dwarf, and therefore in this case it is the time elapsed since the merger). Since it will be useful for our photoionization model (see Section~\ref{sec:photmodel}), we also derive the number of ionizing photons emitted from the white dwarf for the best-fitting model:
\begin{equation}
    L_{\rm ion} = 4\pi R_{\rm WD}^2 \int^{\infty}_{\nu_0} \frac{F_\nu}{h\nu} d\nu = (2.4\pm0.1)\times10^{42}~{\rm s}^{-1}
\end{equation}
where $F_\nu$ is the flux from the white dwarf model as a function of frequency and $\nu_0$ is the photoionization threshold for hydrogen (13.6 eV). 
All the parameters derived from the SED fitting are listed in Table~\ref{tab:SED}. 
The quoted errors on the parameters are purely statistical in nature, and do not reflect the likely much larger systematic uncertainties introduced by the fact that we do not know the magnetic field strength and structure, and we just employ a fixed simple magnetic field structure in our models, 
while we know that the SED 
is significantly affected by the magnetic field strength and geometry. In future work, we plan to obtain an accurate identification of the metal absorption lines, which will yield a much more precise constraint on the field strength and geometry, and therefore we will be able to provide more accurate physical parameters for the white dwarf. 

\begin{table}
	\centering
	\caption{SED fitting.}
	\label{tab:SED}
	\begin{tabular}{l|c} 
		\hline
            \hline
            $R_{\rm WD}$\,[km] & $4,800\pm300$ \\
            $T_{\rm eff}$\,[K] & $35,500\pm300$ \\
            E(B$-$V) & $0.042\pm0.001$ \\
            $L_{\rm ion}$\,[s$^{-1}$] & $(2.4\pm0.1)\times10^{42}$ \\
            $M_{\rm WD}$\,[M$_\odot$] & $1.12\pm0.03$ \\
            $t_{\rm cool}$ [Myr] & $60\pm10$
	\end{tabular}
\end{table}

\subsection{No trace of a brown dwarf companion}
\label{sec:no_BD}

Although we rule out the presence of a companion on a 6.6 minute orbital period (see Section~\ref{sec:var}), the Balmer emission, the strange aperiodic variability in the optical light curve, and the possible cyclotron emission in the FUV could be signatures of accretion or, more generally, of ionized material in the relative vicinity of the white dwarf. We thus consider the possibility of a Roche lobe-filling, sub-stellar companion, such as a brown dwarf, to be the mass donor in this system. To verify the presence of a companion brown dwarf, we obtained infrared photometry of the system (see Section\,\ref{sec:phot} and Fig.\,\ref{fig:SED_BD}). The white dwarf is sufficiently hot that a brown dwarf close enough to be filling its Roche Lobe would be heated by irradiation. Additionally, as the brown dwarf would likely be tidally locked, we would expect strong variability at infrared wavelengths on the orbital period as the irradiated face of the brown dwarf comes in and out of our line of sight. To estimate the minimum effective temperature for the irradiated face of the brown dwarf companion, we employ a grid of brown dwarf evolutionary models \citep[the Sonora Bobcat grid;][]{sonorabobcat}, from which we obtain the radius and intrinsic effective temperature of the brown dwarfs as a function of their age. We then assume that the brown dwarf is overfilling its Roche lobe and therefore that it is orbiting the white dwarf at the distance for which its radius is equal to the Roche lobe (assuming a circular orbit). As the Sonora models extend into planetary regimes, we take the lower brown dwarf mass limit to be 13 Jupiter masses (M$_{\rm Jup}$) and consider all models below this limit as planets (see Fig.\ref{fig:SED_BD}); nonetheless, our convention of substellar and planetary regimes does not impact the underlying physics behind the following calculation of the companion irradiated face temperatures.

The effective temperature of a brown dwarf can be calculated from the total flux $F$ leaving the atmosphere as 
\begin{equation}
T = \left( \frac{F}{\sigma} \right)^{\frac{1}{4}}
\end{equation}
where $\sigma$ is the Stefan–Boltzmann constant. If the brown dwarf is irradiated, we can separate the total flux $F$ into $F = F_{e}+F_{i}$, where $F_{i}$ is the flux from the brown dwarf interior and $F_{e}$ is the fraction of the white dwarf flux that is absorbed and re-emitted uniformly by the brown dwarf \citep{jermyn2017}. The former of the two contributions is obtained from the intrinsic, non-irradiated effective temperature $T_{\rm BD}$ of the brown dwarfs in the Sonora Bobcat grid as $F_{i}=\sigma T_{\rm BD}^4$. The latter contribution is related to the white dwarf luminosity by 
\begin{equation}
F_{e} = \frac{L_{WD}}{16\pi a_{orbit}^2} (1-A_{B}) = \frac{\sigma R_{\rm WD}^2T_{\rm WD}^4}{4a_{orbit}^2} (1-A_{B})
\end{equation}
\citep{jermyn2017}, where $L_{WD} = 4\pi R_{\rm WD}^2 \sigma T_{\rm WD}^4$ is the white dwarf luminosity and $a_{orbit}$ is the orbital separation between the white dwarf and brown dwarf. We also added the $(1-A_{B})$ factor, where $A_{B}$ is the Bond albedo \citep[see e.g.,][]{karttunen2017} of the brown dwarf, to account for the fraction of the incoming radiation from the white dwarf which is scattered into space and does not contribute to the re-emission by the brown dwarf. Finally, we used a conservatively large value of 0.5 for the Bond albedo (implying a lower absorbed flux fraction and hence a lower irradiated temperature), as per the study of \citet{marley1999} and in agreement with the study by \citet{hernandez-santisteban2016}.

The resulting effective temperatures of the brown dwarf and planetary irradiated faces are plotted in the lower panel of Fig.\,\ref{fig:SED_BD}, against both the orbital separation at which the companions fill their Roche lobes and the corresponding orbital period. The different tracks show distinct model ages, the marker sizes are proportional to their radii and the colourbar shows the brown dwarf masses. Models corresponding to the planetary regime ($< 13$\,M$_{\rm Jup}$) are shown in black. It can be seen that the minimum brown dwarf irradiation temperature attainable is $\approx$\,1700\,K, as marked by the golden horizontal line, for an orbital period of $\approx$\,160\,minutes. We also mark in black the lowest mass planet in the Sonora grid ($\approx0.5$\,M$_{\rm Jup}$), which would fill its Roche lobe at a period of $\approx$\,1180\,minutes with a temperature of $\approx$\,900\,K. The purple solid line in the upper panel of Fig.\,\ref{fig:SED_BD} shows the expected combined spectrum for the white dwarf with a 1700\,K companion brown dwarf. We can see that the infrared photometry is inconsistent with such a hot brown dwarf. The coldest brown dwarf/planet that would be consistent with the WIRC IR photometry within three standard deviations has a temperature of  $\approx$\,800\,K (solid pink line): this is significantly below the lower limits on irradiation temperatures for any brown dwarf and for planets above 0.5\,M$_{\rm Jup}$. The WIRC J and H data was obtained over a span of two consecutive hours; furthermore, we have an additional 2-hour long J and K observation taken a few days later and J does not show any hint of variability between the two epochs. Given the possible brown dwarf orbital periods, it is highly unlikely that we have only observed the cold, non-irradiated face of a potential brown dwarf. Hence, we confidently rule out the presence of a brown dwarf companion and, implicitly, that of accretion due to ongoing mass transfer as part of a cataclysmic variable evolution scenario. It is however still possible for a cold planet to be the mass donor in the system, given the orbital period at which it would be filling its Roche lobe. 

\subsection{Doppler tomography of the Balmer emission}
\label{sec:Balmer_loc}

From Fig.\,\ref{fig:optspectrum} we can see that the H$\alpha$ emission presents Doppler shifts as high as $\approx$\,$\pm$2000\,km\,s$^{-1}$, and that there is a narrow peak that jumps from blue-shifted to red-shifted values at $\approx$\,1700\,km\,s$^{-1}$.  These velocity features as a function of phase are directly related to the spatial geometry of the emitting gas \citep[see e.g.][]{marsh88}; in order to visualize this geometry, we employ the technique of Doppler tomography \citep{marsh2001}.
Doppler tomography, first introduced by \citet{marsh88}, is a technique designed to reconstruct the spatial geometry of emitting sources (e.g., circumstellar material) from the period-resolved variations of spectral features associated with them. Numerically, Doppler tomography is usually performed via a maximal entropy algorithm: individual values representing emission densities are assigned to each pixel among a 2D spatial canvas via the maximisation of an appropriately defined associated entropy. This is done under the constraint that the resulting distribution of emission sources is a solution to (i.e., can produce) the phase-resolved variation of its emission features. The need to define an entropy associated to the source distribution arises from the spatial degeneracy of the solution and ensures that the final result is minimally assumptive, i.e., it is reconstructed making use of no information other than that contained by the variations of the spectral features.
However, as these variations only encode information regarding the emitting sources' radial velocity with respect to the observer,  Doppler tomography only recovers the relative distribution of emission density in velocity $Vsin(i)$ space (where \textit{i} is the inclination of the white dwarf rotation axis with respect to the line of sight). This velocity-space solution must then be appropriately interpreted as one of multiple, doubly degenerate, real-space distributions of emitting sources.

\begin{figure}[tb!]
\centering
\vspace{10pt}
\includegraphics[width = \columnwidth, trim = 3cm 13cm 5.5cm 2cm]{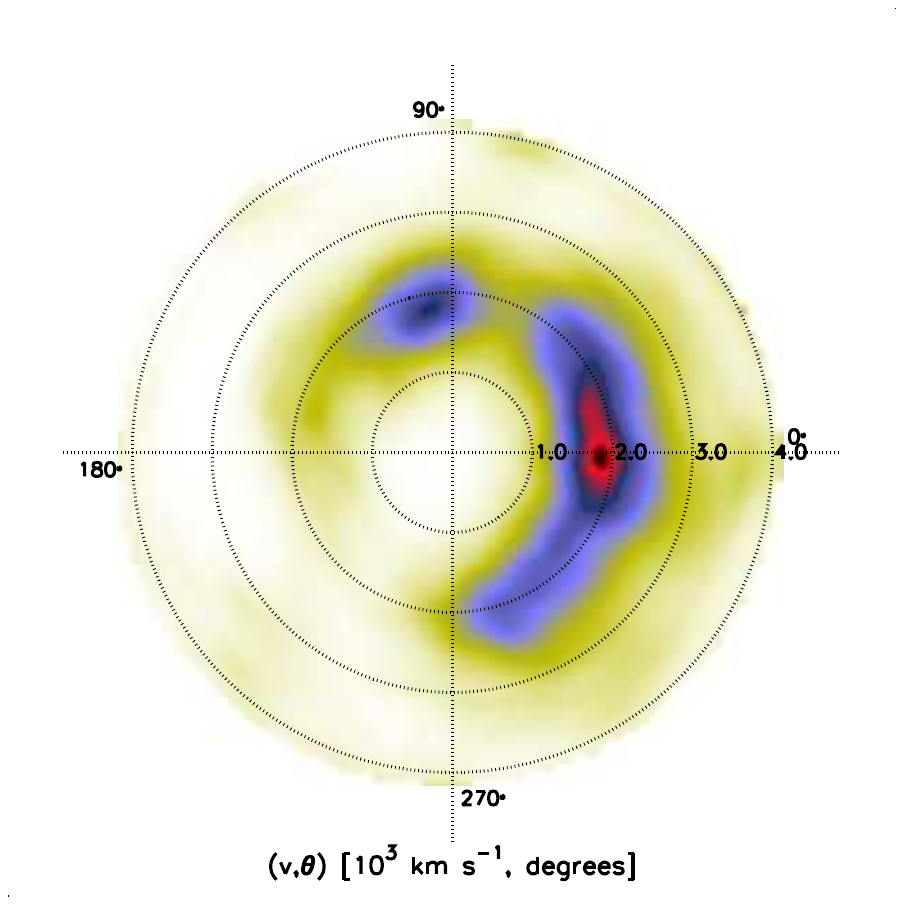}

\caption{Doppler tomogram of the H$\alpha$ emission feature: the minimally assumptive H$\alpha$ emission density distribution, extrapolated from the emission line variation on the 6.6-minute white dwarf spin period, is shown in the $v=Vsin(i)$ velocity space ($i = $\,inclination of the white dwarf rotation axis). The white dwarf is located at the origin at the coordinate system $(v, \theta)$; the evolution of the obtained source distribution with rotation phase can be recovered by rotating the coordinate system uniformly and clockwise over one variation cycle. The concentric dotted lines mark regions of constant velocity separated by 1000\,km/s from each other, while the colours illustrate the density of emission. The emission distribution resembles a half-ring, with velocities ranging between $\approx1500$ and $\approx2500$\,km/s and peaking at a velocity of $\approx1700$\,km/s.
\label{fig:tomo}}\vspace{-5pt}
\end{figure}

We have used the Doppler tomography software DOPTOMOG 2.0 \citep{kotze2015,kotze2016} to map the emission density from the shape of the H$\alpha$ line variability. Fig.\,\ref{fig:tomo} depicts the obtained solution, in the inclination-adjusted velocity space. The values of velocities depicted in the tomogram are as seen by the observer, with the white dwarf lying at the origin of the $[V_{x}-V_{y}]sin(i)$ coordinate system (it is stationary with respect to the observer). The tomographic image in Fig.\,\ref{fig:tomo} is shown at rotation Phase 0; the amplitudes and phase evolution of the velocity vectors in the observer's frame can be recovered by rotating the coordinate system over the obtained distribution, uniformly and clockwise over one variation cycle. For a more detailed explanation of tomography, see e.g., \citet{kotze2015,kotze2016}. The tomogram shows that the material is distributed in a half-circular structure centered around a velocity amplitude of $\approx$\,1700\,km\,s$^{-1}$, dominated by a slightly asymmetric over-dense region. 

Accounting for inclination, we note that a circular-arc distribution of emitting material in velocity space (i.e., spanning a single velocity amplitude) may straightforwardly (but not necessarily) correspond to a circular-arc distribution in real-space; the velocities of both circular Keplerian movement and rigid corotation only depend on the radial coordinate in the orbital plane. As we explain in more detail below, the velocities are too high to be explained by Keplerian orbits (the material would be close enough to the star to be forced to corotate with the strong magnetic field and we would be able to detect Zeeman splitting in the emission line). We therefore suggest that the velocities reflect rigid corotation of the emitting material in the magnetosphere of the white dwarf; as corotation velocities increase with distance from the star, in this case, the tomogram also provides an accurate representation of emission density in real-space, in a frame corotating with the white dwarf.

\begin{figure*}[tb!]
\centering
\vspace{10pt}
\includegraphics[width = 0.9\columnwidth]{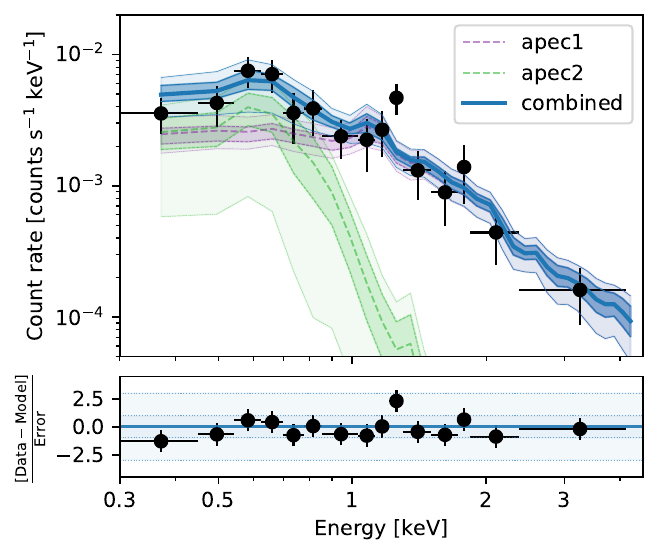}
\includegraphics[width = 0.9\columnwidth]{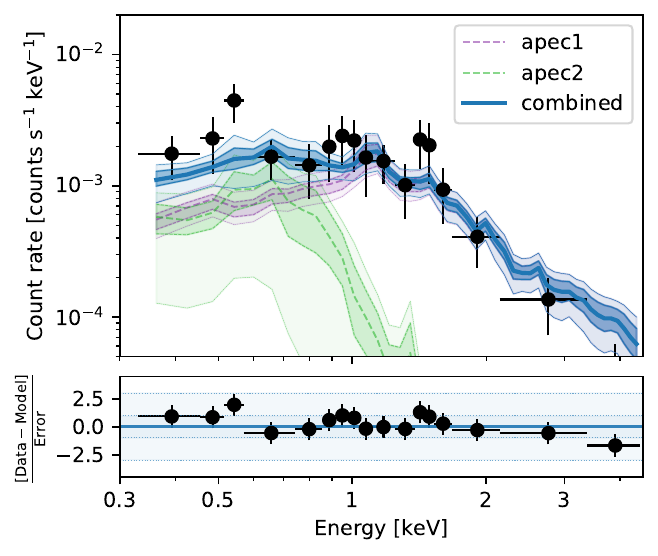} \\
\includegraphics[width = 0.9\columnwidth]{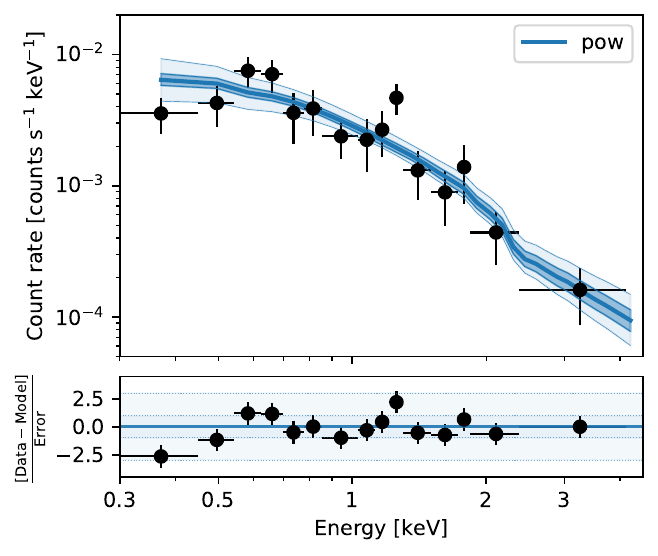}
\includegraphics[width = 0.9\columnwidth]{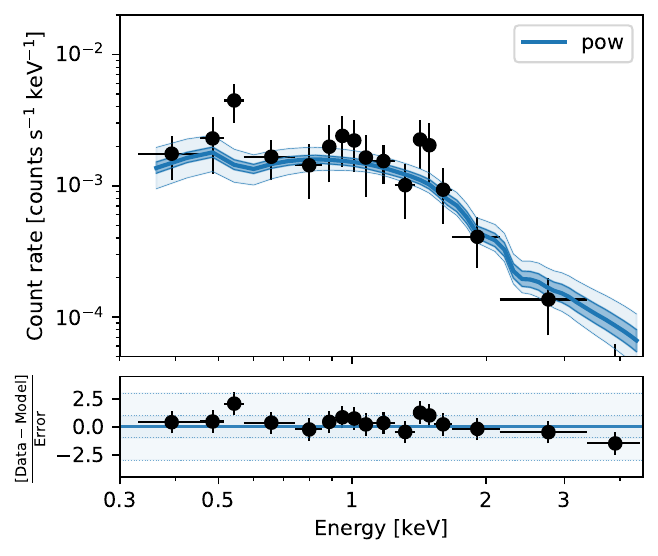} \\
\caption{XMM spectral fit: The X-ray spectra are shown as photon count rate as a function of energy (black data points), for the XMM PN detector (left column) and the combined M1 and M2 detectors (right column). The best fitting two-temperature APEC and power law emission models, convolved with the instrumental response, are shown in blue on the top and bottom rows respectively, with the 3-$\sigma$ contours shown as shaded regions. In the case of the two-temperature APEC model, the two individual temperature components are shown in green and purple, along with their 3-$\sigma$ contours. The residuals of the fitted models, in both detectors, are shown on the bottom of their respective plots.  
\label{fig:bxa-fit}}
\end{figure*}

\subsection{X-ray Analysis}
\label{sec:xray_analysis}
We conduct an analysis of the X-ray spectral data using the Bayesian X-ray Analysis package \cite[BXA;][]{buchner2016-bxa}, which integrates the nested sampling algorithm \texttt{ULTRANEST} \citep{buchner2016-bxa} with an X-ray spectral fitting platform. For the spectral modeling, we use \texttt{XSPEC} \citep{arnaud1996XSPEC} accessed through its Python interface, \texttt{PyXSPEC} . Our procedure involves jointly fitting the background-subtracted data from all three XMM EPIC cameras (PN, MOS1 and MOS2). To ensure reliable results in the low-count regime, we employ the Cash (C-) statistic for fitting \citep{cash1979}. Before fitting, we combined the spectra using the \texttt{SAS} command \texttt{epicspeccombine} and binned according to the optimal binning algorithm described by \citet{kaastra2016-optimal-binning}. We combine the PN spectra for both epochs, and all MOS (M1 and M2) for both observations, resulting in two optimally-binned spectra which we fit simultaneously. We tested two different spectral models: i) a simple power-law and ii) a two-temperature, optically thin plasma model (\texttt{APEC}) with Solar abundance. Both models incorporate an absorption component to represent Galactic $n_{\rm H}$ absorption, as well as any intrinsic absorption within the system. Rather than rely on dust maps to estimate the Galactic absorption, we use the value of extinction derived from our optical and UV SED fitting (see Section~\ref{sec:SED}). We convert this to hydrogen column density according to Equation 1 of \citet{guver2009}. This has the advantage of implicitly accounting for both Galactic absorption and absorption intrinsic to the system, assuming that the X-ray and UV emission originates from the same place. From our SED analysis, we determine an extinction of $E(B-V)=0.042\pm0.001$, from which we find a hydrogen column density of $n_{\rm H}=(2.74 \pm 0.69)\times10^{20}\,\rm{cm^{-2}}$. We adopt a Gaussian prior on the absorption, centred on this value with a standard deviation equal to the uncertainty. 

For the two components of the optically thin plasma model, we find temperatures of $kT_1=3\pm1$\,keV and $kT_2=0.23^{+0.04}_{-0.03}$\,keV. For the power law, we recover a photon index of $\Gamma=2.2\pm0.1$. The Bayes factor for the comparison of the two models is given by $K=Z_{\rm pow}/Z_{\rm apec}$, where $Z$ denotes the Bayesian evidence (see e.g., \citealt{buchner2016-bxa}) for each model, from which we find a value of $K\approx2$ which is sufficiently small that there is no Bayesian evidence that either model is favoured.
We check whether the two best-fitting models provide a reasonable fit to the data using the \texttt{XSPEC} \textit{goodness} routine, which uses Monte Carlo simulations to estimate the fraction of synthetic datasets that would yield a better fit statistic than the observed data. For the 2-temperature model, from 1000 samples we find that 58.4\% of the simulated spectra have lower, or better, fit statistics compared to our best-fitting model, implying that the model is a reasonable description of the observed data. For the power law model, we find 66.6\% of the simulated spectra to have favourable fit statistics compared to our best-fitting model. We thus conclude that both the power law and 2-temperature \texttt{APEC} models provide an adequate fit for the measured X-ray spectroscopic data. In both cases, the posterior on the absorption, $n_{\rm H}$, closely resembles the prior informed by UV/optical extinction. For both models, we fit the combined spectra, but also the spectra from the two epochs separately. The top and bottom panels of Fig.\,\ref{fig:bxa-fit} show the best-fitting model, convolved with the instrumental response, for the \texttt{APEC} and power law models, respectively. The posterior distribution of fitted parameters are shown in the Appendix in Figs.\,\ref{fig:bxa-corners-plaw}\,\&\,\ref{fig:bxa-corners-apec}. In black in the Appendix plots, we show the posterior distributions resulting from the fit of the combined spectra, whilst in blue and orange we show the fits to each epoch separately. From this analysis, we conclude that there is no evidence of variability in the X-ray spectral properties over the 2-day period separating the two epochs. 

\begin{figure*}[tb!]
\centering
\vspace{10pt}

\includegraphics[width = 0.9\columnwidth]{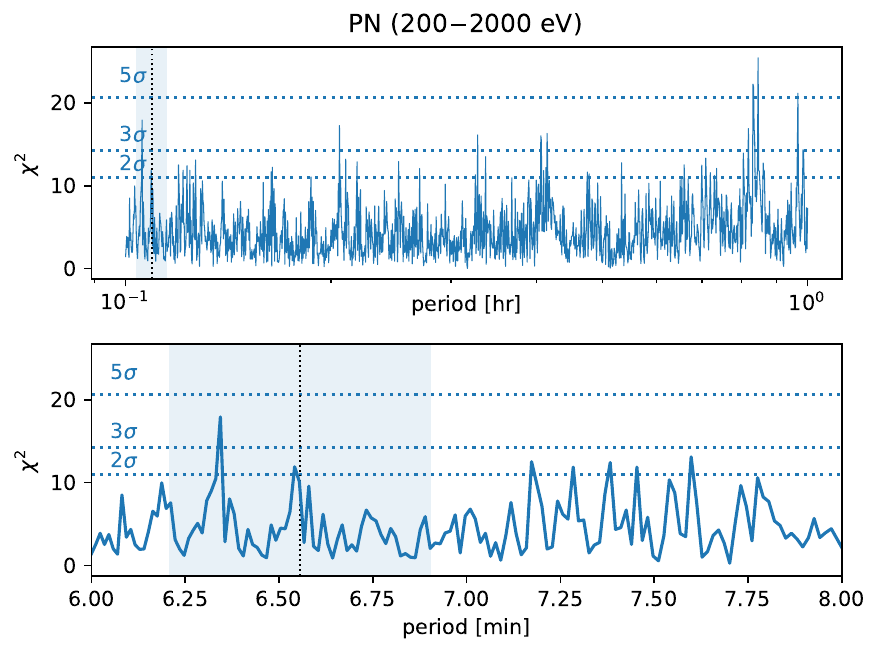}
\includegraphics[width = 0.9\columnwidth]{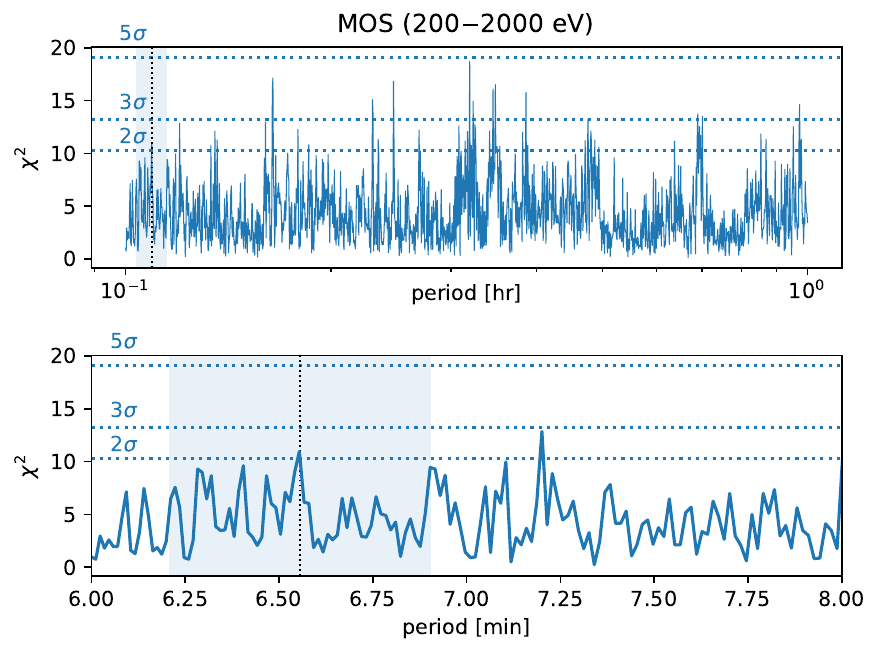}

\caption{Period search in XMM-Newton EPIC data using a $\chi^2$ test at a range of trial periods to search for a periodic signal. We show the resulting periodagram for the EPIC-pn and MOS data on the left and right, respectively. The vertical dotted line shows the period of the white dwarf at the epoch of the observation.
\label{fig:lc-xray-chisq}}
\end{figure*}

\begin{figure}[tb!]
\centering
\vspace{10pt}

\includegraphics[width = 0.9\columnwidth]{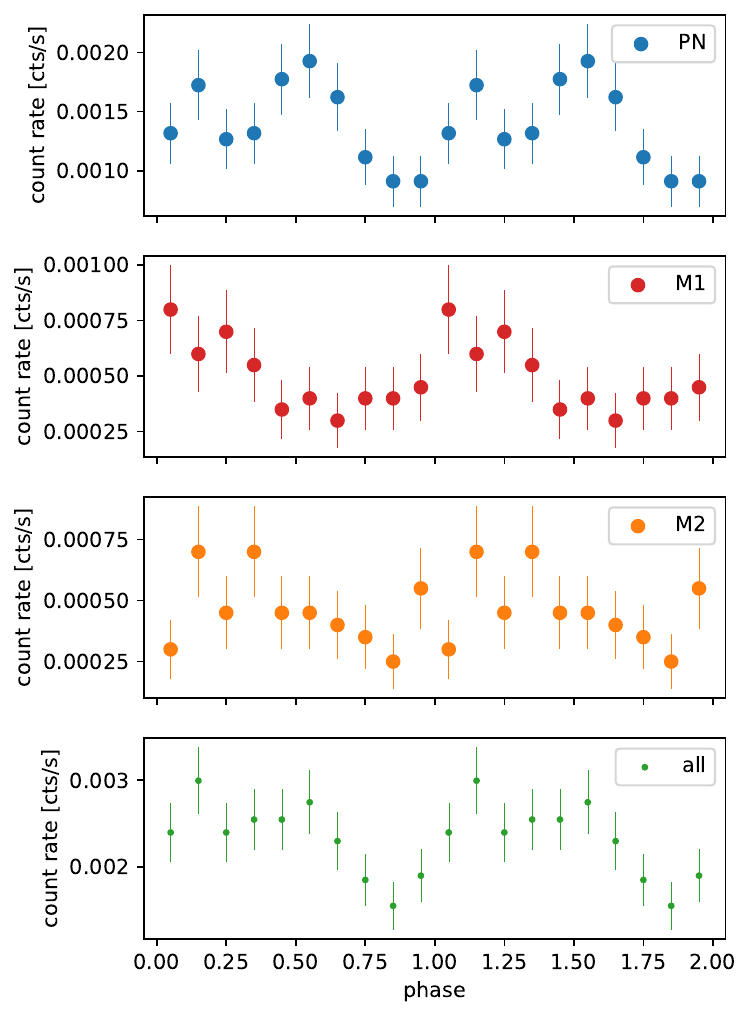}
\caption{Phase folded light curve for the XMM-Newton EPIC-pn and MOS data. Barycenter-corrected events have been folded on the photometric rotation period, with the ephemeris defined for the epoch of the XMM observations, such that phase zero is consistent with all other plots in the paper (the minimum in the optical light curve).
\label{fig:lc-xray}}
\end{figure}

We perform a period search in the XMM-Newton EPIC data by phase-folding and binning the event times at a range of trial periods from 0.1--1 hours, with 5 phase bins. We combine both epochs of XMM data, making use of the barycenter-corrected event times, selecting events with energies in the range 0.2--2.0\,keV to maximize the signal-to-noise. We fit each trial lightcurve with a constant, flat line and assess the goodness-of-fit using the $\chi^2$-statistic, such that large values of $\chi^2$ indicate a lightcurve less-well described as being constant. Fig.\,\ref{fig:lc-xray-chisq} shows the results of this period search for EPIC-pn and MOS (M1 and M2 combined). We indicate the 3 and 5 standard deviations in $\chi^2$ value with the horizontal dashed line. We marginally recover the white dwarf rotation period (shown in the vertical dotted line) with a confidence of 2$\sigma$, which is inadequate to confirm the white dwarf rotation period signal from the current X-ray observations. Fig.\,\ref{fig:lc-xray} shows the phase-folded and phase-binned X-ray events in the 0.2--2.0\,keV band, folded on the rotation period. Although we can see a hint of variability on the rotation period (mostly in the more sensitive PN), a longer, deeper observation is needed to confirm it.

\subsection{RSG 5 cluster membership}
\label{sec:cluster}
ZTF\,J2008+4449 was identified by \citet{Miller2025} as a high-probability white dwarf member of the open cluster RSG 5. The identification was part of a search for high-mass cluster white dwarfs using the \citet{2023A&A...673A.114H} cluster catalog and the Gaia EDR3 white dwarf catalogue from \citet{gentile-fusillo2021}, with the goal of refining the initial-final mass relation for massive white dwarfs. However, they did not include ZTF\,J2008+4449 in their cluster-based white dwarf initial-final mass relation for two reasons: because of its likely nature as a merger remnant and because the white dwarf appears too old to be part of the cluster according to the cluster age derived in \citet{2023A&A...673A.114H}, and therefore they classify it as a likely interloper. It is worth mentioning that ZTF\,J2008+4449 was also independently identified as a candidate member of RSG 5 by \citet{2022AJ....164..215B}.

\begin{figure}
\centering
\includegraphics[width=1.0\columnwidth]{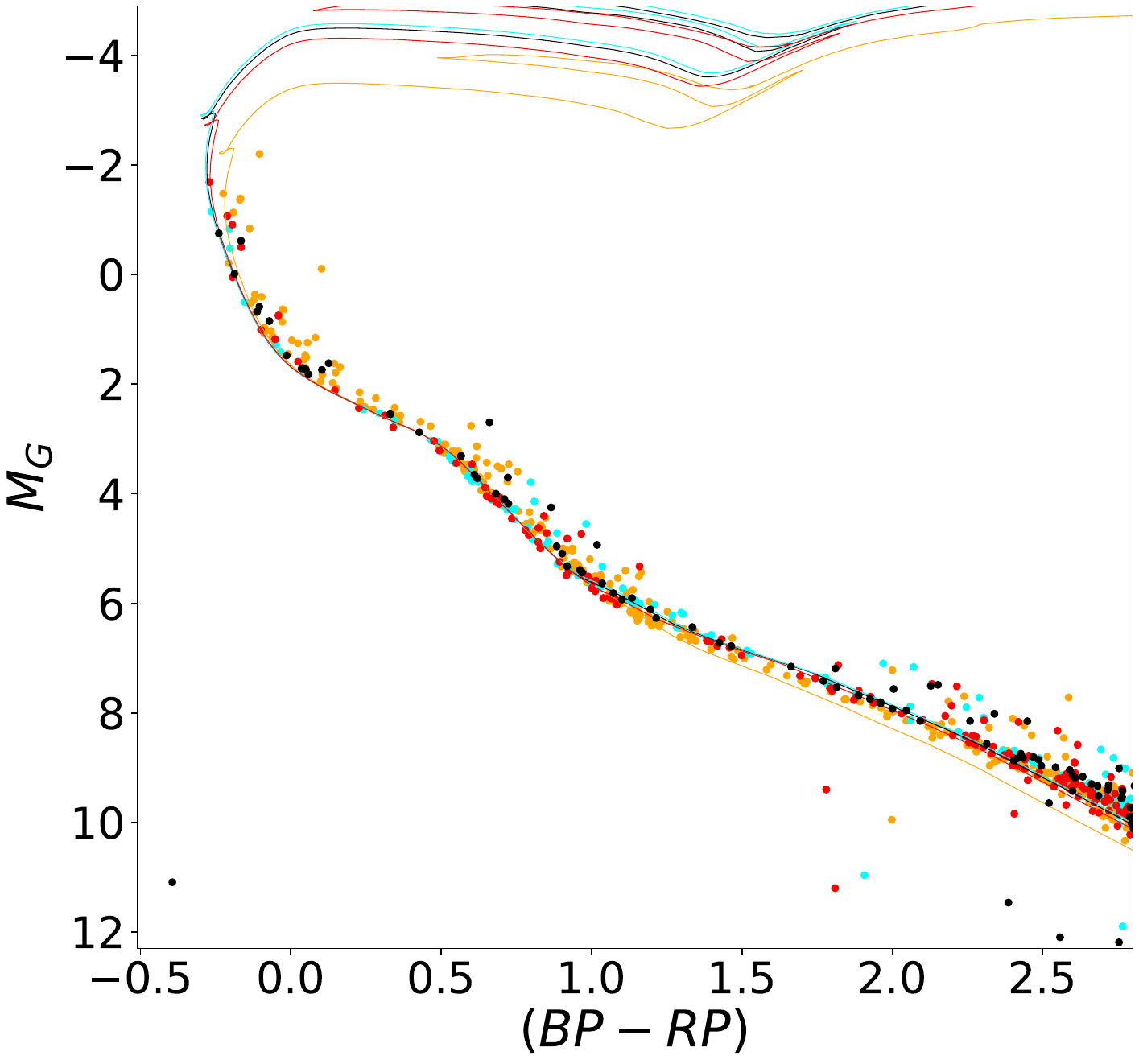}
\caption{Dereddened Gaia DR3 CMD for RSG 5 (black), compared with the CMDs of three reference clusters: Alpha Persei (81\,Myr, orange), IC 2391 (43\,Myr, cyan), and NGC 2451A (50\,Myr, red). The members of each cluster are those with at least a $50\%$ membership probability in the \citet{2023A&A...673A.114H} Milky Way cluster catalogue, with clusters dereddened using the catalogues mean $A_v$. Overlaid are PARSEC isochrones \citep{2012MNRAS.427..127B,2014MNRAS.445.4287T,2014MNRAS.444.2525C,2018MNRAS.476..496F,2022A&A...665A.126N} for our best-fit cluster age of RSG 5 (45\,Myr), as well as the kinematic ages from \citet{2021arXiv211004296H} for the comparison clusters, with the colors matching those of the CMDs.}
\label{fig:RSG5_age}
\end{figure}

Although it is not relevant to the rest of the analysis presented in this paper, the possibility of cluster membership for the white dwarf would be interesting because it would provide a constraint on the total age of the system and therefore on the evolution of the binary before the merger. In order to explore further this possibility, we re-derive the age of RSG 5 with PARSEC 2.0 isochrones to better characterize its uncertainty \citep{2012MNRAS.427..127B,2014MNRAS.445.4287T,2014MNRAS.444.2525C,2018MNRAS.476..496F,2022A&A...665A.126N}. We use stars with at least $50\%$ probability of being cluster members in \citet{2023A&A...673A.114H}, and we apply the heuristic isochrone-fitting technique of \citet{Miller2025}, which uses a visual fit to the main-sequence turnoff (MSTO) to select the best age and then derives $1\sigma$ uncertainties via a weighted $\chi^2$ calculation that also heavily emphasizes the MSTO. Our fit yields a solar metallicity cluster with an age of $45^{+8}_{-11}$\,Myr, in strong agreement with the pre-main sequence age of $46^{+9}_{-7}$\,Myr reported by \citet{2022AJ....164..215B}. The cluster CMD and best-fit isochrone are shown in Fig.~\ref{fig:RSG5_age}.

Since the MSTO of RSG\,5 is sparsely populated, we also overlaid its CMD onto those of IC 2391, NGC 2451A, and Alpha Persei, each of which have estimated kinematic ages, independent of isochrone fitting, from \citet{2021arXiv211004296H}. By visually comparing the lower main sequences, we find that RSG 5 falls squarely between IC 2391 (43\,Myr) and NGC 2451A (50\,Myr), and is clearly offset from Alpha Persei (81\,Myr), as illustrated in Fig.~\ref{fig:RSG5_age}. Given this agreement among ages from the MSTO, lower main sequence, and pre-main sequence, we conclude that the age estimate is solid and that there is no evidence for the true age of RSG\,5 to substantially exceed $\approx50$\,Myr.

Our estimate for the cooling age of ZTF\,J2008+4449, at $t_{\rm cool}=60\pm10$~Myr, is of the order of the age of the cluster, or slightly higher. However, the cooling age is not the total age of the system, as it only indicates the time since the white dwarf was born as a merger remnant, and does not include the main sequence lifetime of the progenitor stars nor the binary inspiral time due to gravitational wave emission. Supposing ZTF J2008+4449 was the product of single star evolution, its mass ($1.12\pm0.03$\,M$_\odot$) would imply a progenitor mass of $\approx6$\,M$_\odot$ \citep{Miller2025,cunningham2024}, corresponding to a main-sequence lifetime of $\approx70$\,Myr from PARSEC isochrones at solar metallicity, which one would have to add to the cooling age to obtain the total age of the system. As we think the star is the product of a double white dwarf merger, we need to consider the evolution of the binary as well. In some binary evolution scenarios, the merger time delay can be very small \citep{toonen2012}, especially if the progenitor stars are massive; however, even for an 8 M$_\odot$ star, the main sequence lifetime is close to 40 Myr. We therefore conclude that the total age of ZTF\,J2008+4449 is likely higher than 100~Myr and therefore exceeds the age of RSG 5 significantly, indicating that the object is not a cluster member.

\section{Discussion}
\label{sec:discussion}
The rapid spin of the white dwarf, together with the high magnetic field threading its surface, strongly suggest that ZTF\,J2008+4449 was born in a double white dwarf merger, and the absence of infrared excess exclude the possibility of mass transfer from a stellar or brown dwarf companion. However, the detection of Balmer emission and X-rays, together with the measured $\dot{P}$, suggest the presence of circumstellar material. In the next sections, we discuss the origin of the 6.6 minute variability and the presence of circumstellar material trapped and ejected by the spinning magnetosphere, the observational constraints on its properties, and possible scenarios for its origin and for the nature of its interaction with the magnetosphere. 

\begin{figure}[tb]
\centering
\includegraphics[width=0.74\columnwidth]{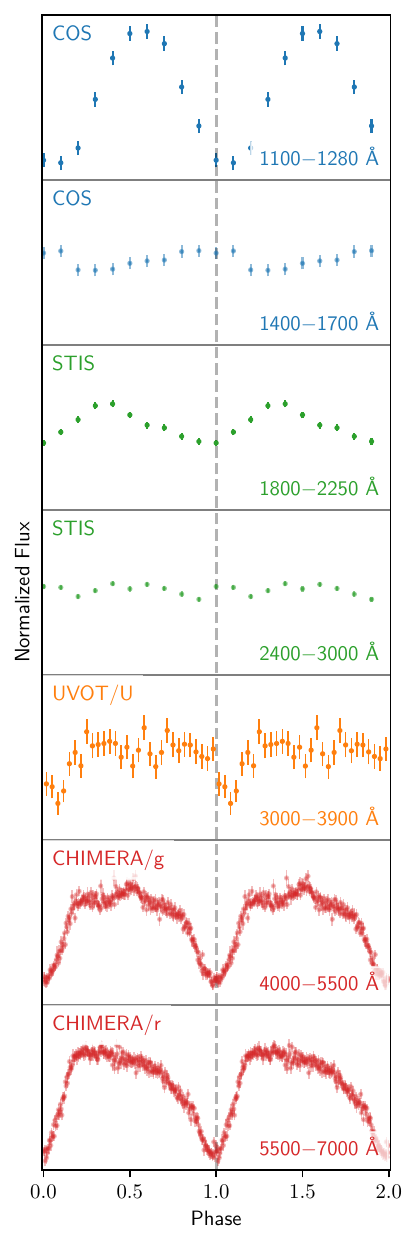}\vspace{-10pt} 
\caption{Optical and UV light curves for ZTF\,J2008+4449, phased at the correct period for each epoch as derived in Section~\ref{sec:period}. Phase 0 corresponds to the minimum in the $r$-band. While for the CHIMERA and XMM/UVOT data the filter is indicated in the upper left corner of each plot, for the STIS and COS data we extracted light curves by averaging the flux in the wavelength range indicated in the bottom right of each plot.}\vspace{-10pt}
\label{fig:all_lc}
\end{figure}

\subsection{Variability}
\label{sec:var}
ZTF\,J2008+4449 shows variability on the 6.6 minute period at all wavelengths from the optical to the FUV, and possibly the X-rays (see Fig.\,\ref{fig:all_lc} and Fig.\,\ref{fig:lc-xray}). The optical light curves present a broad, dip-like feature which differs in depth and in shape between filters: it is deeper in the $r$-band ($\approx$20\%) and it becomes shallower at bluer wavelengths ($\approx$17\% in $g$ and $\approx$10\% in $U$), and the ``ingress'' is slower in the $r$-band than in the $g$-band, while the minimum appears shifted by about 0.1 in phase in the $U$-band. Additionally, the $g$-band displays a ``brightening'' peak at 0.5 phase away from the ``dip''.  The light curve shows aperiodic variability in the bright flat portion of the light curve (between phases 0.1 and 0.7), which results in enhanced scattering in the phase-folded light curve (see also Fig.\,\ref{fig:lc-unbinned}). We currently do not have a model to explain the unusual shape of the light curve, which is morphologically different from the sinusoidal light curves usually observed in magnetic white dwarfs. One possibility would be that some material trapped in the magnetosphere is partially eclipsing the white dwarf; in fact, the optical light curves bear some similarity to the light curves of an M dwarf that shows evidence of material trapped in the magnetosphere, for which this possibility has been suggested \citep{bouma2025}. However, the minimum of the dip-like feature corresponds to the maximum blueshift in the H$\alpha$ emission, indicating that at least the glowing disk (see Section\,\ref{sec:disk}) cannot be responsible for the dip. Material trapped in the magnetosphere further out could be the cause of the dips.

Another issue with the dip scenario is that, as we go to bluer wavelengths, the amplitude of the variability decreases and the dip-like feature disappears. The spectra in the COS and STIS wavelength ranges show very little continuum variation (see Figs.\,\ref{fig:uv-mosaic} and \ref{fig:STIS-mosaic}), with the exception of two humps that show quasi-sinusoidal variations in the 1100--1300\,\AA\ range and in the 1800--2250\,\AA\ range. In Section\,\ref{sec:SED} we point out that this variation could be due to cyclotron emission close to the white dwarf surface as those energies would correspond to the first and second cyclotron harmonics in a magnetic field of about $\approx500$\,MG (see Fig.\,\ref{fig:cyclotron_res}).

Finally, the white dwarf shows strong variability in the Balmer emission lines, which we will analyse in detail in the next few sections, and in one of the absorption lines in the UV. The optical absorption lines are extremely weak and only detectable in the phase-averaged spectrum (see Fig.\,\ref{fig:optspectrum}). The signal-to-noise is too low in the phase-resolved spectra to constrain variability. In the ultraviolet, on the other hand, we detect stronger absorption lines in the 1000--1300\,\AA\ range (see Fig.\,\ref{fig:uv-mosaic}). As we explain in Section\,\ref{sec:SED} and in Appendix\,\ref{sec:metalsUV}, the absorption features most likely correspond to Zeeman components of absorption lines of metals (such as silicon, carbon and nitrogen), shifted by the strong magnetic field. The only exception is the line at 1344\,\AA, which, as we show in Fig.\,\ref{fig:uv-mosaic}, is likely a Zeeman component of Lyman-$\alpha$. None of the candidate metal lines show strong variability (once one subtracts the continuum variation) either in shape or in wavelength, while the Lyman-$\alpha$ component varies strongly in equivalent width, completely disappearing at Phase 0 (see Fig.\,\ref{fig:uv-mosaic}). Strong variation in hydrogen absorption has been observed in other rapidly spinning white dwarfs: the emerging new class of Janus-like double-faced white dwarfs \citep[see e.g.][]{caiazzo2023,Koen2024,cheng2024,moss2024,moss2025}. In these systems, hydrogen and helium on the surface vary in anti-correlation, with the most extreme cases showing one of the two elements disappearing when the other is at maximum. It has been suggested that the variation is due to the presence of a magnetic field threading the surface of the white dwarf and a possible asymmetry in hydrogen content due to a merger origin \citep{bedard2025}; the variation that we see in the Lyman-$\alpha$ component of ZTF\,J2008+4449 could be due to a similar mechanism.

Due to the presence of circumstellar material and the strange shape in the optical light curve, we considered the possibility that the periodic variability on the 6.6 minute is connected to the orbital period of a binary. However, we can exclude this possibility. For a white dwarf in a compact binary, an orbital period of 6.6 minutes would be too short to host a stellar companion (the period minimum for cataclysmic variable systems is $\approx\,80$ minutes). If the 6.6 minute corresponded to an orbital period, it would require the white dwarf companion to be a compact object, such as a black hole, a neutron star or another white dwarf. We can exclude a neutron star or a black hole companion since it would not explain the Balmer emission, as the white dwarf would not be mass transferring (it would not fill its Roche lobe) and because we would see strong radial velocity shifts in the UV absorption lines. A companion white dwarf could be the donor in a 6.6-minute period, like the direct-impact accretor HM Cancri \citep{2003ApJ...598..492I}. However, the donor would have to be a low-mass white dwarf to fill its Roche lobe, and it would outshine the high-mass white dwarf accretor. Additionally, we would expect much lower velocity shifts in the Balmer emission \citep{2010ApJ...711L.138R}. Finally, the modulation in the Lyman-$\alpha$ absorption in phase with 6.6 minutes strongly hints to a rotational rather than orbital modulation. The 6.6 minute periodic variability is therefore most likely associated with the rotation of the white dwarf, rather than the orbital period of a possible binary system.

\begin{figure}[tb]
\centering
\includegraphics[width = \columnwidth]{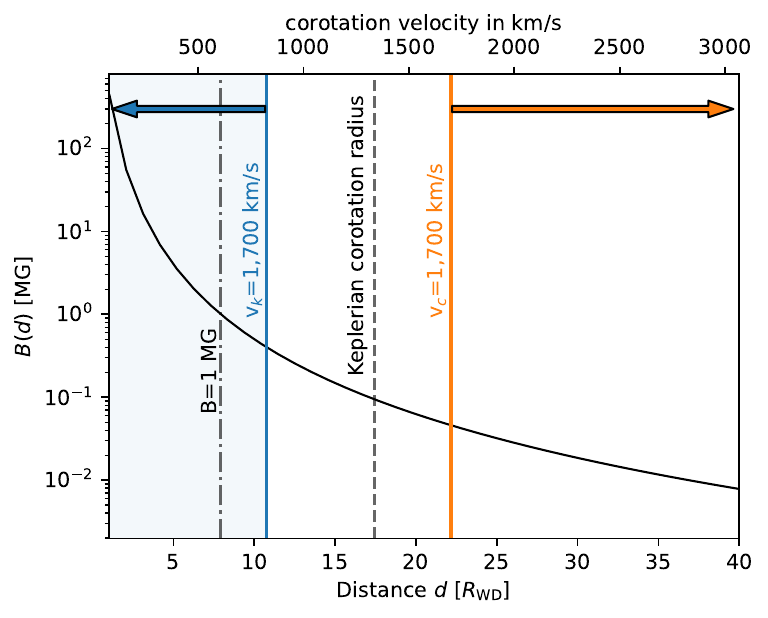}
\caption{Diagram showing the possible location of the line-emitting material. On the x-axis, we plot the distance from the white dwarf ($d$) in white dwarf radii ($R_{\rm WD}\approx 4800$\,km). The black solid line indicates the strength of the magnetic field at the distance $d$, assuming a field of dipole strength at surface of $\approx500$\,MG. We highlight with a dash-dotted vertical line where the field becomes weaker than one MG. The top x-axis marks the corotation velocity corresponding to each radial coordinate. The ratio of centrifugal to gravitational accelerations is unity at the Keplerian corotation radius (vertical dashed line). The vertical blue line indicates the distance at which the Keplerian velocity equals the observed velocity in H$\alpha$; the blue shaded region shows where the emitting material would lie if it was moving in Keplerian orbit, depending on inclination (with the vertical blue line indicating the maximum distance for an inclination $i=90^\circ$). We think the material is instead corotating with the white dwarf, with the vertical orange line indicating the distance at which the corotation velocity equals the observed velocity in H$\alpha$. The orange shaded region shows where we expect the material to lie, depending on inclination (with the vertical orange line indicating the minimum distance for $i=90^\circ$).
}
\label{fig:diag}
\end{figure}

\subsection{Origin of the Balmer emission}
\label{sec:location}
The Balmer emission indicates the presence of hot ionized gas surrounding the white dwarf. The absence of Zeeman splitting, the peculiar jumping behavior between maximally red-shifted and maximally blue-shifted velocities of the peak emission, and the peak velocities can provide hints on the location and geometry of the emitting material. A key parameter of this discussion is the extent of the white dwarf magnetosphere, i.e. the region around the white dwarf in which the magnetic field is strong enough to force material to move along magnetic field lines. The extent of this region is indicated by the magnetospheric radius, which is usually taken as a fraction $\xi$ of the Alfv\'{e}n radius \citep{pringle1972,ghosh1979,papitto2022}:
\begin{align}
    R_{\rm m} &\approx \xi R_{\rm A} \approx \xi \left(\frac{B_p^2 R^6_{\rm WD}}{\dot{M}\sqrt{2GM_{\rm WD}}}\right)^{\frac{2}{7}} \nonumber\\
    &\approx 300~R_{\rm WD} \left(\frac{\xi}{0.5} \right) \left(\frac{B_p}{500\,\rm{MG}}\right)^{\frac{4}{7}} \left(\frac{R_{\rm WD}}{4800\,\rm{km}}\right)^{\frac{12}{7}} \nonumber\\
    & ~~~~~~~~~~~~~~~~~~~~~~~~~ \times \left(\frac{M_{\rm WD}}{1.12\,\rm{M}_\odot}\right)^{-\frac{1}{7}} \left(\frac{\dot{M}}{10^{16}\,\rm{g/s}}\right)^{-\frac{2}{7}}
    \label{eq:ra}
\end{align}
where $B_p$ is the magnetic field at the surface and $\dot{M}$ is the mass flux rate at the magnetospheric radius. Although we do not have a good determination of the magnetic field strength, we are confident that it is of the order of a few hundreds of MG or more (see Section\,\ref{sec:SED}). Employing our best estimate for the field from the absorption lines ($\approx$\,500\,MG), and the white dwarf parameters we derived in Section\,\ref{sec:SED}, we find that the magnetospheric radius is hundreds of white dwarf radii, even for a very high mass accretion rate. 

Two possible scenarios can explain the velocities detected in H$\alpha$: the material could be orbiting the white dwarf at Keplerian velocities, or it could be rigidly corotating with the white dwarf, trapped by the strong magnetic field. Employing the white dwarf mass and radius estimated in Section\,\ref{sec:SED}, we find that a Keplerian velocity ($v_k=\sqrt{GM_{\rm WD}/d}$\,) of 1,700 km\,s$^{-1}$ would indicate, in the case of a perfectly edge-on system, a distance from the white dwarf of $d\approx11\,R_{\rm{WD}}$ (see the blue vertical line in Fig.\,\ref{fig:diag}). For an inclined system, the material would have to be orbiting even closer to the star (blue shaded region highlighted by the blue arrow in Fig.\,\ref{fig:diag}). However, at this distance, we are deep into the magnetosphere of the white dwarf and the magnetic field would be strong enough to force the ionized material into rigid-body corotation with the white dwarf (and therefore move slower than the Keplerian velocity at this distance). Moreover, we would be able to detect the Zeeman splitting in the emission line as the local magnetic field would be higher than 0.1\,MG. We therefore exclude that the line-emitting material is orbiting the white dwarf in a Keplerian disk. If some material is present within the corotation radius, it could fall toward the white dwarf following the field lines, and therefore the line-of-sight velocity that we measure could be a mixture of corotation velocity and infall velocity. The same would apply if the material was moving outward in a wind, following the field lines. However, it would be hard to explain the precise jumping behaviour in the line-of-sight velocity that we detect in both H$\alpha$ and H$\beta$. In the next section, we show that a half ring of ionized gas trapped in corotation with the white dwarf can instead well reproduce the observed H$\alpha$ profile and phase variation.
 
The velocity of material forced into corotation ($v_c=\Omega d$), increases with distance, and a velocity of 1,700\,km\,s$^{-1}$ in an edge-on system would indicate a distance of $d\approx22\,R_{\rm{WD}}$, while the distance would be larger for an inclined system. At this distance, the magnetic field is less than 50\,kG and it can be easily hidden in the low-resolution H$\alpha$ spectrum. As we can see in Fig\,\ref{fig:diag}, at this distance we are outside the Keplerian corotation radius ($r_k=GM_{\rm WD}/\Omega^2$), which is the radius at which the Keplerian and corotation velocities are equal and the centrifugal force and gravity are in balance at the rotational equator. Outside the Keplerian corotation radius, for material in rigid corotation with the white dwarf, the centrifugal force dominates over the gravitational attraction toward the star; however, the magnetic field strength in this region is high enough to trap the material on the closed field lines, preventing it from being ejected.

In several studies on magnetic OB stars and M dwarfs \citep{townsend2005,townsend2005b,townsend2007,townsend2008,bouma2025}, it has been shown that, for a sufficiently large magnetic field strength, ionised material can be trapped in a stellar magnetosphere at radii larger than the Keplerian corotation radius. Within the magnetospheric radius, where ionized gas is forced to move along field lines, gas that is also within the corotation radius, where gravity is stronger than the centrifugal force, will follow field lines toward the star and accrete. On the other hand, if the material is outside the corotation radius, the centrifugal force will push the material outwards as much as the closed field line will allow. Inside the magnetosphere and outside the corotation radius, the maximum distance from the white dwarf at which the material can be outwardly driven by centrifugal forces is set by the shape and finite extent of the field line to which the material is bound. More precisely, spatial potential minima will naturally occur beyond $r_{k}$ in the case of large magnetic fields, as a consequence of magnetic pressure acting on the ionised gas. 
These spatial minima align into accumulation surfaces/volumes into which circumstellar emitting material may settle. This steady accumulation leads to the development of a warped disc, which is supported by the centrifugal force and corotates rigidly with the star \citep{townsend2008}. Although this mechanism was proposed to explain circumstellar emissions of rotating magnetic OB stars, it is not specific only to this class of objects. As such, we consider this to also be the most probable explanation for the consistent radial location of the circumstellar emitting material in ZTF\,J2008+4449.

\subsection{A half ring in corotation}
\label{sec:disk}
The double-peak profile of the Balmer emission in the phase-averaged spectrum of ZTF\,J2008+4449 resembles closely the profile of a thin, photoionized disk \citep[see e.g.][]{rybicki1983,horne1986}. However, when we observe the phase-resolved spectra, we can see that the bulk of the emission shifts abruptly between a maximally red-shifted velocity and maximally blue-shifted one every half a period (see Fig.\,\ref{fig:optspectrum}). 
This behaviour can be explained if the ionized gas is not distributed in a full disk, but is confined in a fraction of approximately $f_d\approx50\%$ of disk that is thin in the radial direction (a half ring). This is also suggested by the Doppler tomogram that we derived in Section~\ref{sec:Balmer_loc}.

We here build a simple toy model to reproduce the observed phase-resolved variations of H$\alpha$ and H$\beta$. We take a half flat ring in corotation with the white dwarf, emitting uniformly, and we indicate the inner and outer radius of the half ring as $R_{\rm in}$ and $R_{\rm out}$. We consider two simple cases for the nature of the emission. If the material is optically thin to H$\alpha$ photons, we expect the emission from each section of the ring to be isotropic. If, on the other hand, the material is optically thick, we employ the Sobolev approximation \citep{hummer1985} and impose an intensity that depends on the velocity gradient (along the line-of-sight) within the ring \citep{rybicki1983}. Under this assumption, and assuming an edge-on inclination, the intensity at any point in the disc is given by 
\begin{equation}
    I(\theta) = I_0\sin\theta\cos\theta
\end{equation}
where $\theta$ is the azimuthal angle along the ring defined with respect to the line-of-sight. In the optically thin case, the intensity is given by simply by $I(\theta)=I_0$, where $I_0$ is the local intensity which in the following we assume has no radial dependence.

\begin{figure}[tb!]
\centering
\includegraphics[width = \columnwidth]{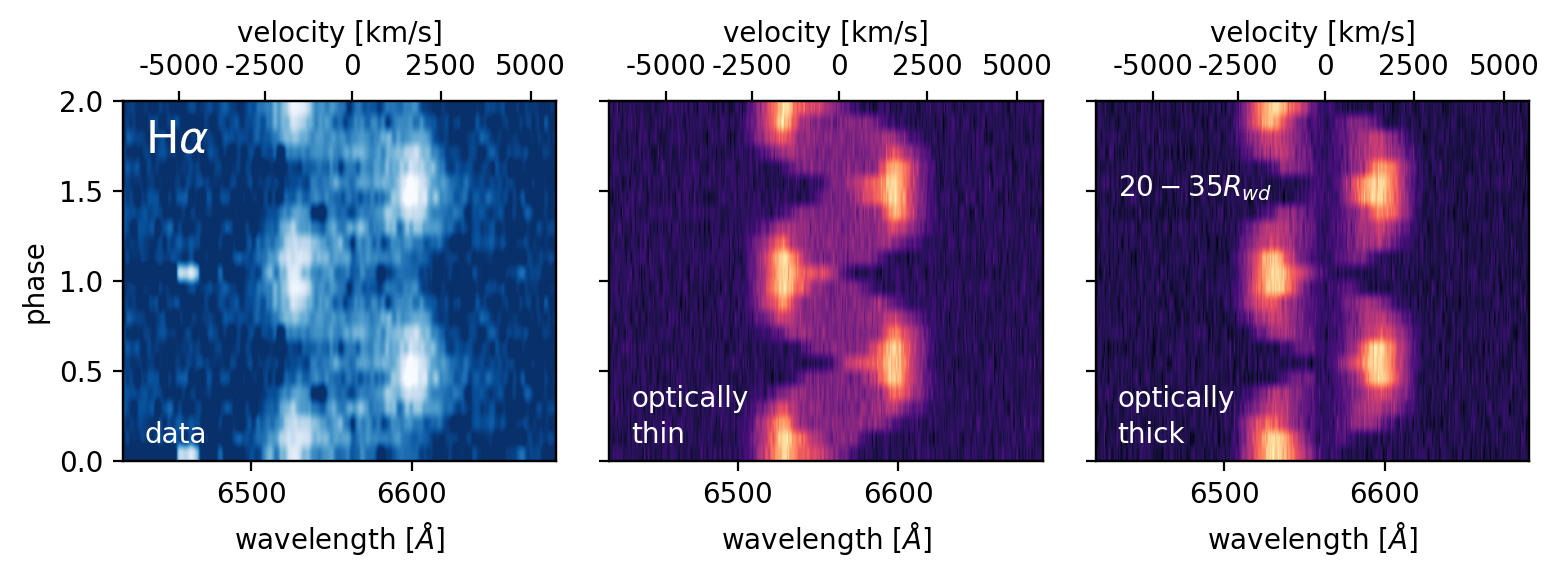}\\
\includegraphics[width = \columnwidth]{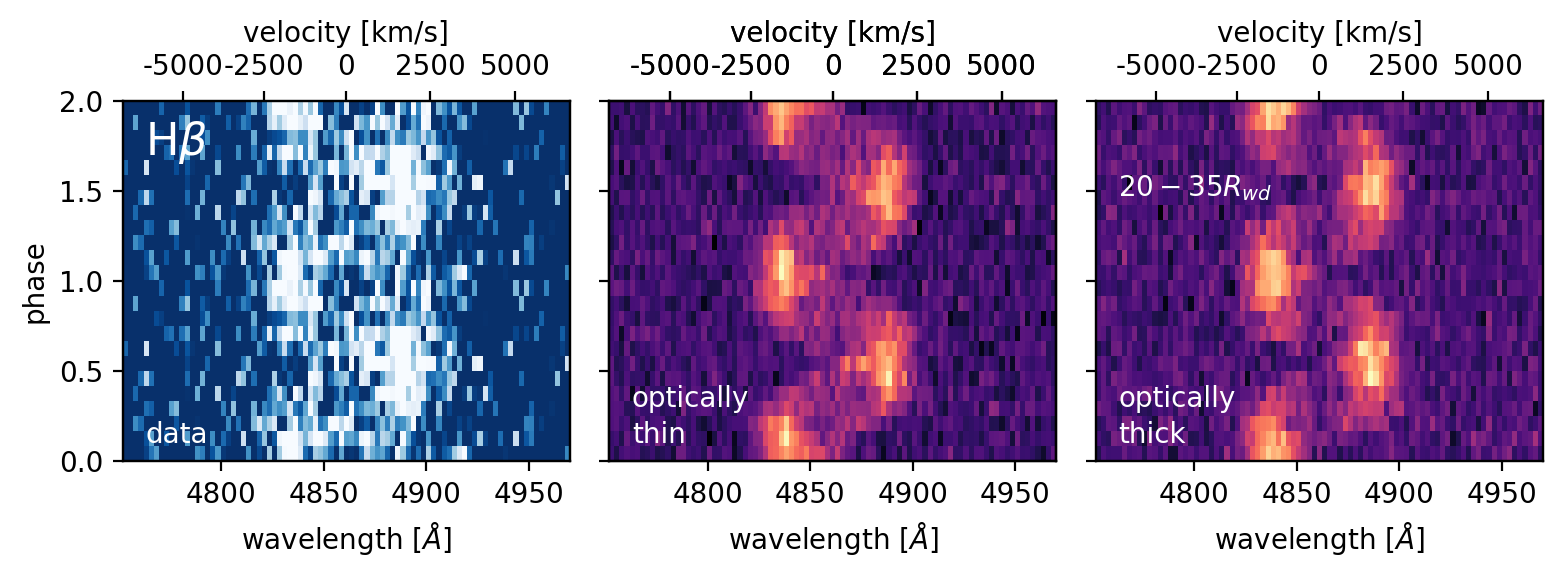}
\caption{Comparison of the trailed spectrogram from LRIS for the H$\alpha$ and H$\beta$ emission of ZTF\,J2008+4449 (upper left and lower left plots respectively, in blue), with our model for optically thin (central plots) and optically thick (right plots) emission from a half ring with $R_{\rm in} = 20 R_{\rm WD}$ and $R_{\rm out} = 35 R_{\rm WD}$.
}
\label{fig:hal_trail}
\end{figure}

We then simulate the emission from the half ring as it rigidly corotates around the white dwarf. In all calculations, we assume an inclination of 90 degrees of the normal to the ring with respect to the line-of-sight; for a lower inclination $i$, the model remains the same, but the inner and outer radii are shifted outwards by a factor $1/\sin i$. Fig.\,\ref{fig:hal_trail} shows the trailed spectrogram for H$\alpha$ and H$\beta$ in ZTF\,J2008+4449 compared with the phase-resolved intensity profile of our model in the optically thin and the optically thick case for a choice of $R_{\rm in} = 20 R_{\rm WD}$ and $R_{\rm out} = 35 R_{\rm WD}$. We can see that the jumping behaviour of H$\alpha$ is well reproduced in both models.

\begin{figure}[tb!]
\centering
\includegraphics[width = \columnwidth]{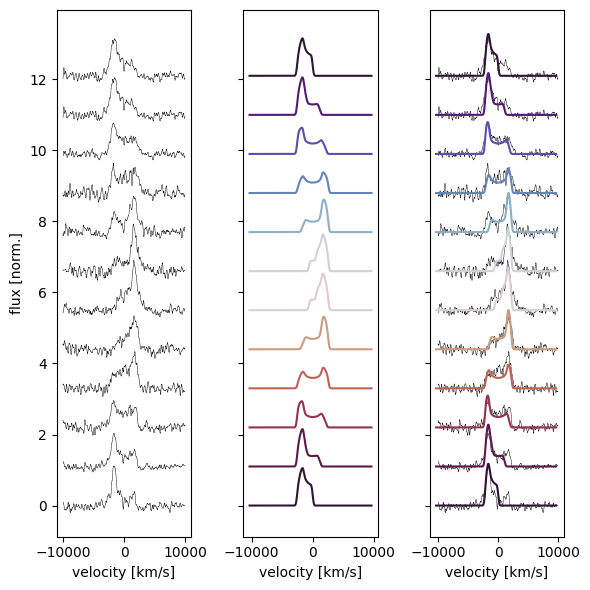}
\caption{\textbf{Left}: observed phase-resolved LRIS spectra for the H$\alpha$ profile of ZTF\,J2008+4449, divided in 10 phases (same as right panel of Fig.\,\ref{fig:optspectrum}). \textbf{Middle}: phase-resolved profile for the best-fit model shown in the upper panel of Fig.\,\ref{fig:hal_hbl_profile}. \textbf{Right}: comparison of the observed phase-resolved profile with the model.
}
\label{fig:hal_ph_res}
\end{figure}

\begin{figure}[tb!]
\centering
\includegraphics[width = \columnwidth]{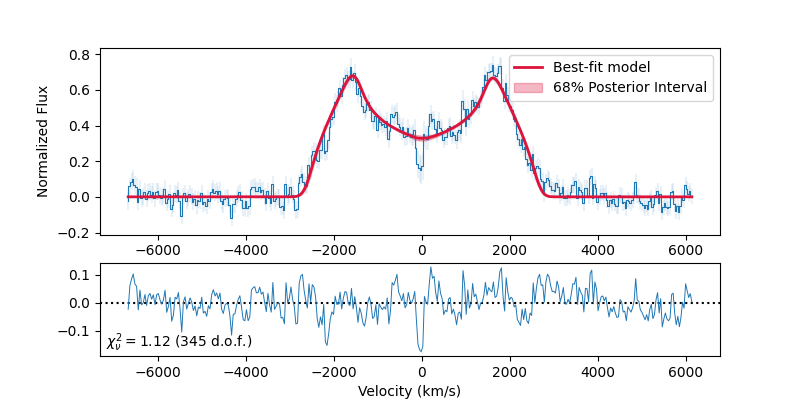} 
\includegraphics[width = \columnwidth]{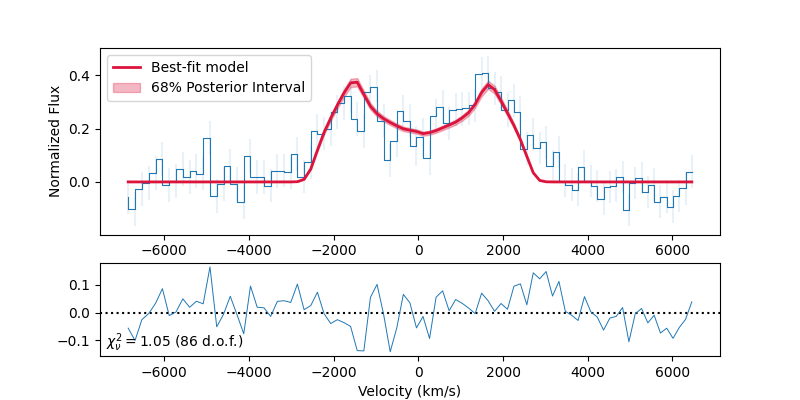} \\
\includegraphics[width = 0.84\columnwidth]{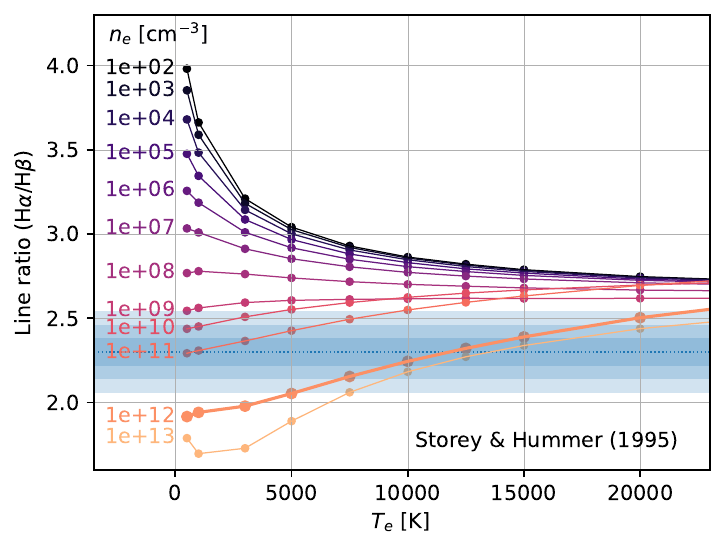} 
\caption{ \textbf{Top}: Five-parameter fit to the H$_{\alpha}$ emission profile employing our half-ring model and a mix of optically thick and optically thin emission, same as right-middle and right-bottom panels of Fig.\,\ref{fig:hal_profile}. The upper panel shows the comparison of our model (in red) to the phase-resolved H$\alpha$ profile observed with LRIS for ZTF\,J2008+4449 (in blue), while the lower panel shows the residuals. \textbf{Middle}: One-parameter fit to the H$_{\beta}$ emission profile for ZTF\,J2008+4449, fixing all parameters to those found for H$_{\alpha}$ (see Fig.\,\ref{fig:corner-disk-thin-sobl}), except the normalisation, which is the only fitted parameter. \textbf{Bottom}: Line ratios of H$\alpha$/H$\beta$ as presented by \cite{storey1995} for a range of gas temperatures and electron densities, shown with points and connecting lines. The measured H$\alpha$/H$\beta$ ratio from the emission line fitting (top two panels) is shown with the horizontal dotted blue line. 
The 1-, 2-, and 3-$\sigma$ contours are shown in the filled horizontal bands. The simulation data (points and connected lines) most closely matching the inferred electron density from our photoionization ring model is plotted with greater thickness.
}
\label{fig:hal_hbl_profile}
\end{figure}

We employ our simple model to fit the phase-averaged profile of the H$\alpha$ emission. In the fit, we leave as free parameters the normalization (which we define as the inverse of the intensity $I_0$), and the inner and outer radius of the ring. Additionally, we also fit the system velocity, $v_{\rm sys}$, allowing the central wavelength of the profile to shift from the rest wavelength of H$\alpha$. We perform a fit for the optically thin case, for the optically thick case and for a combination of both; in this last fit, we allow the fraction of optically thin to optically thick emission to vary freely as an additional parameter. The results of the three fits are shown in Fig.\,\ref{fig:hal_profile} in the Appendix. We find that the best-fitting models are the optically thin one or the mixed model, while the fully optically thick model underpredicts the central section of the profile. Independently of the model, we recover very similar parameters for $R_{\rm in}$, $R_{\rm out}$ and $v_{\rm sys}$. In Fig.\,\ref{fig:hal_hbl_profile}, we show the phase-averaged profile for the best-fit for the mixed model, while we show the posteriors of the fit in the Appendix  Fig.\,\ref{fig:corner-disk-thin-sobl}. The best-fit parameters are: $R_{\rm in}=20.1\pm0.2R_{\rm WD}$, $R_{\rm out}=34.9\pm0.3R_{\rm WD}$ and $v_{\rm sys}=0.8_{-8.9}^{+9.6}$\,km/s. From the profile, we also extract the total flux in H$\alpha$ to be $F_{\rm H \alpha} = (5.02_{-0.05}^{+0.06})\times10^{-16}\,{\rm erg\,s}^{-1}\,{\rm cm}^{-2}$. Fig.\,\ref{fig:hal_ph_res} shows the phase-resolved LRIS spectra of the H$\alpha$ line (left panel) next to the phase-resolved profile of the best-fit for the mixed model (middle panel) and their comparison (right panel). We can see that, although ours is a simple model, it performs extremely well in reproducing the behaviour of the H$\alpha$ profile.

The dominant emission feature detected in this system is the H$\alpha$ profile. The only other detected emission feature is H$\beta$, which is detected at much lower signal-to-noise. Nonetheless, the phase-resolved behaviour of the higher-order Balmer line closely follows that of H$\alpha$ (see Fig.\,\ref{fig:hal_trail}). We therefore also performed a fit to the phase-averaged H$\beta$ profile with the best-fitting H$\alpha$ model in order to estimate the H$\beta$ flux, leaving only the normalization as a free parameter (middle panel of Fig.\,\ref{fig:hal_hbl_profile}). The posterior distribution from the fit yields a best-fit flux of $F_{\rm H\beta}=(2.8\pm 0.1)\times10^{-16}\,\rm{erg~s^{-1}~cm^{-2}}$. Compared to the H$_{\alpha}$ flux, this yields an intensity ratio in photon flux of $R=2.4\pm0.1$. We correct this value for extinction by scaling the measured line fluxes by the best-fit extinction curve found in our SED fitting procedure (see Section\,\ref{sec:SED}), finding a line ratio of $R=2.3\pm0.1$. The bottom panel of Fig.\,\ref{fig:hal_hbl_profile} shows the measured, extinction-corrected line ratio and associated 1-, 2-, and 3-$\sigma$ limits, compared to the simulated line ratios of \citet{storey1995}. 
In Section\,\ref{sec:photmodel}, we present a photoionisation model of the Balmer emission, deriving an estimated electron density of $n_e\approx10^{12}$\,cm$^{-3}$. The corresponding curve in Fig.\,\ref{fig:hal_hbl_profile} is indicated with the thick, orange, solid line. The measured line ratio favours temperatures in the $T_e\approx10000$--15000\,K range, consistent with the expectation for circumstellar photoionized hydrogen \citep[e.g.][]{spitzer1987,cunningham2015}, with a larger range of 5000--20000\,K permitted at 3-$\sigma$. In Section\,\ref{sec:photmodel}, we also estimate the vertical thickness of the ring, from which we derive a temperature, assuming hydrostatic equilibrium and an aligned dipole for magnetic field, of $\approx9000$~K. An improved radiative transfer model, accounting for detailed geometry, will be necessary to better constrain the ring temperature.

\subsection{Hydrogen absorption}
\label{sec:Ha-absorption}
\begin{figure*}[tb!]
\centering
\includegraphics[width = 0.73\columnwidth]{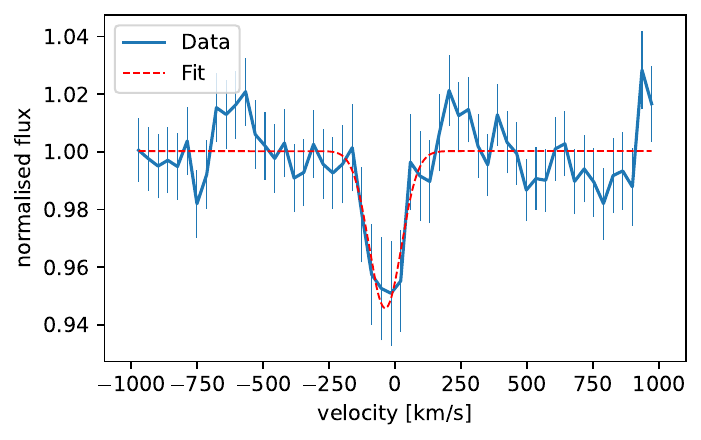}
\includegraphics[width = 1.27\columnwidth]{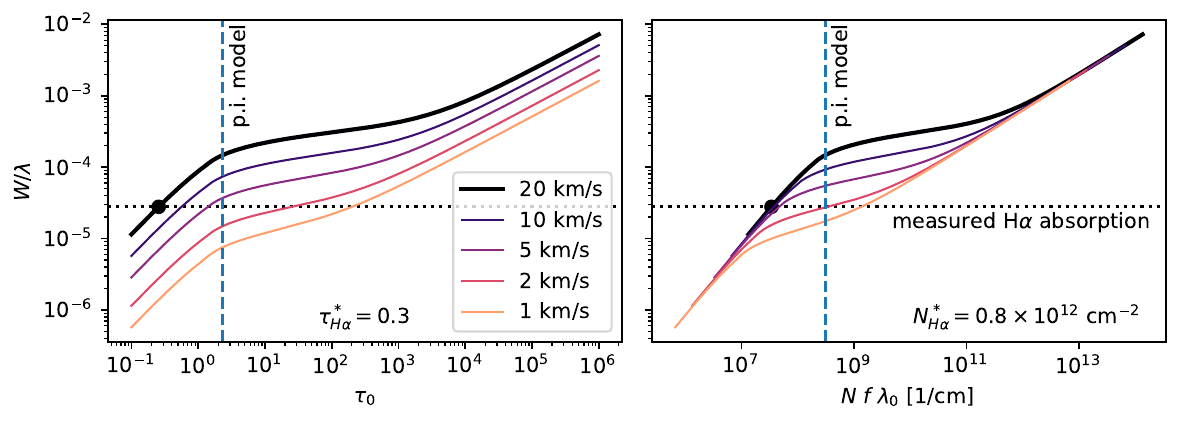}
\caption{ \textbf{Left}: Continuum-normalised spectrum near the rest wavelength of H$_\alpha$. Before the continuum normalisation, the fitted emission feature (Fig.\,\ref{fig:hal_profile}) was first subtracted. The dashed line shows a Gaussian fit to the data. The fitting procedure yields an equivalent width of $EW=8.4 \pm 2.5~\rm{km~s^{-1}}$ and central velocity of $v_0=-37\pm13~\rm{km~s^{-1}}$. \textbf{Middle and right}: Curves of growth for H$_\alpha$. \textbf{Middle}: Evolution of equivalent width ($EW$) as a function of optical depth. \textbf{Right}: Similar to middle panel, but as a function of column density. The dotted horizontal line indicates the measured equivalent width from the left panel. The vertical dashed line indicates the optical depth and column density inferred from the photoionization model.
}
\label{fig:ha_absorption}
\end{figure*}

The phase-averaged spectroscopy reveals a narrow absorption feature near the rest wavelength of H$\alpha$. In the left panel of Fig.\,\ref{fig:ha_absorption}, we show a Gaussian fit to the narrow H$\alpha$ absorption feature. We find an equivalent width of $EW=8.4\pm2.5$\,km\,s$^{-1}$. In the same figure, we also show the curves of growth of the H$\alpha$ line as a function of optical depth and column density in the middle and right panels, respectively \citep[see e.g., equation 9.27 of][]{draine2011}. 
The measured equivalent width implies that the absorbing medium is optically thin to H$\alpha$ for thermal broadening $v>5$\,km\,s$^{-1}$ ($\tau_{\rm H\alpha}<2$). This value, however, is only a lower limit to the optical depth, because the half ring could be inclined with respect to our line of sight and it is likely thin enough that it does not entirely occult the star.  
For comparison, we also indicate the optical depth derived from the photoionisation model (in the next Section, we find an optical depth in emission of $\approx2$, vertical dashed line in the middle panel of Fig.\,\ref{fig:ha_absorption}). 

In the COS UV spectrum, we also detect narrow absorption lines of carbon and silicon at their zero-field rest wavelengths (cyan lines in Fig.~\ref{fig:uv-mosaic}), which cannot be photospheric. These could be due to the interstellar medium or also originate from the trapped circumstellar material, similar to the H$\alpha$ absorption. However, the uncertainty on the continuum level and blending with photospheric absorption lines does not allow us to derive a confident equivalent width. Future higher resolution observations could help understand if the lines are indeed from the half ring, allowing to better constrain its composition.

\subsection{Properties of the glowing ionized gas}
\label{sec:photmodel}
In this section, we estimate the properties of the material glowing in H$\alpha$ and H$\beta$ by assuming that the gas is ionized by the hot white dwarf and by assuming a half-ring geometry. Also, for simplicity, we assume a pure-hydrogen composition. The Balmer emission is very weak, and we cannot exclude the presence of other elements. Helium, for example, can be easily hidden in emission because the number of ionizing photons from the white dwarf is significantly lower than for hydrogen (from the best-fitting magnetic model, the number of ionizing photons is 20 time smaller for hydrogen than helium). Other elements, like carbon, might be present in absorption in the UV spectrum, but we do not have a strong constraint on their abundance. The magnetically broadened photospheric metal lines observed in absorption in the UV spectrum do not provide additional information on the composition of the accreted material, as they could be easily explained by accretion of material at different metallicities (even solar) or by radiative levitation of metals from the interior of the white dwarf.  We leave a detailed photoionization model, that would take into account different compositions, for future studies, and we here focus on deriving a rough estimate of the physical properties of the gas.

For the geometry of the half ring, we impose the inner and outer radius of the ring ($R_{\rm in}$ and $R_{\rm out}$) to be the ones derived in Section~\ref{sec:disk}, based exclusively on the velocity profile. For the volume of the ring, we use the simple expression
\begin{equation}
    V=2\pi f_d H(R_{\rm out}^2-R_{\rm in}^2)
    \label{eq:disk-volume}
\end{equation}
where $H$ is half of the vertical thickness of the ring and $f_d$ is the fraction of the ring, that we assume to be a half, based on the trailed spectrogram and Doppler tomogram.
The actual geometry of the trapped material could be much more complex as it is likely determined by the morphology of the magnetic field and by the inclination of the magnetic axis with respect to the rotation axis \citep[see ][]{townsend2005}; however, with our simple model, we can still obtain an estimate of the properties of the trapped gas, in particular its density and optical depth.

The inner section of the ring, where the bulk of the H$\alpha$ is emitted, is likely almost fully ionized by the incoming radiation, while at the outer edge of the ring, we expect a thin transition region between mostly ionized material and mostly neutral material (the edge of a ``Str\"omgren disk''). Additional material might be trapped in the magnetosphere at larger radii (because of the likely presence of some small ionization fraction which would couple the material to the field), but we do not detect it in H$\alpha$ because the ionization fraction is too low. We can write the equilibrium between ionization and recombination in the volume $V$ of the ring as 
\begin{align}
    Q_H&=V\alpha_{\rm B} n_{\rm p}n_{\rm e}=V\alpha_{\rm B} n_e^2 \,,
    \label{eq:strom}
\end{align}
where $Q_H$ is the rate of ionizations in the ring and $\alpha_{\rm B}(T) \approx2.56\times 10^{-13}  T_{4}^{-0.83}\,\rm{cm^3\,s^{-1}}$ is the recombination rate coefficient in the so-called Case B \citep[which assumes that all Lyman-line photons are absorbed in the medium and converted into lower-series photons,][]{osterbrock2006}; the assumption of Case B also involves the ``on-the-spot'' approximation, in which recombinations to the ground state do not contribute to the recombination rate as they produce photons just above the hydrogen ionisation edge, which are reabsorbed locally. $n_p$ and $n_e$ are the number densities of protons and electrons in the ring, assumed to be uniform, and, since we are assuming a pure hydrogen composition, they are equal to each other and to the total ionized density.
As we expect all the ionizing photons to be absorbed within the ring, we can impose that the rate of ionisation is given by the number of ionising photons entering the ring. If we take the total number of ionizing photons emitted by the white dwarf that we derive from our best fitting magnetic model, $L_{\rm ion}$ (see Section\,\ref{sec:SED} and Table\,\ref{tab:SED}), the number of photons entering the ring is given by $L_{\rm ion}$ times the ratio of the areas of the inner rim of the ring and that of the full sphere at $R_{\rm in}$. The area of the inner rim of the ring can be expressed as $A_{\rm rim}=2\pi R_{\rm in} \cdot 2H\cdot f_d$. We obtain
\begin{equation}
    Q_H = \frac{L_{\rm ion}A_{\rm rim}}{4\pi R_{\rm in}^2}  = L_{\rm ion}f_d\frac{H}{R_{\rm in}}
    \label{eq:Q}\,.
\end{equation}

\noindent
From Eq.\,\ref{eq:strom} and Eq.\,\ref{eq:Q}, we can derive the ionized density in the ring, which does not depend on the thickness or the angular extent of the ring, as
\begin{equation}
    n_e = \sqrt{\frac{L_{\rm ion}}{2\pi \alpha_{\rm B}R_{\rm in}(R_{\rm out}^2-R_{\rm in}^2)}} \approx 1\times10^{12}\,{\rm cm}^{-3}\,.
\end{equation}
Here, the value was calculated by employing the $R_{\rm in}$ and $R_{\rm out}$ obtained in Section~\ref{sec:disk}. From the ionized density in the ring, we can derive the neutral density assuming local equilibrium between ionization and recombination as
\begin{equation}
    n_0=\frac{\alpha_{\rm B}  n_e^2}{q_{ion}}\approx 2\times10^{7}\,{\rm cm}^{-3}\,,
\end{equation}
where $q_{\rm ion}=(\int L_{\rm ion}\sigma_{p} d\nu)/(4\pi R_{\rm in}^2)$ is the ionization rate along the photon path in the radial direction, $\sigma_{\rm p} = \sigma_{0\rm p}(\nu/\nu_0)^{-3}$  is the photoionization cross section for hydrogen ($\sigma_{0\rm p}\approx6.4\times10^{-18}$\,cm$^2$) and $\nu_0$ is the photoionization threshold for hydrogen (13.6 eV).

To estimate if the ring is optically thick to H$\alpha$ photons, we need to estimate the population in the second excited level of hydrogen. Since the only relevant transitions are captures of electrons into excited levels and downward radiative transitions to lower levels, we can write the ratio of densities in the second and ground level, adopting the low density, coronal approximation (in which collisional processes may be neglected), as: 
\begin{equation}
    \frac{n_2}{n_1} = \alpha_{\rm B}n_e\frac{1}{A^{\rm eff}_{21}}=\alpha_{\rm B}n_e\frac{\tau_{L\alpha}}{A_{21}}
    \label{eq:n2n1}
\end{equation}
where $\alpha_{\rm B}n_e$ is the recombination rate and $A^{\rm eff}_{21}$ is the effective transition rate, which can be expressed as the transition rate ($A_{21}=4.41\times10^{7}$\,s$^{-1}$) times the escape probability of Lyman-$\alpha$ photons ($\epsilon\approx1/\tau_{L\alpha}$). $\tau_{L\alpha}$ is the optical depth to Lyman-$\alpha$ photons in the ring; if we define $\delta r = R_{\rm out} - R_{\rm in}$ as the radial extent of our ring, we can find the optical depth to Lyman-$\alpha$ photons  in the radial direction as:
\begin{equation}
\tau_{L\alpha}=\delta r~ n_0 ~\sigma_{\rm L\alpha}  
\label{eq:tau-La}
\end{equation}
where $\sigma_{\rm L\alpha}$ is the absorption cross section for Lyman-$\alpha$ photons.

The frequency dependence of the absorption cross-section for Lyman alpha, $\sigma_{\rm L\alpha}$ is well described by the Voigt function, $\phi(x)$ \citep[see e.g.][] {dijkstra2014}, such that
\begin{equation}
    \sigma_{L\alpha}(x)=\sigma_{\rm L\alpha,0}\cdot\phi(x)
\end{equation}
where $\sigma_{\rm L\alpha,0}$ is the absorption cross section at rest frequency, and  $x$ is the dimensionless frequency defined as 
$x \equiv (\nu - \nu_{\alpha}) / \Delta \nu_{\mathrm{D}}$, where thermal broadening is given by 
\begin{equation}
\Delta \nu_D = \nu_0 \sqrt{\frac{2 k T}{m c^2}} \equiv \frac{\nu_0 v_\mathrm{th}}{c}
\end{equation}
The absorption cross section at rest frequency is given by 
\begin{align}
    \sigma_{\rm L\alpha,0} &= \frac{f_{12}}{\sqrt{\pi}\Delta \nu_{\rm D}}\frac{\pi e^2}{m_e c} \\
    &=5.88\times 10^{-14}\left(\frac{f_{12}}{0.416}\right)\left(\frac{\lambda_{0}}{1215.16~\AA}\right)
    \notag   \\
    & ~~~~~~~~~~~~~~~~~~~~~\quad \times \left(\frac{T}{10^4~\rm{K}}\right)^{-1/2} \rm{cm^2} \nonumber \\
    &=3.79\times10^{-14}\left(\frac{f_{12}}{0.416}\right)\left(\frac{\lambda_{0}}{1215.16~\AA}\right)
    \notag  \\
    & ~~~~~~~~~~~~~~~~~~~~~\quad \times \left(\frac{v_{\rm th}}{20~\rm{km/s}}\right)^{-1} \rm{cm^2} \nonumber 
\end{align}
where $f_{\rm 12}=0.416$ is the Lyman-$\alpha$ oscillator strength and $\lambda_{0}$ is the transition rest wavelength. 
The Voigt profile core can be well approximated by $\phi(x)\approx e^{-x^2}$ in the limit that the broadening is dominated by velocity broadening rather than broadening of the intrinsic line profile. We adopt a normalisation such that $\phi(0)=1$, and the integrated profile becomes $\int\phi(x)dx=\sqrt{\pi}$. We can thus express the Lyman-$\alpha$ absorption cross-section as 
\begin{align}
    \sigma_{\rm L\alpha} &= \sqrt{\pi}~\sigma_{\rm L\alpha,0} \\
    &= 6.7\times10^{-14}\left(\frac{f_{12}}{0.416}\right)\left(\frac{\lambda_0}{1215.16~\AA}\right)\notag \\
    & ~~~~~~~~~~~~~~~~~~~~~\quad \times \left(\frac{v_{\rm th}}{20~\rm{km/s}}\right)^{-1} \rm{cm^2}\,.
    \label{eq:sigma_La}
\end{align}

\noindent
From Eqs.\,\ref{eq:n2n1}, \ref{eq:tau-La} and \ref{eq:sigma_La}, we obtain
\begin{equation}
    \frac{n_2}{n_1} =\alpha_{\rm B}n_e\frac{\delta r~n_0~\sigma_{\rm L\alpha}}{A_{21}}\approx 4\times10^{-5} \, , \quad n_2 \approx 7\times10^{2}\,{\rm cm}^{-3}\,.
\end{equation}
We can then derive the optical thickness of the ring in H$\alpha$ as:
\begin{align}
\tau_{H\alpha} &=\delta r~ n_0 ~ \left(\frac{n_2}{n_1}\right)~\sigma_{\rm H\alpha} \nonumber \\
&= (\delta r ~n_0)^2 \alpha_{\rm B} n_e \frac{\sigma_{\rm L\alpha}\sigma_{H\alpha}}{A_{21}} \approx 2\, .
\label{eq:tau-Ha-population}
\end{align}
where the H$\alpha$ absorption cross-section is given similarly to Eq.\,\eqref{eq:sigma_La}, as 
\begin{align}
    \sigma_{\rm H\alpha} &= \sqrt{\pi}~\sigma_{\rm H\alpha,0} \\
    &= 5.5\times10^{-13}\left(\frac{f_{23}}{0.641}\right)
       \left(\frac{\lambda_0}{6562.8~\AA}\right) \notag \\
    & ~~~~~~~~~~~~~~~~~~~~~\quad \times \left(\frac{v_{\rm th}}{20~\rm{km/s}}\right)^{-1} \rm{cm^2}
\end{align}

The value of order unity for the optical depth of the H$\alpha$ line is consistent with the fit to the H$\alpha$ profile derived in Section~\ref{sec:disk}, where we derived that the best-fitting model is a combination of optically thick and optically thin emission. On the other hand, the equivalent width of the narrow H$\alpha$ absorption feature (see Section~\ref{sec:Ha-absorption}), implies an optical depth of $\tau_{\rm H\alpha}\approx0.3$, suggesting the absorbing medium is optically thin. This discrepancy could indicate that the absorbing material has a lower column density than the emitting material. Such a situation could arise, for instance, in a thin, uniform density (partial) ring at a modest inclination of $\approx$1\,deg away from edge-on, since the absorption is likely absorption of the stellar continuum, for which the circumstellar gas must be between the star and the viewer. The Balmer emission, on the other hand, could be seen at any inclination. This discrepancy could also be explained by uncertainties arising from the simplistic approximate treatment of geometry and radiative transfer presented in this study. For example, the gap between the $n_{e}=10^{11}\,$cm$^{-3}$ and $n_{e}=10^{12}\,$cm$^{-3}$ Balmer line ratio curves in the bottom panel of Fig.\,\ref{fig:hal_hbl_profile}, which shows significant overlap with the rough range of expected electron densities of the ionised gas, probably arises due to collisional (de-)excitation effects becoming important. In this analysis, however, we do not account for this effect. Such uncertainties arising from the simplicity of the employed models may be tested via future studies exploring detailed geometric and radiative transfer models, although this is beyond the scope of this paper.

In all calculations until now, we did not have to specify the vertical thickness $H$ or the angular extent $f_d$ of the ring. However, if we want to determine the total mass contained in the ring, we need to derive its volume. One way to estimate the thickness of the ring is to calculate how many ionizing photons need to be captured in the ring to produce the observed H$\alpha$ emission. We can express the total H$\alpha$ luminosity (in photons per second) as:
\begin{align}
    L_{H\alpha} &= 4\pi j_{\rm H\alpha} V \nonumber\\
    &= \left(\frac{4\pi j_{\rm H\beta}}{n_en_p}\right)\left(\frac{j_{\rm H\beta}}{j_{\rm H\alpha}}\right)n_en_pV \nonumber \\
    &= \zeta(T)n_e^2V
    \label{eq:lha}                  
\end{align}
 where $j_{\rm H\alpha}$ and $j_{\rm H\beta}$ are the emissivities of the two lines and V is the volume of the ring. We group the two terms in parentheses into the function $\zeta(T)$, highlighting that they mostly depend on the temperature of the gas, although weakly (they also have an extremely weak dependence on density, which we ignore). The values for these two terms for Case B are tabulated for different temperatures in Table~4.4 of \citet{osterbrock2006}.
 
By combining Eqs.\,\ref{eq:strom}, \ref{eq:Q}, and \ref{eq:lha} we can derive the following expression for $H$
\begin{equation}
\label{eq:H}
    H = \frac{\alpha_{\rm B}(T)}{\zeta(T)} \frac{L_{H\alpha}}{L_{\rm ion}}\frac{r_i}{f_d}
    \approx0.1\,R_{\rm WD}
\end{equation}
where we have used values for $\alpha_{\rm B}$ and $\zeta$ at a temperature of 10,000\,K; however, the two parameters have a similar dependence on temperature, so the ratio of the two is almost independent of temperature. This value of H is small compared to the inner radius of the ring ($\approx20$\,R$_{\rm WD}$), indicating a very thin disk; this is expected because the centrifugal force, which pushes the material toward the apex of each magnetic field line, is strong and acts together with gravity to reduce the thickness of the disk. \citet{townsend2005} derived a modified expression for the hydrostatic scale height of the trapped disk in the case of a dipole aligned with the rotation axis (their equation 28), which includes the effects of the centrifugal acceleration. Using the same expression, we can derive the temperature of the gas that would result in the scale height derived in eq.\,\ref{eq:H}:
\begin{equation}
    T_{\rm ring} = H^2 \frac{\mu}{2k_{\rm B}}\frac{GM_{\rm WD}}{r_K^3} \left(3-\frac{2r_K^3}{R_{\rm in}^3}\right)\approx6000\,\rm{K}
\end{equation}
where $\mu$ indicates the mean molecular weight (that we take as 1 for pure hydrogen) and $k_{\rm B}$ is the Boltzmann constant. Although we do not expect the magnetic field to be a perfectly aligned dipole, we find a consistent temperature to what is expected from circumstellar photoionized gas and with the temperature estimate that we derived in Section\,\ref{sec:Ha-absorption} from the H$\alpha$ absorption.

Finally, we estimate the total mass in the ring to be
\begin{equation}
    M_{\rm ring} \sim n_e V m_p \ \approx 2\times 10^{-17}\,{\rm M}_\odot\,.
\end{equation}
We note that this mass estimate serves as a lower limit, since this corresponds to the mass of Balmer-emitting hydrogen. The true total mass could be considerably higher when accounting for the presence of non-emitting hydrogen (either due to low ionization fraction, or low temperatures), and the presence of heavier elements such as helium and metals.

\subsection{Mass depletion timescale in the half ring}

In the model presented in the previous Section, the ionized gas glowing in H$\alpha$ is currently trapped in a stable configuration in corotation with the white dwarf because ionized particles strongly feel the effect of the magnetic field. Upon recombination, however, neutral hydrogen atoms can decouple from the magnetic field line to which they were previously bound and move outwardly according to their orbital velocity at the time of recombination (i.e., tangentially to the circular trajectory of the ionised arc) until they are reionized. This effect will create a drift between neutral and ionized particles, and therefore an outward motion that will slowly deplete the reservoir of ionized gas. In reality, the neutral particle will feel a drag from the charged particles around it, so its motion will not be truly ballistic. The ion-neutral drag is dominated by charge transfer and elastic collisions.  At densities below $10^{11}$, there are no collisions within the time for photoionization, and the motion is expected to be ballistic.  At densities above $10^{14}$, on the other hand, there are thousands of collisions in a photoionization time, and the motion is diffusive.  The density of $10^{12}$ that we just derived is in the more difficult intermediate range; calculating an accurate estimate of the timescale of outward motion of the gas would require an estimate of the strength of the drag between ionized and neutral particles in this intermediate regime, which is beyond the scope of this paper. However, we expect the timescale to be shorter than the cooling age of the white dwarf ($\approx60$ Myr), and therefore the ionized gas trapped in the magnetosphere would have had to be replenished by material entering the magnetosphere from outside or from outflows from the white dwarf itself.

\subsection{Period increase: a wind or a propeller mechanism}
\label{sec:xray-origin}
In the previous Sections, we focused only on the ionized gas glowing in H$\alpha$. The X-ray emission of ZTF\,J2008+4449 and the detected increase in its spin period, however, highlight the presence of additional material interacting with the rapidly spinning white dwarf. In particular, the observed $\dot{P}=(1.80\pm0.09)\times10^{-12}$\,s/s is much larger than the period increase expected from magnetic dipole radiation:
\begin{equation}
    \dot{P}_{\rm dipole} = \frac{8 \pi^2}{3c^3}\,\frac{sin^2\alpha R_{\rm WD}^6 B_p^2}{I_{\rm WD}P}\approx5\times10^{-14}\,\mathrm{s/s}\, ,
\end{equation}
where $I_{\rm WD}=(\beta R_{\rm WD})^{2}M_{\rm WD}$ is the moment of inertia of the white dwarf, $\beta$ is the gyration radius \citep{claret2023}, and $\alpha$ is the inclination of the magnetic field to the rotational axis. We have used a value of $\beta \approx 0.39$, as per the models of \citet{claret2023} and for $\alpha$ we have taken the mean of $\sin^2\alpha=0.5$. This value of $\dot{P}_{\rm dipole}$ implies that the change in spin period is most likely driven by interaction with the circumstellar material, and, in particular, material being ejected from the system, carrying away angular momentum. Furthermore, as we commented in the previous Section, the ionized gas trapped in the magnetosphere likely needs to be replenished over smaller timescales than the age of the white dwarf.

If ionized gas can be extracted from the surface of the white dwarf, as for example in a magnetically driven wind, some could get trapped in the magnetosphere, populating the half ring, and some could escape, carrying away angular momentum. Alternatively,
if diffuse infalling material is present outside the magnetosphere of the white dwarf (in Section~\ref{sec:material-origin} we discuss the possible origins of this material), the loss of angular momentum necessary to produce the observed spin-down rate may be driven by the binding (and subsequent outward propelling) of this material into rigid body corotation with the white dwarf once it makes contact with the magnetosphere (i.e., at the magnetospheric radius, Eq.\,\ref{eq:ra}). A similar propelling mechanism has been suggested for the white dwarf propellers like AE Aquarii \citep{wynn1997,eracleous1996} and the so-called transitional pulsars \citep[see][and references therein]{campana2018,papitto2022}. The amount of angular momentum carried away by the ejected material in this mechanism depends on how deeply it can plunge within the magnetosphere and on how efficiently it is accelerated before being ejected. Similarly, in the case of a wind, the loss of angular momentum depends on how far from the star the material detaches from the field.

One way to estimate the minimum infalling or outflowing mass flux $\dot{M}$ would be by assuming that the material is accelerated into rigid corotation and ejected at the magnetospheric radius. If we equate the angular momentum transfer necessary to bring the infalling magnetospheric radius to the angular momentum losses of the white dwarf, we can express $\dot{P}$ as
\begin{equation}
\label{eq:pdot}
     \frac{d ln(P)}{dt} = \frac{\dot{P}}{P} \approx \frac{\dot{M}R^2_A}{I_{\rm WD}} \approx \frac{\dot{M}}{\beta^2 M_{\rm WD}}\left(\frac{R_{\rm m}}{R_{\rm WD}}\right)^2 \; .
\end{equation}  
Since the magnetospheric radius depends on $\dot{M}$, we could in principle substitute the equation for the magnetospheric radius (eq.\,\ref{eq:ra}) in eq.\,\ref{eq:pdot} and derive an $\dot{M}\approx 2\times10^{-18}~{\rm M_\odot~yr}^{-1}$. However, at such a small mass loss rate, the magnetospheric radius lays beyond the light cylinder, which is the distance from the white dwarf at which the corotation velocity becomes equal to the speed of light. A significant transfer of angular momentum between the star and the circumstellar material could only happen if the infalling material reaches within the light cylinder \citep{papitto2022}. Similarly, a wind would have to detach from the field at or within the light cylinder. Therefore, the ejected mass has to be higher to explain the observed $\dot{P}$. A better way to estimate the mass loss rate is to calculate how much material would have to be ejected at the speed of light at the light cylinder to carry away enough angular momentum from the star to explain the $\dot{P}$. We can derive this value by using the expression for the light cylinder radius ($R_{\rm LC}=cP/(2\pi)$, where $c$ is the speed of light) in eq.\,\ref{eq:pdot} instead of $R_{\rm m}$. We obtain $\dot{M}\gtrsim 2\times10^{-15}~{\rm M_\odot~yr}^{-1}$. In reality, the material will likely plunge or become detached much deeper in the magnetosphere and will be ejected at much lower velocities, so this is only an absolute minimum to $\dot{M}$. A third way of estimating $\dot{M}$ is to impose that the magnetospheric radius is smaller than the light cylinder radius, which yields a limit $\dot{M}\gtrsim 3\times10^{-14}~{\rm M_\odot~yr}^{-1}$. This is likely the more realistic and stringent constraint. For both estimates, only a tenth or less of the spin-down luminosity of the white dwarf (that we estimate at $L_{\rm sd} = I_{\rm WD}\omega\dot{\omega} \approx 9\times10^{31}$\,erg/s) would be converted into the kinetic energy carried away by the ejecta.

If we compare these $\dot{M}$ limits with the estimate  of the mass trapped in the half-ring that we derive in Section~\ref{sec:photmodel}, we see that the material glowing in H$\alpha$ is only a very small fraction of the material that is being ejected further out. So even in the case of a very efficient propeller, the ring could be replenished by material entering the magnetosphere from outside. Close to the rotation axis, the centrifugal acceleration for material in rigid corotation with the white dwarf is weak out to large radii; some material could therefore trickle down to field lines much closer to the white dwarf than the magnetospheric radius from above the rotation axis. Some of this material can be captured by field lines that extend right outside the corotation axis, and therefore replenish the ring that we see glowing in H$\alpha$, while some material could reach field lines within the corotation radius, and accrete onto the white dwarf. 

\subsection{X-ray emission mechanism}
The observed X-ray emission ($L_X=(2.3\pm0.4)\times10^{29}$\,erg~s$^{-1}$) could be powered by a fraction of the spin-down luminosity, by the gravitational energy of material infalling into the potential well of the white dwarf, or by a combination of both. The X-ray spectrum is equally well described by a power-law model and by a two-temperature optically thin plasma model (see Section~\ref{sec:xray_analysis}), indicating that the emission could be either thermal or non-thermal. If the X-rays are produced by the interaction of the magnetic field with the infalling (or outflowing) material at the edge of the magnetosphere, a thermal X-ray emission could be produced by shock-heated plasma where the material is being accelerated, as it has been suggested for the soft X-ray component of the white dwarf propeller AE Aquarii \citep{oruru2012,kitaguchi2014}. Alternatively, high-velocity material trapped in the field at the edge of the magnetosphere could deform the field lines, causing continuous local reconnections; these magnetic reconnections could heat the plasma to high temperatures or accelerate particles to relativistic speed and lead to synchrotron emission \citep[as it has been suggested for the pulsed X-ray emission of the binary white dwarf pulsar AR Scorpii,][]{takata2018}.

Another possible origin for the X-ray emission is the region close to the surface of the white dwarf. If some material can reach field lines within the corotation radius, it can accrete onto the surface of the white dwarf; the observed X-ray emission could therefore be due to shock-heated plasma close to the surface of the white dwarf or by hot spots, similar to what is observed in polars. The $\dot{M}$ needed to power this emission would be of the order of $10^{-14}~{\rm M_\odot~yr}^{-1}$. This scenario can also explain what could be cyclotron humps in the UV spectrum.  Alternatively, Alfvén waves excited by the interaction of the material with the field at the magnetospheric radius can carry energy down to the white dwarf surface, giving rise to hotspots or local magnetic reconnections.

Variability of the X-ray emission could provide some important clues on the origin and nature of the emission. As we show in Section~\ref{sec:xray_analysis}, the XMM light curves present a hint of variability on the period of the white dwarf. This could indicate that part of the emission is eclipsed by the white dwarf, hinting at an emission region close to the white dwarf surface, or that the X-ray emission is beamed, indicating the presence of high velocities in the emitting material. Deeper X-ray observations could provide a higher signal-to-noise X-ray light curve, needed to confirm the variability and constrain its amplitude.

\subsection{Origin of the circumstellar material}
\label{sec:material-origin}
Despite the host of evidence for the presence of circumstellar material, in Section\,\ref{sec:no_BD} we have shown that the presence of a stellar or brown dwarf companion filling its Roche lobe and, therefore, of ongoing mass transfer as the source of this material, can be excluded. There is still the possibility, as we show in Section\,\ref{sec:no_BD}, for a Roche-lobe filling gas giant planet to be the source of mass transfer; however, because of the origin of the star as a double white dwarf merger, we find it unlikely that a planet would survive the evolution of the star and find itself at just the right distance from the white dwarf. Also, accretion from the interstellar medium would not be enough to explain the observed spin down and X-ray emission \citep{wesemael1979}. The trapped material glowing in H$\alpha$, as well as the infalling (or outflowing) material which is at the origin of the spin down and of the X-rays, must therefore have a different origin.

One possible origin for the circumstellar material is the tidal disruption of a planetary body, like a comet, asteroid or planet. Accretion of planetary material onto white dwarfs is a common occurrence, as over a quarter of all white dwarfs show the presence of metals in their atmospheres \citep{koester14,zuckerman03,zuckerman10}, and recently, X-rays produced by the accretion of planetary material have been detected on the prototypical polluted white dwarf G\,29--38 \citep{cunningham2022}. Due to the high surface gravity of white dwarfs, heavy elements sink quickly out of the observable layers, so the presence of these metals indicate recent accretion of debris from planetary bodies which have been scattered towards the central stellar remnant and tidally disrupted \citep[e.g.][]{zuckerman07,xu17,vanderburg2015,gaensicke2019,manser2019}. The scattering of planetary bodies to high eccentricities, where they can be tidally disrupted by the white dwarf, is thought to be caused by the chaotic evolution of the orbits of the planetary system after its central star loses mass in the last stages of the asymptotic giant branch, on its way to becoming a white dwarf \citep[for a review, see][]{veras2021b,veras2024}. In the case of ZTF\,J2008+4449, if indeed the white dwarf is the product of a double white dwarf merger, the system has undergone two planetary nebulae and a merger event. No studies have been done on the effects of these combined events on the dynamics of a planetary system; however, we can imagine that, if a reservoir of planetary bodies has survived the previous stages of evolution of the system without being ejected or disrupted, the merger event might have perturbed the orbits of the remaining objects in the system, leading some of them to high enough eccentricities to be tidally disrupted by the white dwarf. 

One possible caveat of this scenario is the composition of the infalling material. In the half-ring we only see emission of hydrogen and, although we cannot exclude the presence of other elements (see discussion in Section~\ref{sec:Ha-absorption}), it is unlikely for the composition of the ionized gas to be hydrogen poor. Planetary debris observed around polluted white dwarfs usually present a rocky composition \citep[see e.g.][]{xu2019}. In those instances in which a gaseous, ionized disk is detected around polluted white dwarfs, it usually shows emission of calcium and other metals, but not hydrogen \citep[e.g.][]{gaensicke2006,manser2020}. There is one notable exception, WD\,J0914+1914, a white dwarf accreting from an evaporating ice giant; however, even in this case, the hydrogen emission is accompanied by emission in silicon, oxygen and iron \citep{gaensicke2019}. 

Since ZTF\,J2008+4449 is the remnant of a double white dwarf merger, another possible explanation for the detected circumstellar gas is the fallback of material ejected during the merger event. Studies such as \citet{rosswog2007} and \citet{ishizaki2021} have proposed the fallback of gravitationally bound ejecta following the merger of compact binary systems, in order to explain the long-lasting X-ray activity of young binary neutron star mergers such as GW\,170817. Furthermore, most simulations of double white dwarf mergers predict that right after the merger, the remnant white dwarf is surrounded by a thick disk and an extended tidal tail \citep{guerrero2004,loren2009,dan2014}. Most of the material in the disk will be accreted rather quickly, on a viscous timescale, while material at high eccentricity in the bound tails accretes at longer timescales \citep{rosswog2007,loren2009}. 
Most simulations, however, only explore the first few minutes or days after the merger \citep[see for example][]{lu2023}, and no studies have been performed to explore if a very low level of accretion could still be ongoing 60 Myr after the merger. Furthermore, as in the planetary scenario, we expect the composition of the fallback material to be dominated by carbon, oxygen and other metals, with very little hydrogen.  

A third possibility would be a wind of ionized gas flowing from the surface of the white dwarf. Radiatively driven winds are observed in hot pre-white dwarfs at low surface gravities \citep{werner1995,dreizler1995}, and for one system it has been suggested that some of the wind might be trapped in the magnetosphere of the star \citep{reindl2019}. In the case of ZTF\,J2008+4449, however, because of its high surface gravity ($\log g\approx 8.8$) and relatively low temperature, a sizeable wind cannot be driven by radiation pressure alone, even if metals are present in the atmosphere \citep[see e.g.][]{vennes1988,unglaub2008}.  As the star is rapidly spinning and highly magnetized, it is possible that the strong magnetic field could power a wind, as plasma is expected to flow outwards on open field lines; moreover, if infalling material is creating stresses on the magnetosphere, Alfv\'{e}n waves could induce reconnections close to the surface, accelerating material into a wind (in which case we would still need an external source of circumstellar material). Our estimate for the mass trapped in the magnetosphere is very small ($2\times10^{-17}$\,M$_\odot$), and, even assuming a fast depletion timescale, only a weak wind would be required to replenish the half ring; the mass loss required to explain the period increase, on the other hand, is orders of magnitude larger. Current studies of magnetically driven winds have mainly been focused on main sequence and evolved stars and mostly on how the magnetic field accelerates winds that are powered by other mechanisms, as radiation or reconnections \citep{weber1967,belcher1976,goldreich1970,thirumalai2010,vidotto2014}. Further studies are needed to explore if a wind can be extracted from such an extreme system in terms of surface gravity, stellar rotation and magnetic field strength as ZTF\,J2008+4449.

\subsection{Evolutionary history of the rotation period}
\label{sec:spin_ev}

From the period and period derivative measured in Section\,\ref{sec:period}, we measure a spin-down age of $P/\dot{P}\sim7$\,Myr, which is an order of magnitude lower than the cooling age of the white dwarf. However, we do not necessarily expect the period derivative to have been constant in the past.
White dwarfs polluted by planetary material are believed to undergo episodic accretion, in which a small planetary body is scattered within the Roche radius of the white dwarf, where it is tidally disrupted and ground to dust, forming a debris disk around the white dwarf \citep{veras2021,cunningham2025}. The lifetime of these disks is uncertain, but it is expected to be of the order of one Myr \citep{girven2012,cunningham2021}.

If, on the other hand, the circumstellar material originated from the fallback of post-merger ejecta, the spin evolution is expected to change depending on the accretion rate \citep{sousa2022}. At early times, when the high accretion rate pushes the magnetospheric radius within the corotation radius, the accretion of material onto the surface of the white dwarf can transfer angular momentum onto the white dwarf, resulting in a negative period derivative (the white dwarf spins up). As the accretion rate diminishes, the period derivative increases until the magnetospheric radius becomes larger than the Keplerian corotation radius. At this point, the material is unable to accrete and is instead ejected, carrying away angular momentum and braking the spin of the white dwarf \citep{campana2018,papitto2022,sousa2022}.

\begin{table*}
	\centering
	\caption{Summary of measured and derived parameters for ZTF\,J2008+4449.}
	\label{tab:summary}
	\begin{tabular}{l|cc}
            \hline
            \hline
            \rule{0pt}{2.5ex}
            Origin & Parameter & Value\\[4pt]
            \hline
            \rule{0pt}{2.5ex}
            \multirow{3}{*}{From Gaia DR3} &
            Gaia ID & 2082008971824158720 \\
            &Parallax & $2.9149\pm0.1349$ mas \\
            & Distance \citep{bailer-jones2021}  & $350\pm20$ pc \\[4pt]
            \hline
            \rule{0pt}{2.5ex}
             \multirow{6}{*}{From SED fitting} & 
            Radius of the WD  & $R_{\rm WD} =4,800\pm300$ km \\
            &Temperature of the WD & $T_{\rm eff}=35,500\pm300$ K \\
            & Mass of the WD & $M_{\rm WD}=1.12\pm0.03$ M$_\odot$ \\
            & Cooling age of the WD & $t_{\rm cool}=60\pm10$ Myr \\
            &Extinction toward the WD & E(B$-$V)$=0.042\pm0.001$ \\
            & Ionizing luminosity from the WD & $L_{\rm ion} =(2.4\pm0.1)\times10^{42}$  s$^{-1}$ \\[5pt]
            \hline
            \rule{0pt}{2.5ex}
            \multirow{1}{*}{From SED and UV and optical spectroscopy} &
            \multirow{1}{*}{Average WD surface magnetic field} & $B_{\rm p}\sim\,$400--600\,MG \\
            [2pt]
            \hline
            \rule{0pt}{2.5ex}
            \multirow{6}{*}{From light curve analysis} &
            \multirow{2}{*}{Spin period of the WD} & $P_0=6.55576688 \pm 0.00000008$ min \\
            & & at $T_0=59401.76$ MJD$_{\rm BDT}$ \\[4pt]
            & Period derivative & $\dot{P} = (1.80\pm0.09)\times10^{-12}$ s/s \\
            & Spin-down age & $P/\dot{P}= 7.0\pm0.4$\,Myr \\
            & Spin-down luminosity & $L_{\rm sd}\approx 2\times10^{32}$\,erg~s$^{-1}$ \\[4pt]
            \hline
            \rule{0pt}{2.5ex}
            \multirow{6}{*}{From X-rays analysis} & X-ray luminosity in the & \multirow{2}{*}{$L_X = (2.3\pm0.4)\times 10^{29}\,\rm{erg\,s^{-1}}$}  \\
            &0.25--10.0\,keV band & \\[4pt]
            & Power law index & $\Gamma=2.2\pm0.1$ \\ [4pt]
            & \multirow{2}{*}{2-temperature model} & $kT_1=3.6^{+1.0}_{-0.7}$\,keV \\
            & & $kT_2=0.23^{+0.04}_{-0.03}$\,keV \\
            [4pt]
            \hline
            \rule{0pt}{2.5ex}
            & Ring inner radius & $R_{\rm in }=20.1\pm0.2\,R_{\rm WD}$\\
            & Ring outer radius & $R_{\rm out }=34.9\pm0.3\,R_{\rm WD}$ \\
            & Ring radial thickness & $\delta r=14.8\pm0.4\,R_{\rm WD}$ \\
            From fit to the Balmer emission  & System velocity & $v_{\rm sys}=0.8^{+9.6}_{-8.9}\,\rm{km\,s^{-1}}$\\
            with the half ring model & Flux in H$\alpha$ & $F_{\rm H\alpha}=(5.02^{+0.06}_{-0.05})\times10^{-16}\,\rm{erg\,s^{-1}\,cm^{-2}}$\\
            & Flux in H$\beta$ & $F_{\rm H\beta}=(2.8\pm 0.1)\times10^{-16}\,\rm{erg\,s^{-1}\,cm^{-2}}$\\
            & Photon ratio$^a$ H$\alpha$/H$\beta$ & $R=2.3\pm0.1$\\
            & Luminosity in H$\alpha$ & $L_{\rm H\alpha}=(5.02^{+0.06}_{-0.05})\times10^{-16}\,\rm{erg\,s^{-1}\,cm^{-2}}$\\ [4pt]
            \hline
            \rule{0pt}{2.5ex}
            \multirow{6}{*}{From photoionization model of the ring}
            & Ionized density in the ring & $n_e\approx 1\times10^{12}$ cm$^{-3}$ \\
            & Neutral density in the ring & $n_0\approx 2\times10^{7}$ cm$^{-3}$ \\
            & Optical depth to H$\alpha$ & $\tau_{\rm H \alpha}\approx 2$ \\
            & Ring vertical thickness & $H\approx0.1\,R_{\rm WD}$ \\
            & Temperature in the ring & $T_{\rm ring}\approx9000$\,K \\
            & Mass in the ring & $M_{\rm ring} \approx 2\times10^{-17}$ M$_\odot$ \\[4pt]
            \hline
            \hline
	\end{tabular} \\
    \vspace{10pt}
    \footnotesize 
    $^a$ Photon ratio includes a correction for extinction \vspace{10pt}
\end{table*}

\subsection{Comparison with ZTF\,J1901+1458}
\label{sec:comparison}
ZTF\,J2008+4449 is not the only merger remnant to display the signature of circumstellar material in the absence of a companion star or brown dwarf. In a companion paper \citep{desai2025}, we show that another merger remnant, ZTF\,J1901+1458, presents very similar characteristics to ZTF\,J2008+4449 as well the presence of circumstellar material through similar X-ray emission properties.

Just as the white dwarf treated in this work, ZTF\,J1901+1458 is massive (nearing the Chandrasekhar mass at $\approx1.3M_{\odot}$), with a radius of $\approx2600$\,km, and rapidly rotating on a very similar period of $\approx 6.9$ minutes. These characteristics, together with its strong magnetic field of $\approx800\,$MG
point towards the fact that ZTF\,J1901+1458 is also most likely a double white dwarf merger remnant \citep{caiazzo2021}.
Most significantly, both systems are sources of soft X-ray emissions, and their X-ray spectra are strikingly similar to each other: when fitted by a two-temperature optically thin plasma model, we recover similar temperatures (the soft component is $kT=0.23^{+0.04}_{-0.03}$ for ZTF\,J2008+4449 and $kT=0.23\pm0.03$ for ZTF\,J1901+1458, while the hard component is $kT=3.6^{+1.0}_{-0.7}$ for ZTF\,J2008+4449 compared to  $kT=2.9^{+1.2}_{-0.8}$ for ZTF\,J1901+1458). Finally, for both systems, we have strong constraints from infrared photometry showing the absence of a stellar or sub-stellar companion.

Although we do not have a strong constraint on the origin of the circumstellar material in the two systems, the striking similarities between ZTF\,J2008+4449 and ZTF\,J1901+1458, in terms of mass, magnetic field, spin period and X-ray emission, hint to the two system being part of the same class of merger remnants interacting with circumstellar material. However, the two systems show some differences. First, the X-ray luminosity is $\approx100$ times larger in ZTF\,J2008+4449 ($L_{X}=(2.3\pm0.4)\times10^{29}$\,erg/s) as compared to that in ZTF\,J1901+1458 ($L_X=1.28^{+0.18}_{-0.15}\times10^{27}$\,erg/s). If the X-ray emission in both systems is at least partly powered by accretion, this could imply a larger mass influx rate in the former. 
Also, ZTF\,J1901+1458 does not display the H$\alpha$ emission seen in the optical spectrum of ZTF\,J2008+4449 and we do not detect a significant spindown, although our observational constraints for the $\dot{P}$ in ZTF\,J1901+1458 are much worse than for ZTF\,J2008+4449, and we can only derive a limit of $\dot{P}<10^{-11}$\,s/s.
These differences could be due to the difference in age between the two systems: the cooling age of ZTF\,J1901+1458, estimated through SED fitting, is $490\pm10$~Myr, about ten times older than ZTF\,J2008+4449. If, for example, the X-ray activity is due to fallback material from the merger, it would be expected for the accretion rate to diminish with time (although it is still to be understood if fallback accretion can happen at such late times). In the case of a mass loss from the surface of the white dwarf, on the other hand, the higher surface gravity of  ZTF\,J1901+1458 might be the reason for a weaker wind.

\section{Summary and outlook}
\label{sec:summary}

We have presented our discovery and characterisation of the hot, massive, and highly magnetized variable white dwarf ZTF\,J2008+4449, offering a description of the data, its analysis, and an interpretation of the physical properties of the white dwarf and surrounding material. In Table\,\ref{tab:summary}, we list the parameters that we have derived for the white dwarf and we summarize our results here:
\begin{itemize}
    \renewcommand\labelitemi{$\bullet$}
    \item The exceptionally high magnetic field ($\approx400-600\,$MG), short rotation period ($P\approx6.6$\,minutes), and large mass ($1.12\pm0.03M_{\odot}$) of the isolated white dwarf ZTF\,J2008+4449 classify it as a likely
    double white dwarf merger remnant.\vspace{1pt}
    \item ZTF\,J2008+4449 shows spectral and photometric variability across a broad range of wavelengths, from the optical to the FUV (and possibly X-rays), on its 6.6 minutes rotation period. Although difficult to find a definitive explanation for all of the variable features, their ensemble points towards the possibility of cyclotron emission, complex magnetic field geometry, and possibly the variation of chemical abundances across the surface of the white dwarf.
    \item The significant spin down detected from the optical light curve, $\dot{P} = (1.80\pm0.09)\times10^{-12}$s/s, along with the detection of Balmer emission lines and X-rays, indicate the presence of circumstellar material in close proximity and interacting with the white dwarf and its magnetosphere.
    \item The Doppler variation seen in the Balmer emission lines is consistent with that of a toroidal half-arc of ionised gas in corotation with the white dwarf at a distance of $\approx 20\,R_{\rm WD}$ and a radial extent of $\approx 15\,R_{\rm WD}$, implying the trapping of circumstellar gas in the magnetosphere at a distance larger than the Keplerian corotation radius.
    \item The detected spin-down can be explained by some material being ejected from the system, carrying away angular momentum from the star. We find that, to explain the spindown of the white dwarf, the massloss is likely higher than $\dot{M}>3\times10^{-14}$\,M$_\odot$/yr. The observed X-rays could be associated with the ejection region, being produced by shocks or magnetic reconnection. Alternatively, if a fraction of the infalling material manages to reach the surface of the white dwarf, the X-rays could be due to accretion onto the magnetic poles of the white dwarf.
    \item Infrared photometry excludes the presence of a Roche lobe filling brown dwarf companion as the source of the material through mass transfer.
    \item We identify three possible sources for the circumstellar material: the tidal disruption of a planetary object (a comet, asteroid or small planet) scattered at high eccentricity toward the white dwarf, fallback accretion of gravitationally bound ejecta from the merger event or a wind from the surface of the white dwarf. 
    \item Although Gaia astrometry of ZTF\,J2008+4449 suggests that the white dwarf might be a member of the young open cluster RSG 5 \citep{Miller2025,2022AJ....164..215B}, we exclude this possibility because our re-derived age (Section~\ref{sec:cluster}) shows that the cluster cannot be substantially older than 50~Myr, which is comparable to the cooling age of the white dwarf ($60\pm10$ Myr) and therefore likely much smaller than its total age.
\end{itemize}

The discovery of circumstellar material in two white dwarf merger remnants (ZTF\,J2008+4449 and ZTF\,J1901+1458) potentially opens 
a novel avenue towards a deeper understanding of compact binary evolution and the physics of compact mergers. Additionally, their study could also provide significant insight into accretion phenomena onto highly magnetic objects, given the likely complex nature of the interaction between the intense magnetic fields of the white dwarfs and their circumstellar material. Lastly, the striking similarities between these two uniquely behaving systems raise the possibility of their mutual membership in a new class of compact merger remnants interacting with circumstellar material. 

A further in depth analysis of both systems is required to confirm their nature and fully determine their physical properties. As such, we intend to conduct a more detailed photoionization study of the half ring of material glowing in H$\alpha$ in ZTF\,J2008+4449, to produce a physically accurate model of the spatial geometry of the emitting region of ionised gas and to constrain its composition. Understanding the geometry and properties of the emission region will be a crucial step in constraining the ongoing accretion processes and relating them to the geometry of the magnetic field, the spin decay and the X-ray emission of ZTF\,J2008+4449. 
A key aspect will be a better determination of the magnetic field strength on the surface of the white dwarf. The calculations reported in this work for the transitions of metals in high magnetic fields show that the observed absorption lines in the UV are likely stationary Zeeman components of elements like carbon, nitrogen and silicon (see Fig.\,\ref{fig:metal_lines}); however, more detailed calculations are needed to correctly identify the lines and determine the magnetic field strength.  An in depth analysis of the phase-resolved UV spectra with bespoke magnetic models of different geometries would also allow to constrain the geometry of the magnetic field, which  may be much more complex than that of a simple dipole, as hinted by the complex photometric variability (see Section\,\ref{sec:var}), and the asymmetric structure of the trapped ionised gas.

Additional observational campaigns will also be crucial in understanding this peculiar white dwarf. Deep observations in the near and mid-infrared would allow us to constrain the presence of a cold disk or cloud of material surrounding the white dwarf, or of planetary objects feeding the accretion. Radio observations would reveal possible non-thermal emission, as the pulsating synchrotron emission observed in white dwarf pulsars. At the other end of the spectrum, deeper X-rays observations could better constrain spectral properties of the white dwarf and reveal if the X-rays are variable on the 6.6 spin period, providing additional information on the emission region. 

Finally, discovering new white dwarfs like ZTF\,J2008+4449 and ZTF\,J1901+1458 will allow us to constrain the origin of these fascinating objects and understand how they evolve. It is likely that many more of these systems are waiting to be discovered in the archives of time-domain surveys like ZTF, ATLAS and TESS, and many more are bound to be discovered with the Vera C. Rubin Observatory.

\begin{acknowledgements}
      We thank Lynne Hillenbrand and Soumyadeep Bhattacharjee for helpful discussions, and Kishalay De for his help with the WIRC reduction pipeline. IC was supported by NASA through grants from the Space Telescope Science Institute, under NASA contracts NASA.22K1813, NAS5-26555 and NAS5-03127.
      TC was supported by NASA through the NASA Hubble Fellowship grant HST-HF2-51527.001-A awarded by the Space Telescope Science Institute, which is operated by the Association of Universities for Research in Astronomy, Inc., for NASA, under contract NAS5-26555.
      This project has received funding from the European Research Council (ERC) under the European Union’s Horizon 2020 research and innovation programme (Grant agreement No. 101020057).
      This work was based on observations obtained with the Samuel Oschin Telescope 48-inch and the 60-inch Telescope at the Palomar Observatory as part of the Zwicky Transient Facility project. ZTF is supported by the National Science Foundation under Grants No. AST-1440341, AST-2034437, and currently Award \#2407588. ZTF receives additional funding from the ZTF partnership. Current members include Caltech, USA; Caltech/IPAC, USA; University of Maryland, USA; University of California, Berkeley, USA; University of Wisconsin at Milwaukee, USA; Cornell University, USA; Drexel University, USA; University of North Carolina at Chapel Hill, USA; Institute of Science and Technology, Austria; National Central University, Taiwan, and OKC, University of Stockholm, Sweden. Operations are conducted by Caltech's Optical Observatory (COO), Caltech/IPAC, and the University of Washington at Seattle, USA.
      This work has made use of data from the European Space Agency (ESA) mission {\it Gaia} (\url{https://www.cosmos.esa.int/gaia}), processed by the {\it Gaia} Data Processing and Analysis Consortium (DPAC, \url{https://www.cosmos.esa.int/web/gaia/dpac/consortium}). Funding for the DPAC has been provided by national institutions, in particular the institutions participating in the {\it Gaia} Multilateral Agreement.
        The Pan-STARRS1 Surveys (PS1) and the PS1 public science archive have been made possible through contributions by the Institute for Astronomy, the University of Hawaii, the Pan-STARRS Project Office, the Max-Planck Society and its participating institutes, the Max Planck Institute for Astronomy, Heidelberg and the Max Planck Institute for Extraterrestrial Physics, Garching, The Johns Hopkins University, Durham University, the University of Edinburgh, the Queen's University Belfast, the Harvard-Smithsonian Center for Astrophysics, the Las Cumbres Observatory Global Telescope Network Incorporated, the National Central University of Taiwan, the Space Telescope Science Institute, the National Aeronautics and Space Administration under Grant No. NNX08AR22G issued through the Planetary Science Division of the NASA Science Mission Directorate, the National Science Foundation Grant No. AST–1238877, the University of Maryland, Eotvos Lorand University (ELTE), the Los Alamos National Laboratory, and the Gordon and Betty Moore Foundation.
        This work made use of Astropy:\footnote{http://www.astropy.org} a community-developed core Python package and an ecosystem of tools and resources for astronomy \citep{astropy:2013, astropy:2018, astropy:2022}
\end{acknowledgements}

%

\bibliographystyle{aa}
\bibliography{sample631}{}

\begin{appendix}
\onecolumn

\section{Zeeman shifted metal lines in the UV spectrum}
\label{sec:metalsUV}

As described in Section\,\ref{sec:UV}, the UV spectrum shows a suite of lines which we do not identify with either Hydrogen or Helium transitions in high magnetic fields. These lines are shaded in red in the topmost panel of Fig.\,\ref{fig:uv-mosaic}, which we show once more in Fig.\,\ref{fig:metal_lines}. We thus consider the possibility that these absorption lines are produced by the presence of light metallic elements in the atmosphere of the white dwarf. In a first attempt to constrain the ionic species present which could contribute to the lines' formation and subsequently place limits on the surface magnetic field strength, we have investigated the behaviour of the most frequently present light element transitions in hot dwarf novae with temperatures similar to that of our white dwarf. For this, we have compiled a list of transitions seen in such hot ($\approx\,30000$\,K) dwarf novae \citep[see e.g.,][]{godon2022} in the approximate wavelength range which contains the aforementioned unidentified absorption lines in our spectrum. These transitions are compiled in Table\,\ref{table:ionic_species} along with their rest wavelengths and belong to carbon, nitrogen, and silicon.

\begin{table}[h!]
\centering
\begin{tabular}{c c} 
 \hline
 Ionic species & Transition rest wavelengths [\AA]\\ 
 \hline\hline
 N V & 1240 [\AA]\\ 
 \hline\hline
 C II & 1335 [\AA]\\ 
 \hline
 C III & 977, 1175 [\AA]\\
 \hline
 C IV & 1549 [\AA]\\
 \hline\hline
 Si II & 1190, 1260, 1249, 1304 [\AA]\\
 \hline
 Si III & 1206, 1298 [\AA]\\
 \hline
\end{tabular}
\caption{Metal transitions with zero-field rest wavelengths in the UV for which we calculated the transition energy as a function of magnetic field strength.}
\label{table:ionic_species}
\end{table}

We then proceeded to calculate the transition energy of the Zeeman components of our compiled lines as a function of magnetic field from perturbation theory, including linear and quadratic operators in the Paschen-Back regime \citep{1975Ap&SS..36..459K,2024MNRAS.527.3122V}. The calculations assume $<r^2>$ expectation values, or mean square radii \citep{1981tass.book.....C}, involved in the calculation of the quadratic term for individual levels in a zero-field environment \citep[see ][]{2024MNRAS.527.3122V}. Self consistent calculations of atomic wave functions and $<r^2>$ expectation values at large magnetic fields should be performed to improve these results. Our current results are shown in Fig.\,\ref{fig:metal_lines} for line components with relatively significant oscillator strengths at a field of 500 MG. However, we remark that these results may not be fully accurate at the highest of fields, as a result of the perturbative methods utilised, which may break down at high magnetic field values. Unfortunately, more accurate calculations at high fields \citep[see ][]{2023MNRAS.520.3560H} are not currently available for these line transitions.

Our calculations show that the $\pi$ Zeeman components of each line (the transitions with no change in the $m$ quantum number, $\Delta m=0$) do not deviate far from their zero-field  wavelength, and they all cluster in the UV wavelength range  where we detect the broad absorption lines. The other Zeeman components, on the other hand, vary rapidly with field and, thus, could be much more strongly affected by magnetic broadening, making them harder to detect. As such, it is plausible that the observed features are produced by some of the central components in these transitions, indicating that light metals could indeed be present at the surface of the white dwarf, in particular carbon and silicon.

\clearpage

\section{XMM data and data statistics tables}
\label{sec:XMM_tables}

\begin{table}[h!]
\def\arraystretch{1.1}
   \centering
   \resizebox{\columnwidth}{!}{
   \begin{tabular}{|c|c|c|c|c|c|}
   \hline
      Instrument & Mode & Filter & Start & End & Duration(s) \\
      \hline 
      \multicolumn{6}{|c|}{EPIC (ObsID 0952990101)} \\
      \hline
      MOS1 & Full Frame & MEDIUM & 2025-05-06 22:07:19 & 2025-05-07 06:32:16 & 30297 \\
      MOS2 & Full Frame & MEDIUM & 2025-05-06 22:07:39 & 2025-05-07 06:31:49 & 30250 \\
      pn   & Full Frame & THIN1  & 2025-05-06 22:32:58 & 2025-05-07 06:28:29 & 28531 \\
      \hline
      \multicolumn{6}{|c|}{OM (ObsID 0952990101, Filter: U, 7 exposures)} \\
      \hline
      OM & Image + Fast & U & 2025-05-06 22:15:39 & 2025-05-07 06:20:59 & 26376 \\
      \hline
      \multicolumn{6}{|c|}{EPIC (ObsID 0952990201)} \\
      \hline
      MOS1 & Full Frame & MEDIUM & 2025-05-08 21:46:41 & 2025-05-09 06:25:13 & 31112 \\
      MOS2 & Full Frame & MEDIUM & 2025-05-08 21:47:03 & 2025-05-09 06:24:59 & 31076 \\
      pn   & Full Frame & THIN1  & 2025-05-08 22:12:22 & 2025-05-09 06:20:03 & 29261 \\
      \hline
      \multicolumn{6}{|c|}{OM (ObsID 0952990201, Filter: UVM2)} \\
      \hline
      OM & Image + Fast & UVM2 & 2025-05-08 21:55:01 & 2025-05-09 06:12:42 & 27718 \\
      \hline
   \end{tabular}
   }
   \caption{
      Table of \textit{XMM-Newton} (EPIC, OM, and RGS; ObsIDs 0952990101, 0952990201; PI Caiazzo) observations of ZTFJ2008+4449.\newline
      \textbf{EPIC:} \newline
      MOS1 total: \textbf{30297 + 31112 = 61409}s; MOS2 total: \textbf{30250 + 31076 = 61326}s; pn total: \textbf{28531 + 29261 = 57792}s.\newline
      \textbf{OM:} U filter (0952990101): \textbf{26376}s; UVM2 filter (0952990201): \textbf{27718}s.\newline
      \textbf{RGS:} RGS1: \textbf{30627}s; RGS2: \textbf{30617}s.
   }
   \label{tab:XMMobs}
\end{table}

\begin{table}[h!]
\def\arraystretch{0.8}
\centering
\caption{Event counts and statistics for all XMM instruments for the first observation, second observation, and both observations combined. The source (src) counts are those measured in a 20 arcsec source aperture centred on the Gaia position of the target. The expected background counts are the measured background counts in the 80 arcsec aperture, scaled to the source aperture size. The 90\% confidence limits (CL$_{\rm 90}$) on the count rates are shown, with associated detection significance.}
\begin{tabular}{l l l r r r r r}
\hline
Instr. & Obs & Band & src & bkg & \multicolumn{2}{c}{CL90} & Sig. \\
   &     & [keV] & [cts] & [exp. cts] & low & high & \multicolumn{1}{c}{$\sigma$} \\
\hline
\multirow{15}{*}{PN} & Obs 1 & 0.2--0.5 & 33 & 18.2 & 24 & 43 & 3.0 \\
 &  & 0.5--2.0 & 104 & 33.1 & 88 & 121 & 9.8 \\
 &  & 2.0--4.5 & 22 & 8.6 & 15 & 30 & 3.7 \\
 &  & 4.5--7.5 & 8 & 7.2 & 4 & 13 & 0.2 \\
 &  & 0.2--10.0 & 169 & 72.9 & 148 & 191 & 9.6 \\
\cline{2-8}
 & Obs 2 & 0.2--0.5 & 36 & 19.9 & 26 & 46 & 3.2 \\
 &  & 0.5--2.0 & 110 & 40.3 & 93 & 128 & 9.0 \\
 &  & 2.0--4.5 & 16 & 12.4 & 10 & 23 & 0.9 \\
 &  & 4.5--7.5 & 11 & 9.8 & 6 & 17 & 0.3 \\
 &  & 0.2--10.0 & 177 & 88.4 & 155 & 199 & 8.3 \\
\cline{2-8}
 & Combined & 0.2--0.5 & 69 & 38.1 & 56 & 83 & 4.4 \\
 &  & 0.5--2.0 & 214 & 73.4 & 190 & 238 & 13.3 \\
 &  & 2.0--4.5 & 38 & 21.1 & 28 & 48 & 3.3 \\
 &  & 4.5--7.5 & 19 & 17.1 & 12 & 26 & 0.4 \\
 &  & 0.2--10.0 & 346 & 161.2 & 316 & 377 & 12.6 \\
\hline
\multirow{15}{*}{M1} & Obs 1 & 0.2--0.5 & 7 & 2.7 & 3 & 12 & 2.1 \\
 &  & 0.5--2.0 & 42 & 9.4 & 32 & 53 & 7.7 \\
 &  & 2.0--4.5 & 6 & 3.0 & 2 & 10 & 1.4 \\
 &  & 4.5--7.5 & 2 & 2.4 & 0 & 5 & -- \\
 &  & 0.2--10.0 & 59 & 18.7 & 47 & 72 & 7.4 \\
\cline{2-8}
 & Obs 2 & 0.2--0.5 & 8 & 4.8 & 4 & 13 & 1.2 \\
 &  & 0.5--2.0 & 44 & 14.6 & 33 & 55 & 6.1 \\
 &  & 2.0--4.5 & 8 & 3.8 & 4 & 13 & 1.7 \\
 &  & 4.5--7.5 & 4 & 3.1 & 1 & 8 & 0.3 \\
 &  & 0.2--10.0 & 66 & 28.6 & 53 & 80 & 5.9 \\
\cline{2-8}
 & Combined & 0.2--0.5 & 15 & 7.4 & 9 & 22 & 2.3 \\
 &  & 0.5--2.0 & 86 & 24.0 & 71 & 102 & 9.7 \\
 &  & 2.0--4.5 & 14 & 6.8 & 8 & 20 & 2.3 \\
 &  & 4.5--7.5 & 6 & 5.5 & 2 & 10 & 0.1 \\
 &  & 0.2--10.0 & 125 & 47.2 & 107 & 144 & 9.3 \\
\hline
\multirow{15}{*}{M2} & Obs 1 & 0.2--0.5 & 8 & 1.5 & 4 & 13 & 3.6 \\
 &  & 0.5--2.0 & 35 & 8.7 & 26 & 45 & 6.6 \\
 &  & 2.0--4.5 & 2 & 2.3 & 0 & 5 & -- \\
 &  & 4.5--7.5 & 2 & 1.8 & 0 & 5 & -- \\
 &  & 0.2--10.0 & 48 & 15.2 & 37 & 60 & 6.6 \\
\cline{2-8}
 & Obs 2 & 0.2--0.5 & 11 & 3.6 & 6 & 17 & 3.0 \\
 &  & 0.5--2.0 & 38 & 11.9 & 28 & 48 & 5.9 \\
 &  & 2.0--4.5 & 9 & 3.9 & 4 & 14 & 2.1 \\
 &  & 4.5--7.5 & 6 & 3.0 & 2 & 10 & 1.4 \\
 &  & 0.2--10.0 & 66 & 24.1 & 53 & 80 & 7.0 \\
\cline{2-8}
 & Combined & 0.2--0.5 & 19 & 5.1 & 12 & 26 & 4.6 \\
 &  & 0.5--2.0 & 73 & 20.6 & 59 & 87 & 8.9 \\
 &  & 2.0--4.5 & 11 & 6.2 & 6 & 17 & 1.6 \\
 &  & 4.5--7.5 & 8 & 4.8 & 4 & 13 & 1.2 \\
 &  & 0.2--10.0 & 114 & 39.3 & 97 & 132 & 9.6 \\
\hline
\end{tabular}
\label{tab:allinst_counts_both}
\end{table}

\clearpage

\section{Additional plots}
\label{sec:add_plots}

\begin{figure}[h!]
\centering
\includegraphics[width =0.5\columnwidth]{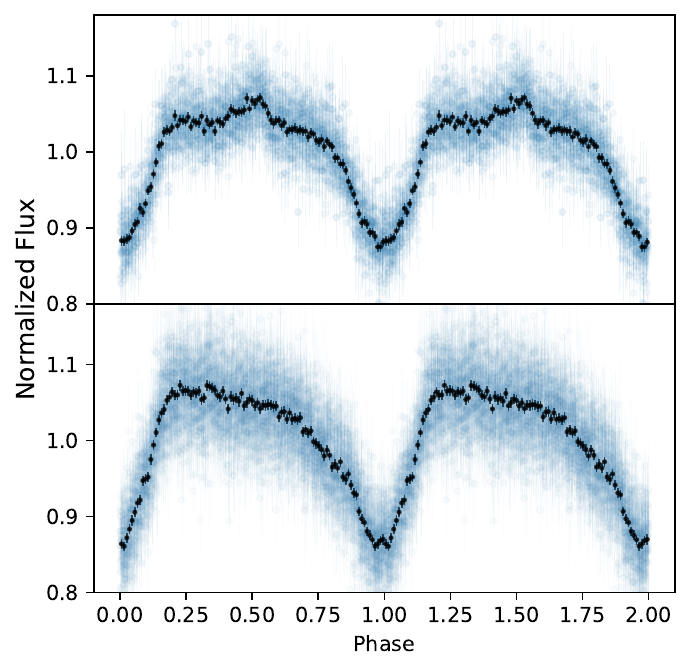}
\caption{CHIMERA light curve for ZTF\,J2008+4449. In blue, we show the unbinned data for the $g$ (upper) and $r$ (lower) filters, phase-folded using the correct period and period derivative for each epoch as derived in Section~\ref{sec:period} In black, we show a binned light curve.
}
\label{fig:lc-unbinned}
\end{figure}

\begin{figure*}[h!]
\centering
\includegraphics[width = \columnwidth]{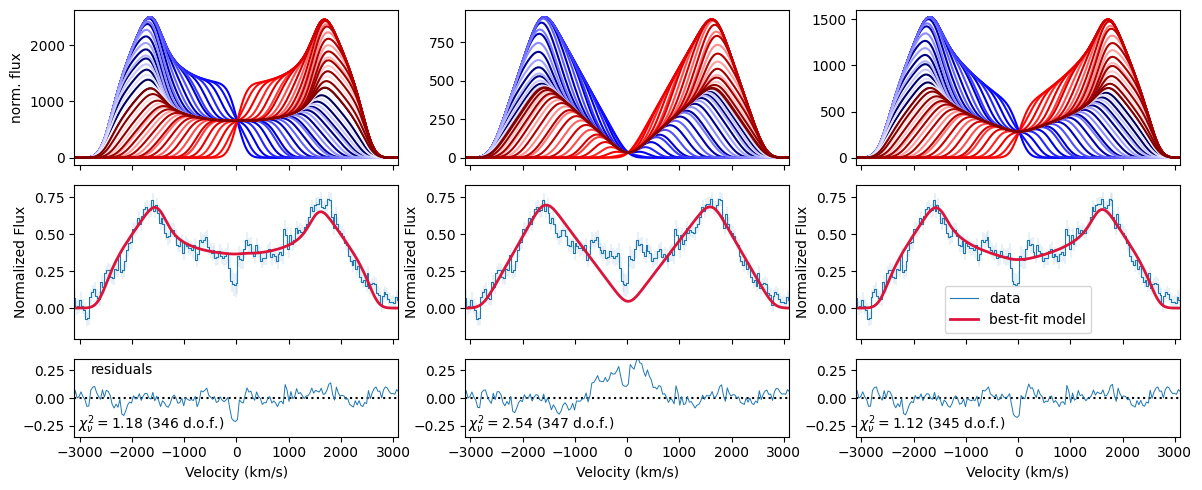}
\caption{Fit of our toy model for an half ring to the phase-averaged H$\alpha$ profile for ZTF\,J2008+4449. The left plots are for optically thin emission, the middle are for optically thick and the right ones are for a mix of optically thin and optically thick emission. The upper panels show the phase-resolved profiles for each model, the middle panels show the best fit phase-averaged compared to the data and the lower panels show the residuals and the reduced $\chi^2$ for each fit. We can see that the mixed model is a marginally better fit to the data than the optically thin model, and both are much better fits than the fully optically thick model.
}
\label{fig:hal_profile}
\end{figure*}

\clearpage

\section{Corner Plots}
\label{sec:corners}
\begin{figure}[ht!]
\centering
\vspace{10pt}
\includegraphics[width = 0.9\columnwidth]{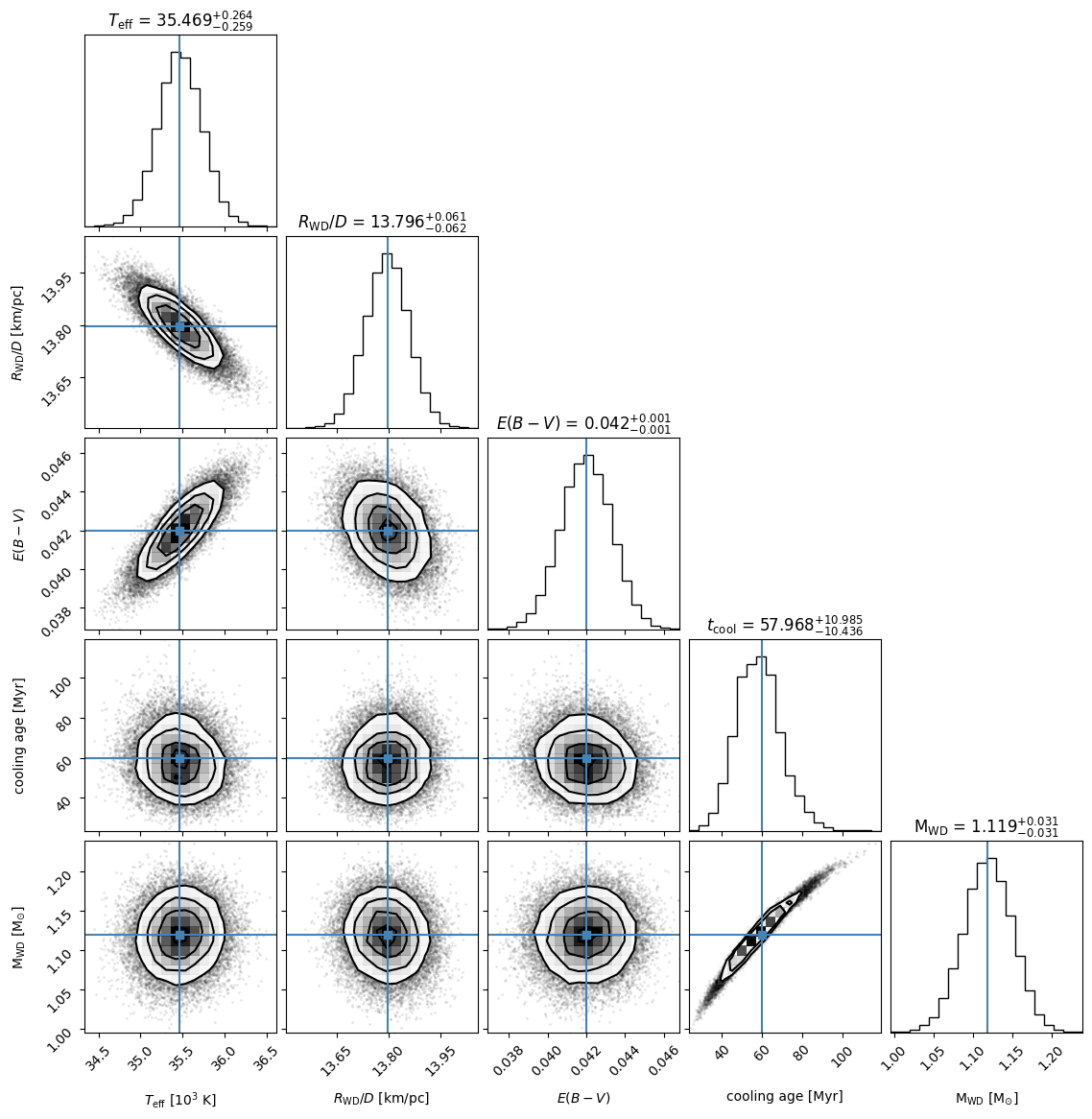}
\caption{Corner plot for the SED fitting presented in Section~\ref{sec:SED} for ZTF\,J2008+4449. As both the distance and the radius of the white dwarf act as a normalization factor on the atmosphere models, we fit for the ratio $R_*/D$, where $R_*$ is the radius of the star in km and $D$ is the distance to the star in parsecs. We then convert the value obtained to a radius for the white dwarf by employing the parallax measured by Gaia DR3 and factoring in the error on the parallax. The other two parameters, effective temperature and reddening, are fitted directly. We employed an uniform prior on all parameters.
\label{fig:corner}}
\end{figure}

\begin{figure}[ht!]
\centering
\includegraphics[width = 0.6\columnwidth]{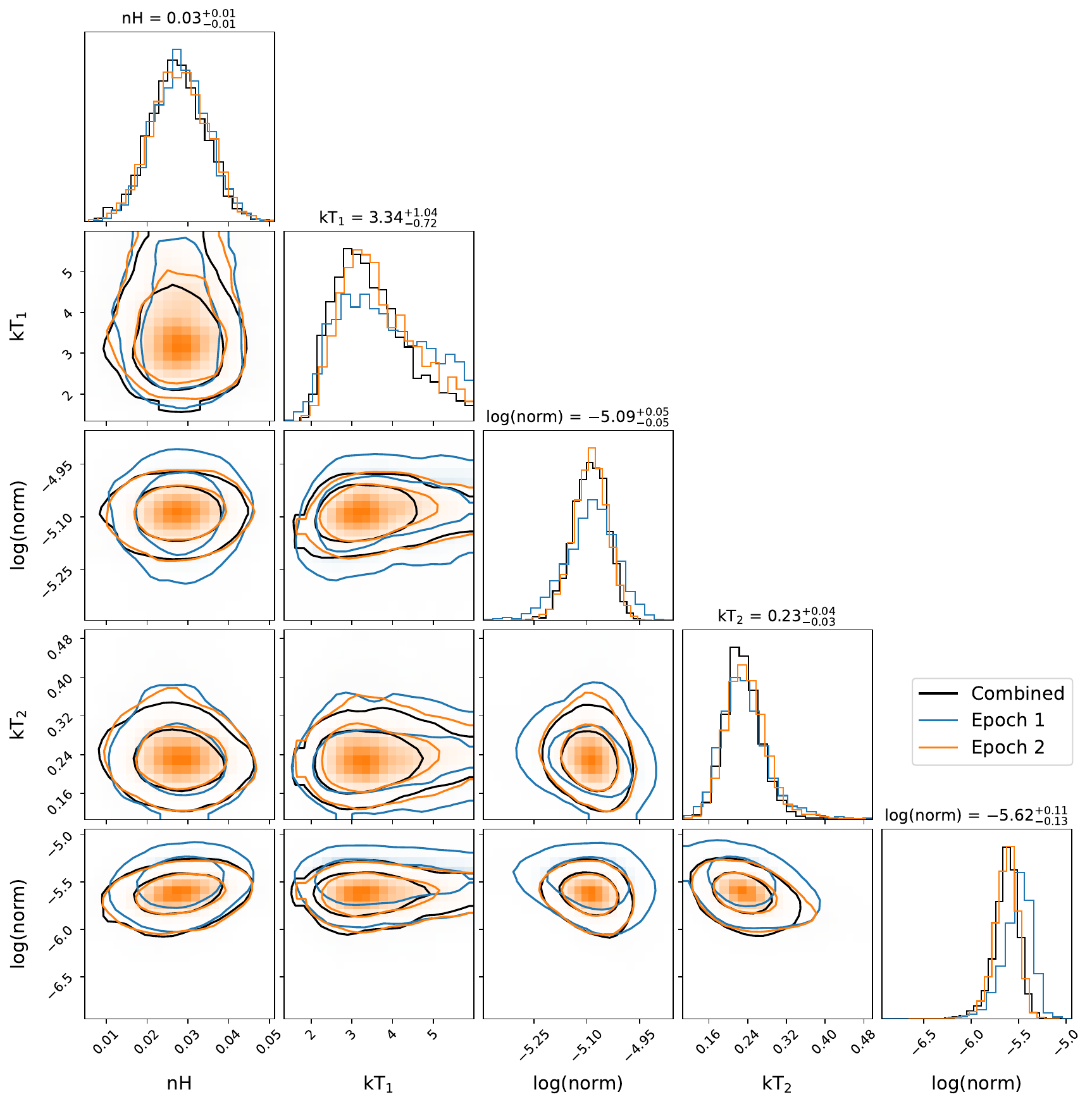} 
\caption{Corner plot for the X-ray spectral fit presented in Section~\ref{sec:xray_analysis} for a two-temperature Apec model. The different colors indicate fits on the two epochs and on the combined spectra; the data show no evidence of variability.
\label{fig:bxa-corners-apec}}
\end{figure}

\begin{figure}[tb!]
\centering
\includegraphics[width = 0.5\columnwidth]{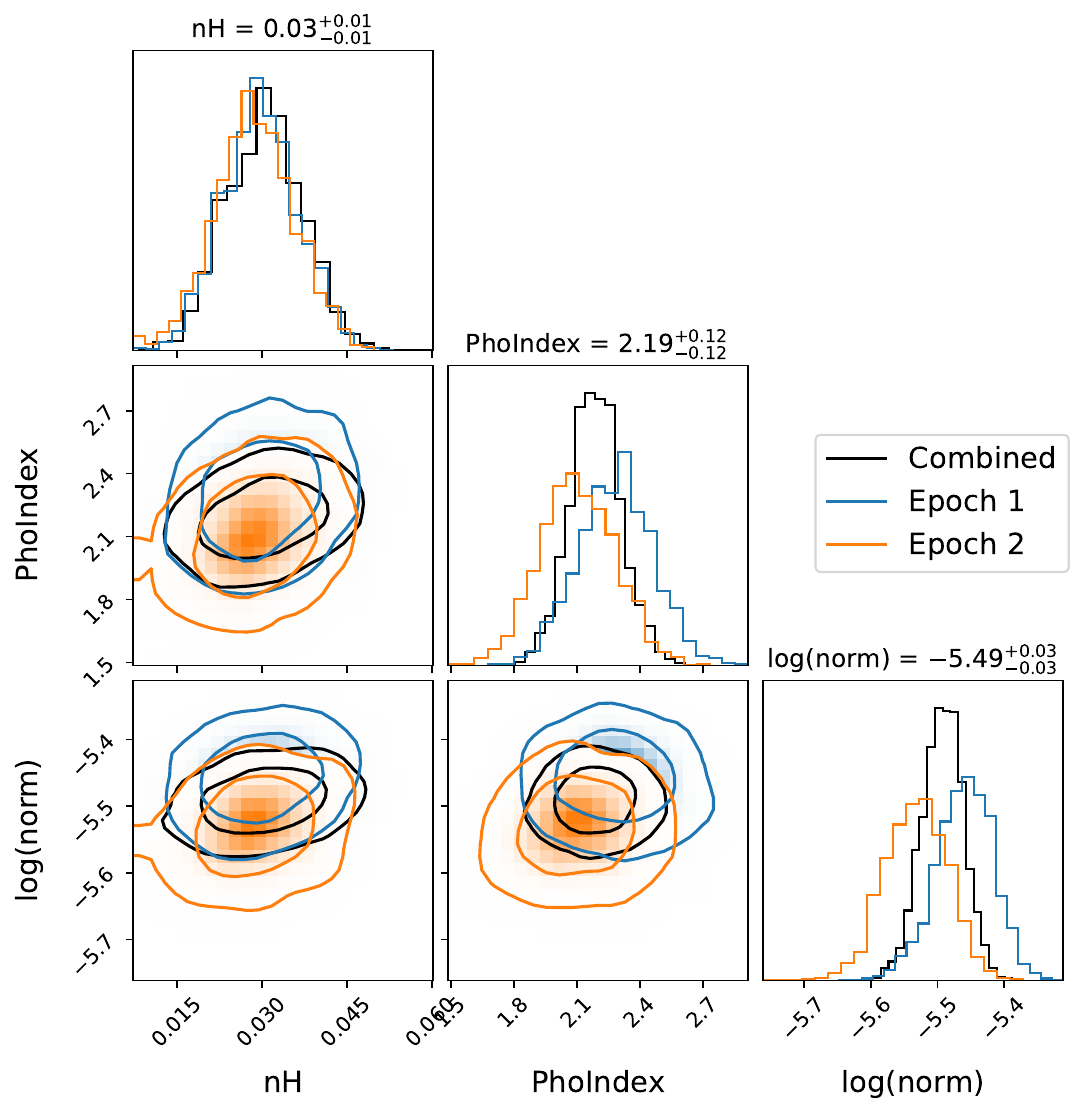} 
\caption{Corner plot for the X-ray spectral fit presented in Section~\ref{sec:xray_analysis} for a power-law model. The different colors indicate fits on the two epochs and on the combined spectra; the data show no evidence of variability.
\label{fig:bxa-corners-plaw}}
\end{figure}

\begin{figure*}[ht!]
\centering
\includegraphics[width = \textwidth]{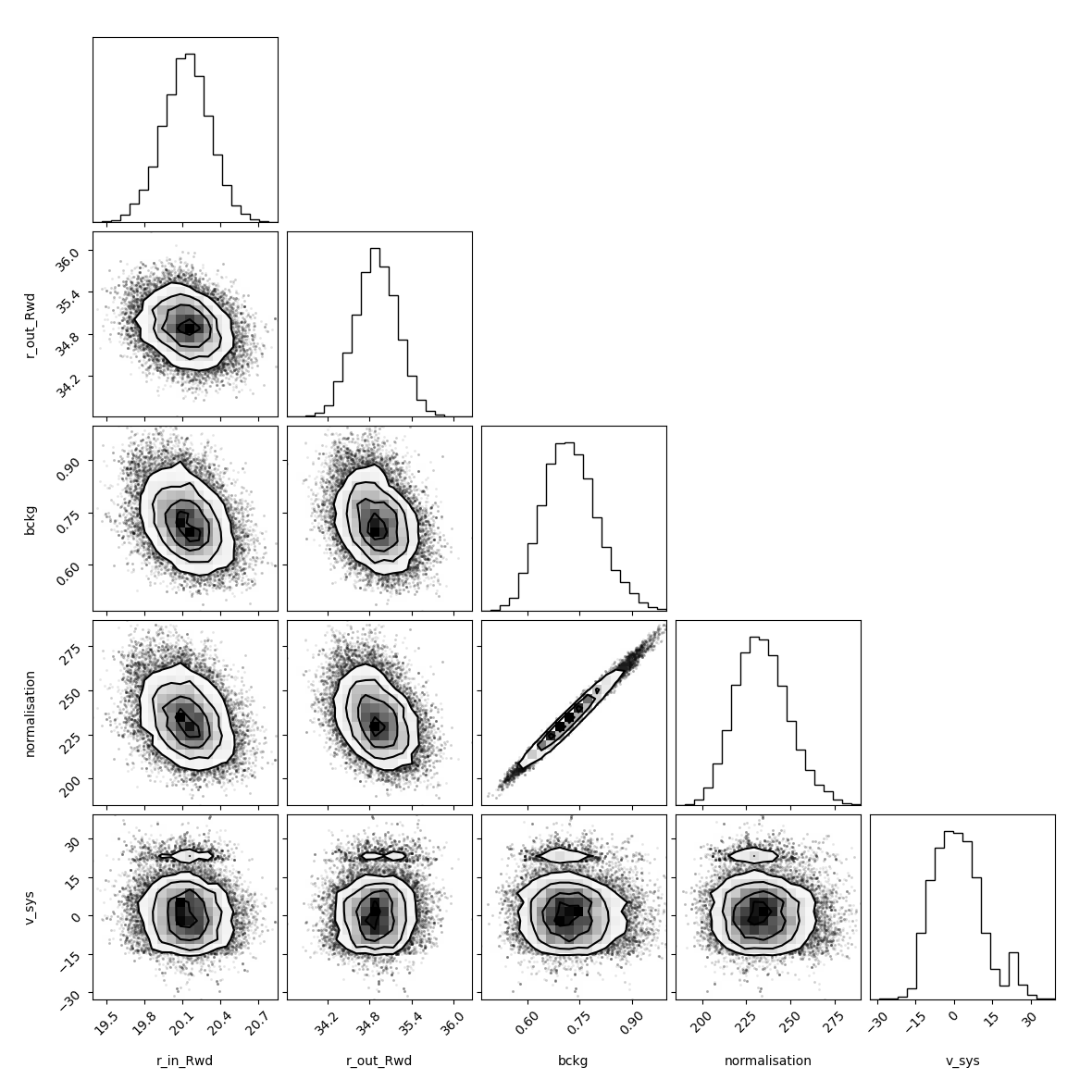}
\caption{Posterior distributions for the MCMC fit to the phase-averaged H$_\alpha$ emission profile. The fitting procedure include 5 free parameters; 1) inner ring radius, 2) outer ring radius, 3) fractional contribution from optically-thin emission, 4) absolute normalisation, and 5) system velocity. The posteriors on the inner and outer ring radii yield a constraint on the radial extent of the (partially-emitting) ring of $\delta r=14.8\pm0.4~R_{\rm WD}$. The fitting procedure favoured a contribution dominated by optically-thin emission. The system velocity was found to be consistent with zero.}
\label{fig:corner-disk-thin-sobl}
\end{figure*}

We here show the corner plots for different analysis in the paper. The first corner plot in Fig.\,\ref{fig:corner} shows the results of the SED fitting discussed in Section~\ref{sec:SED}. 
Then we show the corner plots for the X-ray analysis presented in Section~\ref{sec:xray_analysis}. The first is for the X-ray spectral fit with a two-temperature Apec model (Fig.\,\ref{fig:bxa-corners-apec}) and the second is for the power law model (Fig.\,\ref{fig:bxa-corners-plaw}). Finally, the last corner plot (Fig.\,\ref{fig:corner-disk-thin-sobl}) shows the posterior distributions for the fit of the half ring model to the H$\alpha$ emission profile presented in Section~\ref{sec:disk}.


\end{appendix}

\end{document}